\documentclass[12pt]{article}
\pdfoutput=1

\usepackage{hyperref}
\hypersetup{
  colorlinks=true,
  allcolors=darkpurple,
  pdfborder={0 0 0},
  linktocpage=false
}

\usepackage[normalem]{ulem}
\usepackage[utf8]{inputenc}
\usepackage[left=2.55cm, right=2.55cm, top=2.55cm, bottom=2.55cm]{geometry}
\usepackage{amsmath,amssymb}
\usepackage{nccmath}
\usepackage{slashed}
\usepackage{xcolor}
\usepackage{graphicx}
\usepackage{url}
\usepackage{cancel}
\usepackage{cite}
\usepackage{tabularx,booktabs}
\usepackage{multicol}
\usepackage{units}
\usepackage{xspace}
\usepackage[labelfont=bf]{caption}

\DeclareUnicodeCharacter{2192}{$\leftarrow$}
\DeclareUnicodeCharacter{3B1}{$\alpha$}
\DeclareUnicodeCharacter{3C0}{$\pi$}
\DeclareUnicodeCharacter{3B4}{$\delta$}
\DeclareUnicodeCharacter{2212}{$-$}
\DeclareUnicodeCharacter{3C3}{$\sigma$}

\usepackage{tikz}
\usepackage[compat=1.1.0]{tikz-feynman}

\usetikzlibrary{calc}
\usetikzlibrary{shapes.misc}
\tikzset{
	>=latex,
    photon/.style={decorate, decoration={snake}, draw=black, thick},
    fermionnoarrow/.style={draw=black, postaction={decorate}, thick},
    scalar/.style={draw=black, postaction={decorate}, decoration={markings,mark=at position .55 with {\arrow{>}}}, thick, dashed},
    scalarnoarrow/.style={draw=black, postaction={decorate},  thick, dashed},
    fermion/.style={draw=black, postaction={decorate},decoration={markings,mark=at position .55 with {\arrow{>}}}, thick},
    gluon/.style={decorate, draw=black, decoration={coil,amplitude=4pt, segment length=5pt}, thick},
    vertex/.style={draw,shape=circle,fill=black,minimum size=3pt,inner sep=0pt},
    fillvertex/.style={draw,shape=circle,fill=black,minimum size=5pt,inner sep=0pt},
    openvertex/.style={draw,shape=circle,minimum size=5pt,inner sep=0pt},
    blob/.style={draw=red,shape=circle,fill=red,minimum size=6pt,inner sep=0pt},
    redvertex/.style={draw=red,shape=circle,fill=red,minimum size=3pt,inner sep=0pt},
    cross/.style={cross out, draw=black,thick, minimum size=5pt, inner sep=0pt, outer sep=0pt}
}

\definecolor{darkred}{rgb}{0.6,0,0}
\definecolor{darkpurple}{rgb}{0.5,0,0.5}

\newcommand{\hc}{\mathrm{h.c.}}

\newcommand{\nn}{\nonumber}
\newcommand{\mueg}{$\mu \rightarrow e \gamma$}
\newcommand{\taueg}{$\tau \rightarrow e \gamma$}
\newcommand{\taumug}{$\tau \rightarrow \mu \gamma$}

\def\gsim{\raise0.3ex\hbox{$\;>$\kern-0.75em\raise-1.1ex\hbox{$\sim\;$}}}
\def\lsim{\raise0.3ex\hbox{$\;<$\kern-0.75em\raise-1.1ex\hbox{$\sim\;$}}}

\begin{document}

\vspace*{-2cm}
\begin{flushright}
IFIC/20-17 \\
\vspace*{2mm}
\end{flushright}

\begin{center}
\vspace*{15mm}

\vspace{1cm}
{\Large \bf 
Minimal 3-loop neutrino mass models \\ and charged lepton flavor violation
} \\
\vspace{1cm}

{\bf Ricardo Cepedello$^{\text{a}}$, Martin Hirsch$^{\text{a}}$, Paulina Rocha-Mor\'an$^{\text{b}}$, Avelino Vicente$^{\text{a,c}}$}

 \vspace*{.5cm} 
$^{\text{a}}$Instituto de F\'{\i}sica Corpuscular (CSIC-Universitat de Val\`{e}ncia), \\
 C/ Catedr\'atico Jos\'e Beltr\'an 2, E-46980 Paterna (Val\`{e}ncia), Spain

 \vspace*{.3cm} 
$^{\text{b}}$Bethe Center for Theoretical Physics and
Physikalisches Institut der Universit\"at Bonn,
Nussallee 12, D-53115 Bonn, Germany

 \vspace*{.3cm} 
$^{\text{c}}$Departament de F\'{\i}sica Te\`{o}rica, Universitat de Val\`{e}ncia, 46100 Burjassot, Spain

 \vspace*{.3cm}
 \href{mailto:ricepe@ific.uv.es}{ricepe@ific.uv.es},
 \href{mailto:mahirsch@ific.uv.es}{mahirsch@ific.uv.es},
 \href{mailto:procha@th.physik.uni-bonn.de}{procha@th.physik.uni-bonn.de},
 \href{mailto:avelino.vicente@ific.uv.es}{avelino.vicente@ific.uv.es}
\end{center}

\vspace*{10mm}
\begin{abstract}\noindent\normalsize
  We study charged lepton flavor violation for the three most popular 3-loop Majorana neutrino mass models. We call these models ``minimal'' since their particle content correspond to the minimal sets for which genuine 3-loop models can be constructed. In all the three minimal models the neutrino mass matrix is proportional to some powers of Standard Model lepton masses, providing additional suppression factors on top of the expected loop suppression. To correctly explain neutrino masses, therefore large Yukawa couplings are needed in these models. We calculate charged lepton flavor violating observables and find that the three minimal models survive the current constraints only in very narrow regions of their parameter spaces.
\end{abstract}



\newpage

\section{Introduction}
\label{sec:intro}

One could understand the smallness of the observed active neutrino
masses, in principle, if they are generated radiatively. It is
therefore not surprising that loop models of neutrino masses have a
rather long
history~\cite{Zee:1980ai,Cheng:1980qt,Zee:1985id,Babu:1988ki}.
Systematic classifications of loop models have been published for
1-loop~\cite{Bonnet:2012kz},
2-loop~\cite{Sierra:2014rxa,Cepedello:2019zqf} and, recently, even
3-loop~\cite{Cepedello:2018rfh} diagrams.  For a detailed discussion
we refer to the review~\cite{Cai:2017jrq}.

In this work, we will study how upper limits on charged lepton flavor
violating (CLFV) observables constrain 3-loop neutrino mass models. We
will focus on some particular, well-known models, which we consider
``minimal'' models. The term ``minimal'' here refers to the fact that
for models at 3-loop level at least three different {\em types} of
particles beyond the Standard Model (SM) particle content are needed,
in order to avoid lower order diagrams.~\footnote{{\em Types} of
  particles refers to the fact, that in case one of the new particles
  is a fermion, usually at least two copies (``families'') of fermions
  are needed for a realistic neutrino mass matrix.}  The three models
that we will study in this paper are the so-called
cocktail~\cite{Gustafsson:2012vj}, Krauss-Nasri-Trodden
(KNT)~\cite{Krauss:2002px} and Aoki-Kanemura-Seto
(AKS)~\cite{Aoki:2008av} models.

These three models are probably the best-known 3-loop models in the
literature, and a number of other papers have studied them (or some
variations thereof). The cocktail model, for example, has been studied
also in~\cite{Geng:2014gua}. There are also versions of the cocktail
model in which the $W$ bosons are replaced by
scalars~\cite{Kajiyama:2013lja,Hatanaka:2014tba,Alcaide:2017xoe}. For
the AKS model, one can find some discussion on phenomenology and
vacuum stability constraints
in~\cite{Aoki:2009vf,Aoki:2010aq,Aoki:2011zg}, while a variant of the
AKS model with doubly-charged vector-like fermions and a scalar
doublet with hypercharge $Y=3/2$ (plus the singlets of the AKS model)
can be found in~\cite{Okada:2015hia}.  Other variants of the AKS model
in which the exotic particles are all electroweak singlets can be
found in~\cite{Gu:2016xno,Ho:2016aye}. Finally, for the KNT model,
different phenomenological and theoretical aspects were studied
in~\cite{Cheung:2004xm,Ahriche:2014cda,Ahriche:2014oda,Ahriche:2015loa,Ahriche:2015taa,Ahriche:2015lqa,Ahriche:2014xra,Ahriche:2015wha}.
There are also variations of the KNT model, like the colored
KNT~\cite{Gu:2012tn,Nomura:2016ezz,Cheung:2016frv,Hati:2018fzc}, or a
model with vector-like fermions added to the KNT
model~\cite{Okada:2016rav}. Other variants can be found
in~\cite{Ng:2013xja,Chen:2014ska}.

Common to all the three minimal models is that their neutrino mass
diagrams are proportional to two powers of SM lepton masses. Together
with the 3-loop suppression of $1/(16 \pi^2)^3$, this results in the
prediction of rather small neutrino mass eigenvalues, unless the new
Yukawa couplings of the models take very large values.  However, in
all models off-diagonal entries for these new Yukawa couplings are
required, since neutrino oscillation experiments have measured large
neutrino angles, see for example~\cite{deSalas:2017kay} for a recent
global fit of neutrino data. Therefore, one expects that CLFV limits
will put severe constraints on these minimal models.  This simple
observation forms the motivation of the current paper.

The rest of this paper is organized as follows. In
Section~\ref{sec:notation} we will set up the notation and briefly
discuss two scalar extensions of the SM. In Section~\ref{sec:cocktail}
we will discuss the cocktail model. We will first introduce the model
and the neutrino mass generation mechanism
in~\ref{subsec:model-cocktail}, and then we will present our numerical
results for this model in~\ref{subsec:results-cocktail}. We start with
the cocktail model, since the flavor structure of the neutrino mass
matrix in this case is the simplest of the three models. We then
discuss in a similar way the KNT model in Section~\ref{sec:KNT} and
the AKS model in Section~\ref{sec:AKS}. We close with a short
discussion. A number of technical aspects on the calculation of the
loop integrals are relegated to Appendix~\ref{app:integrals}.

\section{Notation and conventions}
\label{sec:notation}

In order to make the discussion more transparent for the reader, it is
convenient to adopt a common notation and use the same conventions for
the three models considered here. This is the aim of this section. \\

\begin{table}
\centering
\begin{tabular}{| c c c c c |}
\hline  
 & generations & $\mathrm{SU(3)}_c$ & $\mathrm{SU(2)}_L$ & $\mathrm{U(1)}_Y$ \\
\hline
\hline    
$L$ & 3 & ${\bf 1}$ & ${\bf 2}$ & $-1/2$ \\
$e_R$ & 3 & ${\bf 1}$ & ${\bf 1}$ & $-1$ \\
$Q$ & 3 & ${\bf 3}$ & ${\bf 2}$ & $1/6$ \\
$u_R$ & 3 & ${\bf 3}$ & ${\bf 1}$ & $2/3$ \\
$d_R$ & 3 & ${\bf 3}$ & ${\bf 1}$ & $-1/3$ \\
\hline
\hline
\end{tabular}
\caption{SM fermions and their charges under the SM gauge group.
\label{tab:SMfermions}}
\end{table}

The three minimal 3-loop neutrino mass models studied in this work are
based on the SM gauge group, $\rm SU(3)_c \times SU(2)_L \times
U(1)_Y$. This local symmetry is supplemented by a global
$\mathbb{Z}_2$ parity, which is introduced to forbid the tree-, 1- and
2-loop contributions to the neutrino mass matrix, as explained
below. The SM fermions as well as their charges under the SM gauge
group are given in Table~\ref{tab:SMfermions}. They will all be
assumed to be even under the global $\mathbb{Z}_2$ symmetry. The
particle spectrum of the 3-loop models explored in this paper may
contain new fermions, and these will be fully specified for each model
in the next sections. In what concerns their scalar sectors, they can
be regarded as extensions of three well-known scenarios: the SM scalar
sector, the Two Higgs Doublet Model (2HDM) scalar sector and the Inert
Doublet Model (IDM) scalar sector. We now describe the scalar fields
in these three minimal scenarios, the way the electroweak symmetry
gets broken in each case and the Yukawa interactions with the SM
fermions.

\begin{itemize}
\item {\bf SM scalar sector}
\end{itemize}

\begin{table}
\centering
\begin{tabular}{| c c c c c |}
\hline  
 & generations & $\mathrm{SU(3)}_c$ & $\mathrm{SU(2)}_L$ & $\mathrm{U(1)}_Y$ \\
\hline
\hline
$H$ & 1 & ${\bf 1}$ & ${\bf 2}$ & $1/2$ \\
\hline
\hline
\end{tabular}
\caption{SM scalar sector, containing only the usual Higgs doublet $H$.
\label{tab:SMscalar}}
\end{table}

The SM scalar sector contains only the usual Higgs doublet, $H$, as
shown in Table~\ref{tab:SMscalar}. This doublet can be decomposed in
terms of its $\rm SU(2)_L$ components as
\begin{equation}
H = \left( \begin{array}{c}
H^+ \\
H^0 \end{array} \right) \, .
\end{equation}
The Yukawa couplings of the SM are
\begin{equation}  \label{eq:SMYuk}
- \mathcal L_Y^{\rm SM} = y_e \, \overline L \, H \, e_R + y_u \, \overline Q \, \widetilde H \, u_R + y_d \, \overline Q \, H \, d_R + \hc \, ,
\end{equation}
where we have defined $\widetilde H = i \tau_2 H^\ast$, with $\tau_2$
the second Pauli matrix. We have omitted flavor and $\mathrm{SU(2)}_L$
indices in the previous expression to simplify the notation. The
electroweak symmetry gets spontaneously broken by the Higgs vacuum
expectation value (VEV),
\begin{equation}
\langle H \rangle = \frac{1}{\sqrt{2}} \left( \begin{array}{c}
0 \\
v \end{array} \right) \, ,
\end{equation}
with $v \simeq 246$ GeV. In the three models discussed below, $H$ will
be even under the $\mathbb{Z}_2$ parity.

\begin{itemize}
\item {\bf 2HDM scalar sector}
\end{itemize}

\begin{table}
\centering
\begin{tabular}{| c c c c c |}
\hline  
 & generations & $\mathrm{SU(3)}_c$ & $\mathrm{SU(2)}_L$ & $\mathrm{U(1)}_Y$ \\
\hline
\hline
$\Phi_1$ & 1 & ${\bf 1}$ & ${\bf 2}$ & $1/2$ \\
$\Phi_2$ & 1 & ${\bf 1}$ & ${\bf 2}$ & $1/2$ \\
\hline
\hline
\end{tabular}
\caption{2HDM scalar sector, composed of the two $\mathrm{SU(2)}_L$ scalar doublets $\Phi_1$ and $\Phi_2$.
\label{tab:2HDMscalar}}
\end{table}

The 2HDM scalar sector is composed of two scalar doublets, $\Phi_1$
and $\Phi_2$, with identical quantum numbers under the SM gauge
symmetry, as shown in
Table~\ref{tab:2HDMscalar}. They can be decomposed in terms of their 
$\rm SU(2)_L$ components as
\begin{equation}
\Phi_1 = \left( \begin{array}{c}
\Phi_1^+ \\
\Phi_1^0 \end{array} \right) \quad , \quad \Phi_2 = \left( \begin{array}{c}
\Phi_2^+ \\
\Phi_2^0 \end{array} \right) \, .
\end{equation}
Since both scalar doublets have exactly the same quantum numbers, and
in particular since they will both be assumed to be even under the
$\mathbb{Z}_2$ symmetry, flavor changing neutral current interactions
are in principle present. This dangerous feature can be fixed by
introducing a second (softly broken) $\mathbb{Z}_2$ symmetry, under
which one of the two doublets and some of the SM fermions are
charged. There are several possibilities, and here we will just assume
that this symmetry makes $\Phi_1$ leptophilic, and $\Phi_2$
leptophobic.~\footnote{This is the choice of the authors of the AKS
  model~\cite{Aoki:2008av}, and we will stick to it although the more
  common possibilities of a type-I or type-II 2HDM are equally valid.}
Under this assumption, the 2HDM Yukawa interactions are given by
\begin{equation}  \label{eq:2HDMYuk}
- \mathcal L_Y^{\rm 2HDM} = y_e \, \overline L \, \Phi_1 \, e_R + y_u \, \overline Q \, \widetilde \Phi_2 \, u_R + y_d \, \overline Q \, \Phi_2 \, d_R + \hc \, .
\end{equation}
Again, flavor and $\mathrm{SU(2)}_L$ indices have been omitted for the
sake of clarity. We see that, as explained above, $\Phi_1$ only
couples to leptons, while $\Phi_2$ only couples to quarks. In the
2HDM, both scalar doublets are assumed to take VEVs,
\begin{equation}
\langle \Phi_1 \rangle = \frac{1}{\sqrt{2}} \left( \begin{array}{c}
0 \\
v_1 \end{array} \right) \quad , \quad \langle \Phi_2 \rangle = \frac{1}{\sqrt{2}} \left( \begin{array}{c}
0 \\
v_2 \end{array} \right) \, ,
\end{equation}
such that the usual electroweak VEV $v$ is given by
\begin{equation}
v^2 = v_1^2 + v_2^2 \, .
\end{equation}
We also define the ratio
\begin{equation}
\tan \beta = \frac{v_2}{v_1} \, .
\end{equation}

\begin{itemize}
\item {\bf IDM scalar sector}
\end{itemize}

\begin{table}
\centering
\begin{tabular}{| c c c c c c |}
\hline  
 & generations & $\mathrm{SU(3)}_c$ & $\mathrm{SU(2)}_L$ & $\mathrm{U(1)}_Y$ & $\mathbb{Z}_2$ \\
\hline
\hline
$H$ & 1 & ${\bf 1}$ & ${\bf 2}$ & $1/2$ & $+$ \\
$\eta$ & 1 & ${\bf 1}$ & ${\bf 2}$ & $1/2$ & $-$ \\
\hline
\hline
\end{tabular}
\caption{IDM scalar sector, containing the standard Higgs doublet $H$ as well as a second inert doublet $\eta$ charged under the $\mathbb{Z}_2$ parity.
\label{tab:IDMscalar}}
\end{table}

In the IDM, a second scalar doublet denoted as $\eta$ is
introduced. In contrast to the 2HDM, this doublet is odd under the
$\mathbb{Z}_2$ parity, as shown in Table~\ref{tab:IDMscalar}. The
inert doublet $\eta$ can be decomposed in terms of its
$\mathrm{SU(2)_L}$ components as
\begin{equation}
\eta = \left( \begin{array}{c}
\eta^+ \\
\eta^0 \end{array} \right) \, .
\end{equation}
Since the SM fermions are even under $\mathbb{Z}_2$, $\eta$ does not
couple to them and the IDM Yukawa interactions are exactly the same as
those in the SM, see Eq.~\eqref{eq:SMYuk}. The scalar potential of the
IDM is assumed to be such that only the SM Higgs doublet takes a VEV,
\begin{equation}
\langle H \rangle = \frac{1}{\sqrt{2}} \left( \begin{array}{c}
0 \\
v \end{array} \right) \quad , \quad \langle \eta \rangle = 0 \, .
\end{equation}
Therefore, electroweak symmetry breaking takes place in the standard
way and the $\mathbb{Z}_2$ parity remains exactly conserved. Finally,
one can split the neutral component of the $\eta$ doublet as
\begin{equation}
\eta^0 = \frac{1}{\sqrt{2}} (\eta_R + i \, \eta_I) \, ,
\end{equation}
so that $\eta_R$ and $\eta_I$ are, respectively, the real and
imaginary parts of $\eta^0$. Under the assumption of CP conservation
in the scalar sector, these two states are mass eigenstates, since the
$\mathbb{Z}_2$ symmetry forbids their mixing with the SM neutral
scalar. \\

In the next sections we will completely specify the new fields and
interactions for the three models considered here. In what concerns
the leptonic sector, some comments are in order. Neutrino oscillation
data fixes the mass squared splittings $\Delta m_{\rm Atm}^2$ and
$\Delta m_{\odot}^2$, the three leptonic mixing angles and the
so-called ``Dirac'' phase $\delta$. We will work in the basis in which
the charged lepton mass matrix is diagonal,
\begin{equation} \label{eq:m-E}
\widehat{\mathcal{M}}_e = \text{diag} \left( m_e, m_\mu, m_\tau \right) \, .
\end{equation}
This implies that the unitary matrix $U$ that brings the $3 \times 3$
Majorana neutrino mass matrix $\mathcal{M}_\nu$ to diagonal form as
\begin{equation} \label{eq:m-U}
\widehat{\mathcal{M}}_\nu = \text{diag} \left( m_{\nu_1}, m_{\nu_2}, m_{\nu_3} \right) = U^T \, \mathcal{M}_\nu \, U \, ,
\end{equation}
corresponds to the leptonic mixing matrix that enters the charged
current interactions.  We will use the PDG
parametrization~\cite{Tanabashi:2018oca} and write $U$ as
\begin{equation} \label{eq:Unu}
U = R_{23} \, R_{13} \, R_{12} \, P \, ,
\end{equation}
where $R_{ij}$ are the standard rotation matrices and $P$ is a
diagonal matrix containing the Majorana phases,
\begin{equation} \label{eq:Unu}
P = \text{diag}(1,e^{i \alpha_{12}/2},e^{i \alpha_{13}/2}) \, .
\end{equation}

\section{Cocktail model}
\label{sec:cocktail}

We begin with the so-called cocktail model, introduced
in~\cite{Gustafsson:2012vj}, since the neutrino mass matrix in this
model has the simplest flavor structure.

\subsection{The model}
\label{subsec:model-cocktail}

The cocktail model can be regarded as an extension of the IDM.  In
addition to the IDM fields, the particle content of the cocktail model
includes the two $\mathrm{SU(2)_L}$ singlet scalars $S$ and $\rho$,
singly and doubly charged.  Interestingly, the model does not have any
new fermion, just scalars. The $\eta$ and $S$ scalar fields are taken
to be odd under the $\mathbb{Z}_2$ parity, while the rest of the
fields in the model are even.\footnote{$S$ and $\eta$ need to be odd
  under the $\mathbb{Z}_2$ symmetry, in order to forbid Yukawa
  couplings with the SM leptons. These couplings would otherwise
  generate a 1-loop neutrino mass diagram, as in the Zee
  model~\cite{Zee:1980ai}.} The quantum numbers $S$ and $\rho$ are
given in Table~\ref{tab:cocktail}.  Counting also $\eta$, there are
then three new multiplets in the cocktail model, with respect to the
SM.

\begin{table}
\centering
\begin{tabular}{| c c c c c c |}
\hline  
 & generations & $\mathrm{SU(3)}_c$ & $\mathrm{SU(2)}_L$ & $\mathrm{U(1)}_Y$ & $\mathbb{Z}_2$ \\
\hline
\hline    
$S$ & 1 & ${\bf 1}$ & ${\bf 1}$ & $1$ & $-$ \\
$\rho$ & 1 & ${\bf 1}$ & ${\bf 1}$ & $2$ & $+$ \\
\hline
\hline
\end{tabular}
\caption{New particles in the cocktail model with respect to the IDM.}
\label{tab:cocktail}
\end{table}

The Lagrangian of the cocktail model contains only one additional
Yukawa term with respect to the SM,
\begin{equation} \label{eq:YukCocktail}
- \mathcal L \supset h \, \overline{e_R^c} \, e_R \, \rho + \hc \, ,
\end{equation}
where $h$ is a symmetric $3 \times 3$ matrix. Flavor indices have been
omitted in this expression for the sake of clarity. In addition, the
new scalar potential couplings are given by
\begin{align} \label{eq:PotCocktail}
\mathcal V &\supset M_{S}^2 |S|^2 + M_{\rho}^2 |\rho|^2 + M_{\eta}^2
|\eta|^2 + \frac{1}{2} \lambda_S \, |S|^4 + \frac{1}{2} \lambda_\rho
\, |\rho|^4 + \frac{1}{2} \lambda_\eta \, |\eta|^4 \nn \\ &+
\lambda_{S\rho} \, |S|^2 |\rho|^2 + \lambda_{S\eta} \, |S|^2 |\eta|^2
+ \lambda_{\rho\eta} \, |\rho|^2 |\eta|^2 \nn \\
&+ \lambda_{\rho H} |\rho|^2 |H|^2 + \lambda_{SH} |S|^2 |H|^2
+ \lambda_{\eta H}^{(1)} |\eta|^2 |H|^2
+ \lambda_{\eta H}^{(3)} H^{\dagger}\eta^{\dagger}H\eta
\nn \\
&+ \Big[ \mu_1 \,  H \eta S^\ast  + \frac 12 \mu_2  \rho \, S^\ast S^\ast  
   + \kappa  H \eta S \rho^\ast 
+ \frac{1}{2} \, \lambda_5  (H \eta^\ast)^2 + \hc \Big] \,
.
\end{align}    
We have omitted $\mathrm{SU(2)_L}$ indices to simplify the
notation. The parameters $\mu_1$ and $\mu_2$ are trilinear couplings
with dimensions of mass, $\kappa$ and all $\lambda$'s are dimensionless.
Most important is the term proportional to $\lambda_5$, see discussion
below.

The singly charged scalars $S^+$ and $\eta^+$ mix after electroweak
symmetry breaking, due to the term proportional to $\mu_1$. This leads
to two $\mathcal{H}^+_i$ mass eigenstates, with mixing angle $\beta$,
see Appendix~\ref{app:int-cocktail}. The model also includes the
doubly-charged scalar $\rho^{++}$, with mass
\begin{equation}
m_{\rho^{++}}^2 = M_{\rho}^2 + \frac{1}{2} \, \lambda_{\rho H} \, v^2 \, .
\end{equation}

The cocktail model has other interesting features that will not be
discussed in any detail here. For instance, the $\mathbb{Z}_2$ parity
of the model is conserved after electroweak symmetry breaking, so that
the lightest $\mathbb{Z}_2$-odd state is stable and can in principle
constitute a good DM candidate.

\subsubsection*{Neutrino masses}

\begin{figure}[t]
\centering
\includegraphics[width=0.6\textwidth]{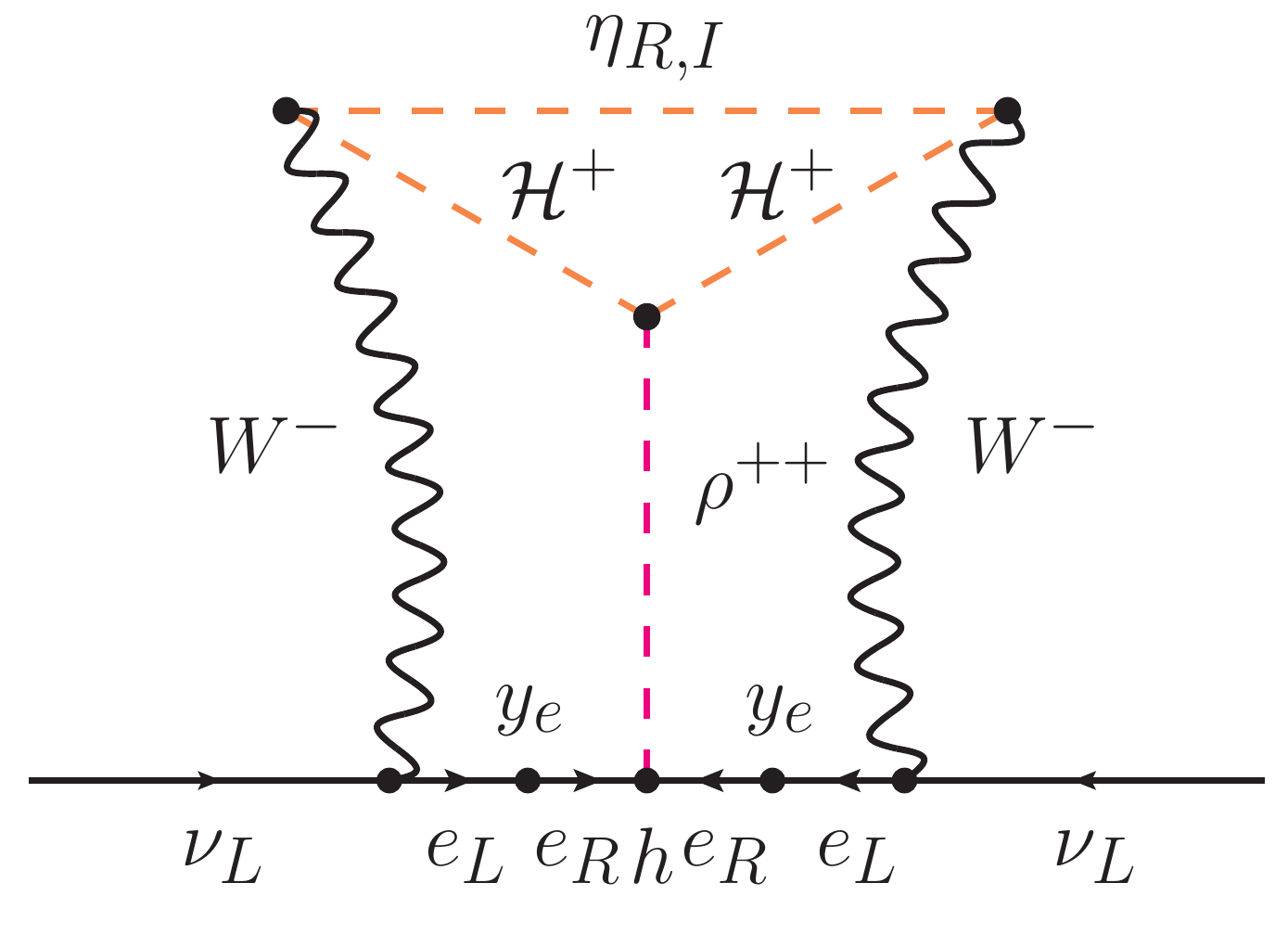}
\caption{3-loop neutrino masses in the cocktail model. The inert
  doublet $\eta$ is split into its real and imaginary parts, $\eta^0 =
  \frac{1}{\sqrt{2}} (\eta_R + i \, \eta_I)$, due to the scalar
  potential terms proportional to $\lambda_5$. $\mathcal{H}^+ \equiv
  \mathcal{H}^+_{1,2}$ represent the singly charged scalars in the
  model, obtained after diagonalizing the mass matrix of the $\left\{
  S^+ , \eta^+ \right\}$ states.
  \label{fig:cocktail}
  }
\end{figure}

The 3-loop diagram leading to neutrino masses in the cocktail model is
shown in Fig.~\ref{fig:cocktail}. In the unitary gauge this diagram is
the only diagram contributing to the neutrino mass matrix. However, in
order to understand how to maximize the contribution of this diagram
to the neutrino mass matrix, it is more useful to calculate all
diagrams in Feynman-'t Hooft gauge. This is discussed in detail in
Appendix~\ref{app:int-cocktail}.

In an analogous way to the well-known scotogenic
model~\cite{Ma:2006km}, the diagram shown in Fig.~\ref{fig:cocktail}
vanishes in the limit $m_{\eta_R}^2 - m_{\eta_I}^2 \propto \lambda_5
\to 0$, since in this limit the model conserves lepton number. We can
then write the neutrino mass matrix in the cocktail model as:
\begin{align} \label{eq:mnu cocktail}
  \left(\mathcal{M}_\nu\right)_{ij} = \frac{\lambda_5}{(16\pi^2)^3}
  \frac{m_i \, h_{ij} \, m_j}{m_{\rho^{++}}} \, F_{\rm Cocktail} \,,
\end{align}
where $m_i$ and $m_j$ are charged lepton masses. Here we have hidden
all the complexities of the calculation in the dimensionless factor
$F_{\rm Cocktail}$. This factor contains the loop integrals, depending
on the masses of the scalars, and prefactors containing coupling
constants, etc, see Appendix~\ref{app:int-cocktail}.

\subsection{Results}
\label{subsec:results-cocktail}

The cocktail model is an example of a \textit{type-II-seesaw-like}
model. In this class of models, the neutrino mass matrix is
proportional to a symmetric Yukawa matrix,
\begin{equation} \label{eq:typeII}
\mathcal{M}_\nu \sim Y \, \frac{v^2}{\Lambda} \, ,
\end{equation}
with $Y_{ij} = Y_{ji}$ and $\Lambda$ some generic mass scale. This
allows one to fit the observed neutrino masses and mixing angles in a
trivial way. Furthermore, the tight relation given in
Eq.~\eqref{eq:typeII} implies very specific predictions for ratios of
CLFV observables and strongly reduces the number of free parameters in
the model. As we will discuss now, this has very important
consequences.

From the experimental data we can reconstruct the neutrino
mass matrix in the flavor basis as
\begin{equation} \label{eq:Mnu}
{\cal M}_{\nu} = U^\ast \, \widehat{\cal M}_\nu \, U^\dagger \, .
\end{equation}
This allows us to calculate the Yukawa $h$ necessary to fit the
experimental data using the expression in Eq.~\eqref{eq:mnu
  cocktail}. We find
\begin{equation} \label{eq:hfit}
h = (16\pi^2)^3 \, \frac{m_{\rho^{++}}}{\lambda_5 \, F_{\rm Cocktail}} \,
      {\widehat{\cal M}_{e}}^{-1} \, {\cal M}_\nu \, {\widehat{\cal M}_{e}}^{-1} \, .
\end{equation}
where $\widehat{\cal M}_{e}$ is the diagonal matrix with the measured
charged lepton masses, see Eq.~\eqref{eq:m-E}.

Let us first make a rough numerical estimate. Choosing normal
hierarchy ($m_{\nu_1}\to 0$) and $\delta=0$ for simplicity and
inserting $m_{\rho^{++}} = 800$ GeV, which is roughly the current
experimental bound from LHC data
\cite{Aaboud:2017qph,CMS:2017pet,Aaboud:2018qcu}, we find
\begin{equation} \label{eq:hexa}
  h \simeq
 \begin{pmatrix}
 46000 & 450 & 5.7 \\
 450   & 8.1 & 0.36 \\
 5.7   & 0.36 & 0.026
 \end{pmatrix}
   \Big(\frac{m_{\rho^{++}}}{\rm 800\hskip1mm GeV}\Big)
   \Big(\frac{1}{\lambda_5}\Big)
   \Big(\frac{1}{F_{\rm Cocktail}}\Big).
\end{equation}
These values are obviously much too large to be realistic. We therefore
searched the parameter space, intending to identify regions, in which
$h$ can fulfill the bounds from perturbativity and lepton flavor
violation searches. This search was done in two steps.

First, we maximize $F_{\rm Cocktail}$ and $\lambda_5$. For $\lambda_5$
we use $\lambda_5=4\pi$, the largest value allowed by perturbativity.
We then scanned all free mass parameters entering in $F_{\rm
  Cocktail}$, for details see appendix
\ref{app:int-cocktail}. Generally speaking, $F_{\rm Cocktail}$ is
maximized when $\mu_1$, $\mu_2$ and $\kappa$ take the largest values
allowed, while the remaining free mass eigenvalues of the model take
the lowest possible values allowed by experimental searches. The
maximal value of $F_{\rm Cocktail}$ found in this numerical scan is
$F_{\rm Cocktail}^{\rm max} \simeq 192$. We will use this number in
all plots below. This choice is conservative in the sense that the
Yukawa couplings $h_{ij}$ will be larger for all other choices, thus
constraints from charged lepton flavor violation searches will only
be more stringent in other parts of the parameter space.

\begin{figure}[t]
\centering
\includegraphics[width=0.48\textwidth]{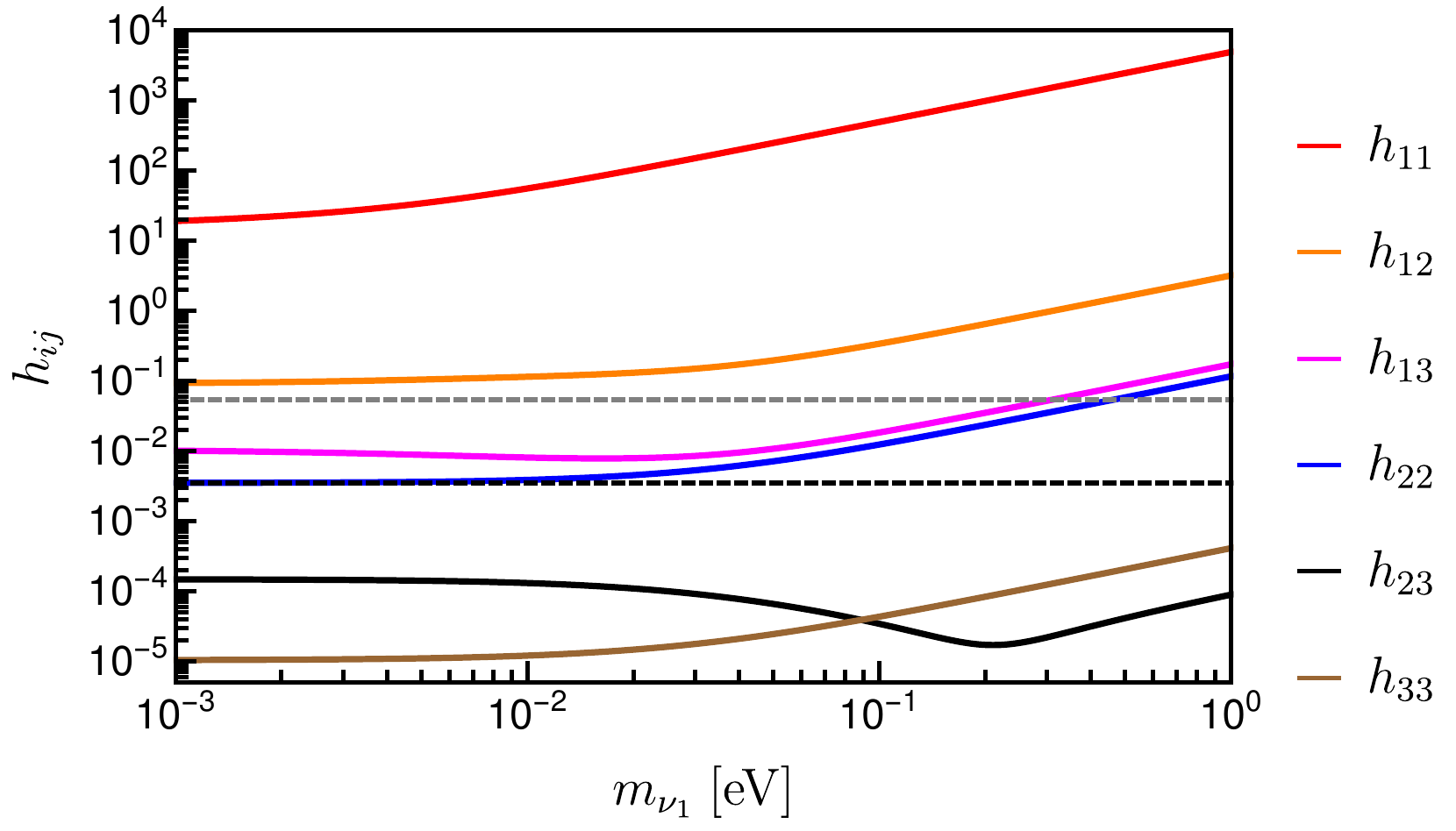}
\includegraphics[width=0.48\textwidth]{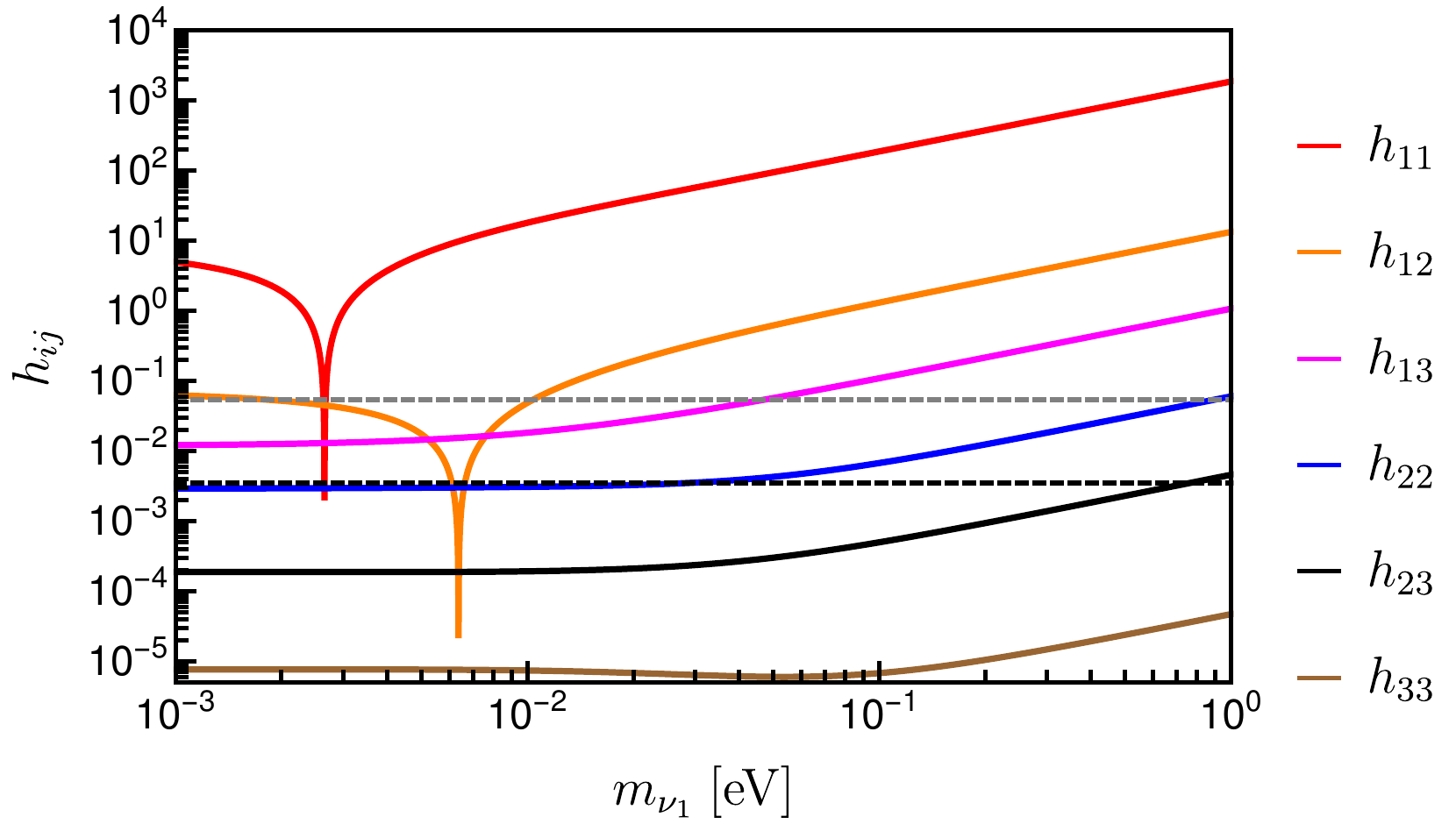}
\caption{Yukawa couplings $h_{ij}$ as function of the lightest
  neutrino mass, calculated with $F_{\rm Cocktail}^{\rm max}$. These
  Yukawas should therefore be understood as {\em lower limits}. In
  both plots we have used the best fit point data from the global
  oscillation fit~\cite{deSalas:2017kay}, except $\delta=0$. The plot
  to the left shows the case $(\alpha_{12}, \alpha_{13})=(0,0)$, the
  plot on the right $(\alpha_{12}, \alpha_{13})=(\pi,0)$.  The dashed
  gray (black) lines in the background are rough estimates for the
  typical size that the Yukawa couplings should have, in order to
  satisfy limits from muon (tau) CLFV decays.  These lines are only
  for orientation.
  \label{fig:Yuks}
  }
\end{figure}

\begin{figure}[t]
\centering
\includegraphics[width=0.55\textwidth]{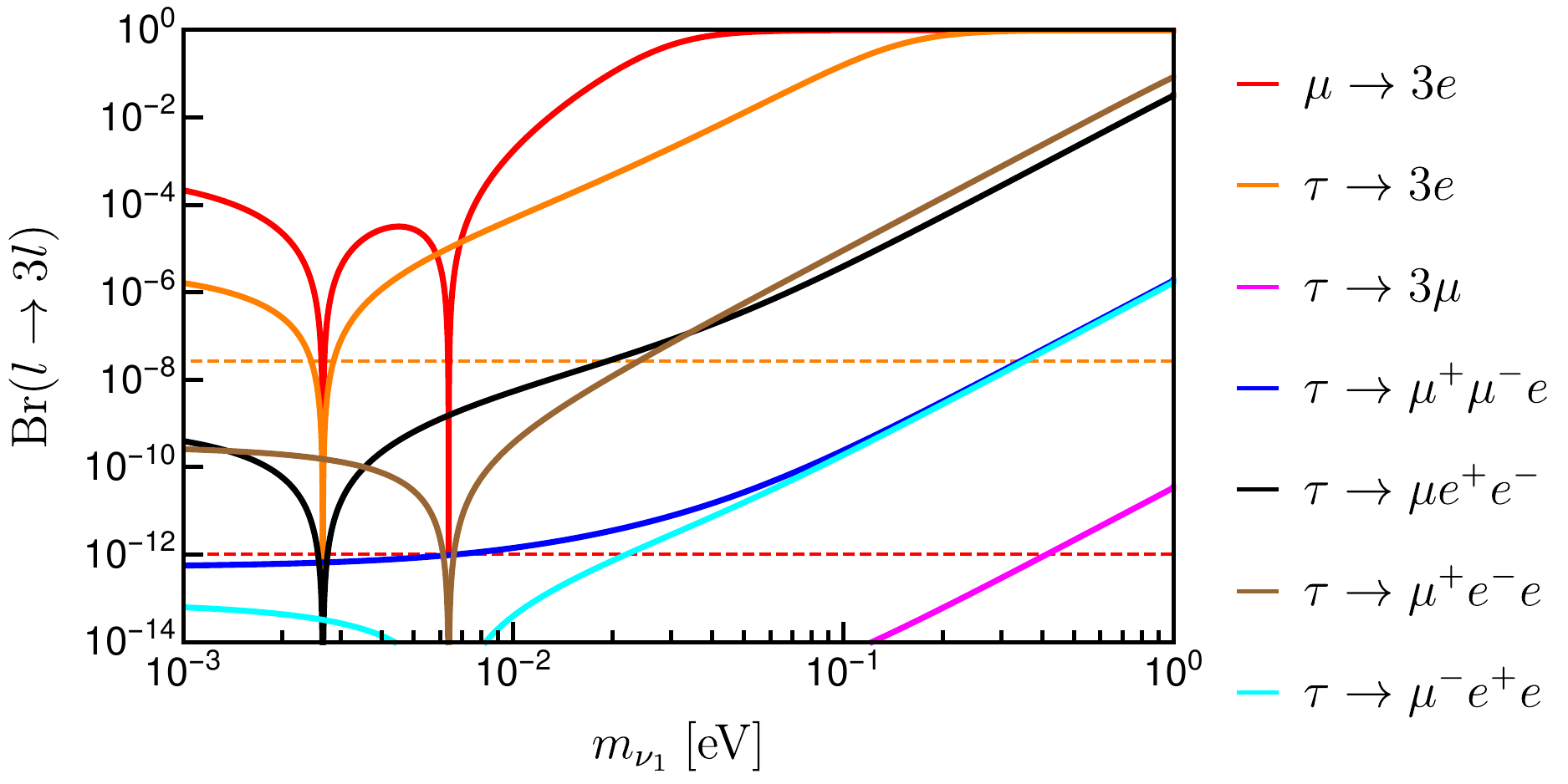}
\includegraphics[width=0.43\textwidth]{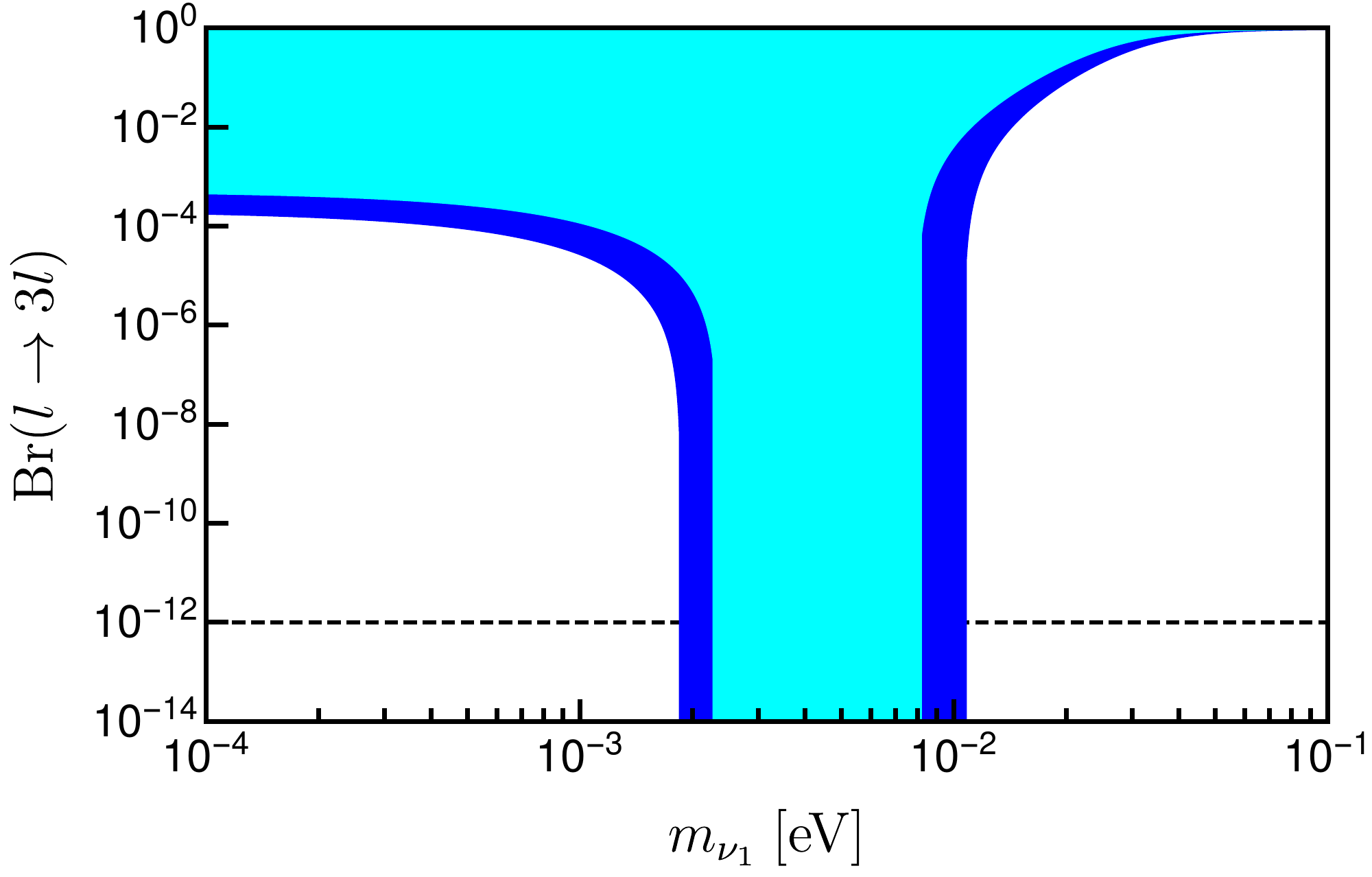}
\caption{Br($l_i \to l_j l_k l_m$) as function of the lightest
  neutrino mass. To the left, all different combinations of lepton
  generations are considered using the b.f.p. of oscillation data and
  $\alpha_{12}=\pi$ and $2\delta-\alpha_{13}=0$.  To the right,
  Br($\mu\to 3 \,e$) scanned over the uncertainty in neutrino
  oscillation data is shown.  The light and dark blue areas correspond
  to the $1 \, \sigma$ and $3 \, \sigma$ uncertainties, respectively.
  This plot scans over the Majorana phases.
  \label{fig:Br}
  }
\end{figure}

Once $F_{\rm Cocktail}^{\rm max}$ is fixed, one can scan over the free
parameters in the neutrino sector. Oscillation
data~\cite{deSalas:2017kay} fixes rather well $\Delta m^2_{\rm Atm}$,
$\Delta m^2_{\odot}$ and all three mixing angles; there is also an
indication for a non-zero value of $\delta$. This leaves us with three
essentially free parameters, the two Majorana phases and $m_{\nu_1}$,
equivalent to the overall neutrino mass scale, for which there are
only upper limits from neutrinoless double beta ($0\nu\beta\beta$)
decay~\cite{KamLAND-Zen:2016pfg,Agostini:2018tnm} and
cosmology~\cite{Aghanim:2018eyx}. We note that there is a slight
preference in the data for normal hierarchy (NH, also called normal
ordering) over inverted hierarchy (IH).

In Fig.~\ref{fig:Yuks} we plot the absolute values of the 6
independent entries in the Yukawa matrix $h$ as a function of
$m_{\nu_1}$. The oscillation data have been fixed at their best fit
point (b.f.p.) values, except $\delta=0$ for simplicity. The plot on
the left was obtained with vanishing Majorana phases, whereas the
one on the right takes $(\alpha_{12}, \alpha_{13})=(\pi,0)$. Given
that we used $F_{\rm Cocktail}^{\rm max}$ in this plot, the numerical
values of $h_{ij}$ are much smaller than in Eq.~\eqref{eq:hexa}, but
$h_{11}$ is still in the non-perturbative region everywhere in the
left plot. In the right plot, however, there are two special points,
where cancellations among different contributions of the neutrino mass
eigenstates lead to a vanishing value for either $h_{11}$ or
$h_{12}$. Such cancellations are well known in studies of
$0\nu\beta\beta$ decay. The effective Majorana mass, $m_{ee}$,
\footnote{$m_{ee}$, also sometimes called $\langle m_{\nu}\rangle$, is
  defined as $m_{ee}=\sum_j U_{ej}^2m_j$.} depends on the Majorana
phases in the same way as $h_{11}$. As in $m_{ee}$, one can therefore
not obtain a cancellation for the cases (i) NH without Majorana
phases, and (ii) IH for any choice of parameters.  The cocktail model
can therefore explain neutrino data only for normal hierarchy and some
particular combination of Majorana phases, as we are going to discuss
now in some more detail.

As Fig.~\ref{fig:Yuks} demonstrates, only in some exceptional points
can $h_{11}$ be small enough to enter the perturbative region. We
therefore scanned over $(\alpha_{12}, \alpha_{13})$ and $m_{\nu_1}$,
in the full $3 \, \sigma$ range of oscillation data. In this scan, we
calculate the CLFV observable Br($l_i \to l_j l_k l_m$), with different
combinations of lepton flavors, for the minimal value of
$m_{\rho^{++}}$ allowed by LHC data.  Fig.~\ref{fig:Br} to the left
shows Br($l_i \to l_j l_k l_m$) for all different combinations of
$i,j,k,m$ using the b.f.p. of neutrino oscillation data. The most
stringent constraint on the model comes from the experimental upper
limit on Br($\mu \to 3 \, e$)$\le 10^{-12}$~\cite{Bellgardt:1987du}.  The
plot to the right then shows the allowed regions in parameter
space, scanning over the complete range of oscillation parameters and
phases. All acceptable points lie in the range $m_{\nu_1} = (2-10)$
meV.

\begin{figure}[t]
\centering
\includegraphics[width=0.495\textwidth]{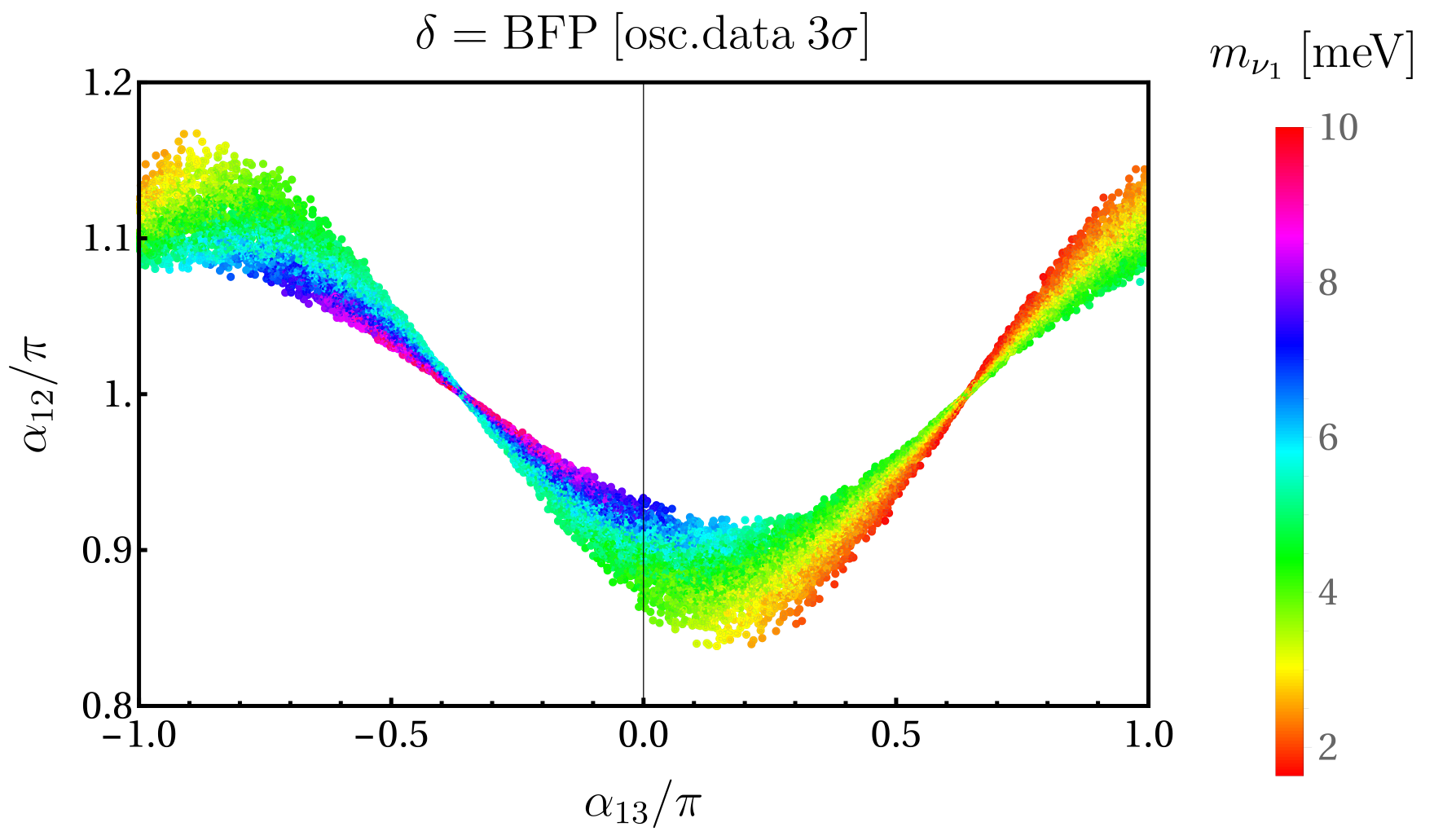}
\includegraphics[width=0.47\textwidth]{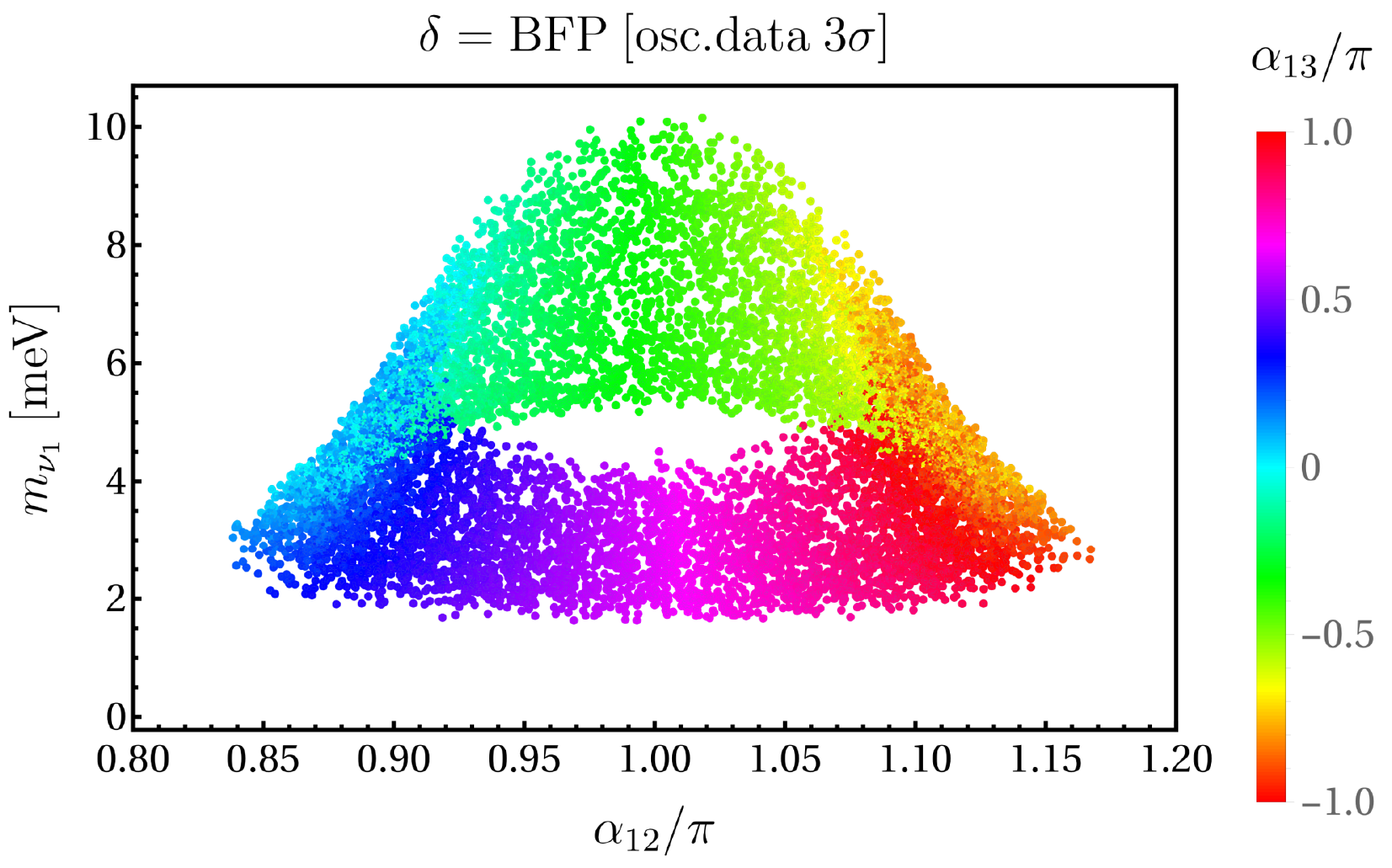}
\caption{Allowed parameter space for $\alpha_{12}$, $\alpha_{13}$ and
  $m_{\nu_1}$. Note that $m_{\nu_1}$ is shown in units of meV.
  Neutrino oscillation data was scanned over the $3 \, \sigma$
  uncertainties, except $\delta$ which is taken at its best fit value
  for simplicity.
  \label{fig:ph}
  }
\end{figure}

Fig.~\ref{fig:ph} shows a scan over the allowed range of Majorana
phases and the lightest neutrino mass. The plot to the left shows the
plane ($\alpha_{13},\alpha_{12}$), the one to the right
($\alpha_{12},m_{\nu_1}$). The model can fulfill the constraint from
Br($\mu \to 3 \, e$) only in a very narrow range of phases. In particular,
$\alpha_{12}$ has to be close to $\pi$ in all points, while also
$m_{\nu_1}$ is fixed in a rather narrow interval.

\begin{figure}[t]
\centering
\includegraphics[width=0.6\textwidth]{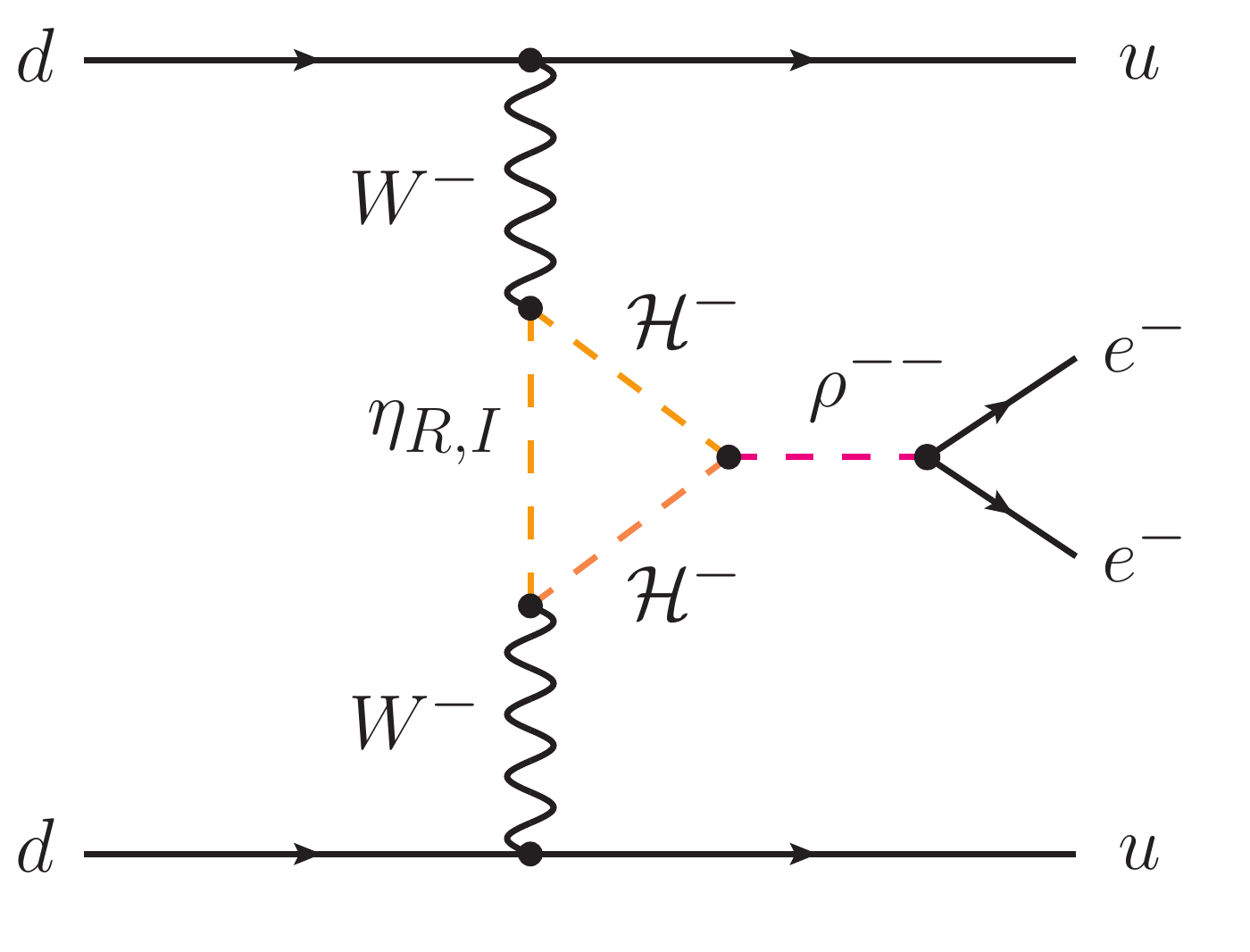}
\caption{1-loop neutrinoless double beta decay diagram in the cocktail
  model.
\label{fig:cocktail_0nbb}
}
\end{figure}

Finally we note that the acceptable points of the model lie in regions
of parameter space where the $0\nu\beta\beta$ decay observable
$m_{ee}$ is unmeasurably small. There is, however, a 1-loop
short-range diagram contributing to $0\nu\beta\beta$ decay in the
cocktail model~\cite{Gustafsson:2014vpa}, see
Fig.~\ref{fig:cocktail_0nbb}. This diagram depends on the same
parameters as the 3-loop neutrino mass diagram in
Fig.~\ref{fig:cocktail}.  In particular, note that the sum over
$\eta_{R,I}$ generates the same dependence on $\lambda_5$ as for the
neutrino mass.

We have calculated this diagram and estimated its contribution to the
$0\nu\beta\beta$ decay half-life, including the QCD running of the
short-range operator~\cite{Gonzalez:2015ady,Arbelaez:2016uto}.  Using
the same mass parameters that maximize the 3-loop diagram, in
particular $m_{\rho^{++}}=800$ GeV, the current limit on the half-life
of $^{136}$Xe~\cite{KamLAND-Zen:2016pfg} imposes a limit on $h_{11}$
of roughly $|h_{11}|\lsim 5 \times 10^{-4}$. This limit is around a
factor $\sim 7$ more stringent than the one obtained from the upper
limit on Br($\mu \to 3 \, e$).~\footnote{See~\cite{Alcaide:2017xoe}
  for a variant of the cocktail model inducing a different 1-loop
  short-range $0\nu\beta\beta$ decay diagram.}

We can conclude that the cocktail model is severely constrained from
perturbativity arguments and from searches for CLVF. The model has
acceptable points only within a narrow window of $m_{\nu_1}$ and for
particular combinations of the Majorana phases.

\section{KNT model}
\label{sec:KNT}

We continue with the KNT model~\cite{Krauss:2002px}. This was the
first radiative neutrino mass model at 3-loop order proposed.

\subsection{The model}
\label{subsec:model-KNT}

In addition to the SM particles, the KNT model contains three copies
of the fermionic singlet $N$ and two singly-charged singlet scalars
$X$ and $S$. A discrete $\mathbb{Z}_2$ symmetry is imposed, under
which $S$ and $N$ are odd and the rest of particles in the model are
even. The quantum numbers of the new particles in the KNT model are
given in Table~\ref{tab:KNT}.

\begin{table}
\centering
\begin{tabular}{| c c c c c c |}
\hline  
 & generations & $\mathrm{SU(3)}_c$ & $\mathrm{SU(2)}_L$ & $\mathrm{U(1)}_Y$ & $\mathbb{Z}_2$ \\
\hline
\hline    
$X$ & 1 & ${\bf 1}$ & ${\bf 1}$ & $1$ & $+$ \\
$S$ & 1 & ${\bf 1}$ & ${\bf 1}$ & $1$ & $-$ \\
\hline
\hline    
$N$ & 3 & ${\bf 1}$ & ${\bf 1}$ & $0$ & $-$ \\  
\hline
\hline
\end{tabular}
\caption{New particles in the KNT model with respect to the SM.}
\label{tab:KNT}
\end{table}

The Lagrangian of the model contains the following pieces
\begin{align} \label{eq:YukKNT}
  -\mathcal L &\supset f \, \overline{L^c} \, L \, X
  + g^\ast \, \overline{N^c} \, e_R \, S
  + \frac{1}{2} M_N \overline{N^c} N + \hc \, ,
\end{align}
where we have omitted $\mathrm{SU(2)_L}$ and flavor indices to
simplify the notation. We note that $f$ is an antisymmetric $3 \times
3$ Yukawa matrix, while $M_N$ is a symmetric $3 \times 3$ Majorana
mass matrix, which we take to be diagonal without loss of
generality. The scalar potential of the model also contains additional
terms besides those in the SM. These are given by
\begin{align} \label{eq:PotKNT}
\mathcal V &\supset M_{X}^2 |X|^2 + M_{S}^2 |S|^2 + \frac{1}{2} \lambda_1 \, |X|^4 + \frac{1}{2} \lambda_2 \, |S|^4 + \lambda_{12} \, |X|^2 |S|^2 \nn \\
&+ \lambda_H^{(1)} \, |H|^2 |X|^2 + \lambda_H^{(2)} \, |H|^2 |S|^2 + \frac{1}{4} \, \left[ \lambda_S \, (X S^\ast)^2 + \hc \right] \, .
\end{align}    
The presence of the $\lambda_S$ quartic coupling precludes the
definition of a conserved lepton number. Indeed, one can easily see
that the simultaneous presence of the Lagrangian terms in
Eqs.~\eqref{eq:YukKNT} and \eqref{eq:PotKNT} breaks lepton number in
two units. The masses of the physical scalar states in the KNT model
are given by
\begin{eqnarray}
m_H^2&=&\lambda \, v^2 \, , \\
m_{s_1}^2&=& M_{X}^2 + \frac{1}{2} \, \lambda_H^{(1)} \, v^2 \, , \\
m_{s_2}^2&=& M_{S}^2 + \frac{1}{2} \, \lambda_H^{(2)} \, v^2 \, .
\end{eqnarray}

We also note that the lightest $\mathbb{Z}_2$-odd state in the KNT
model is completely stable. Assuming the hierarchy $M_{N_1} <
m_{s_1}<m_{s_2}$, this state is the lightest fermion singlet, which then
constitutes a good DM candidate. In fact, the KNT model is
historically the first radiative neutrino mass theory with a stable DM
candidate running in the loop.

\subsubsection*{Neutrino masses}

\begin{figure}[t]
\centering
\includegraphics[width=0.78\textwidth]{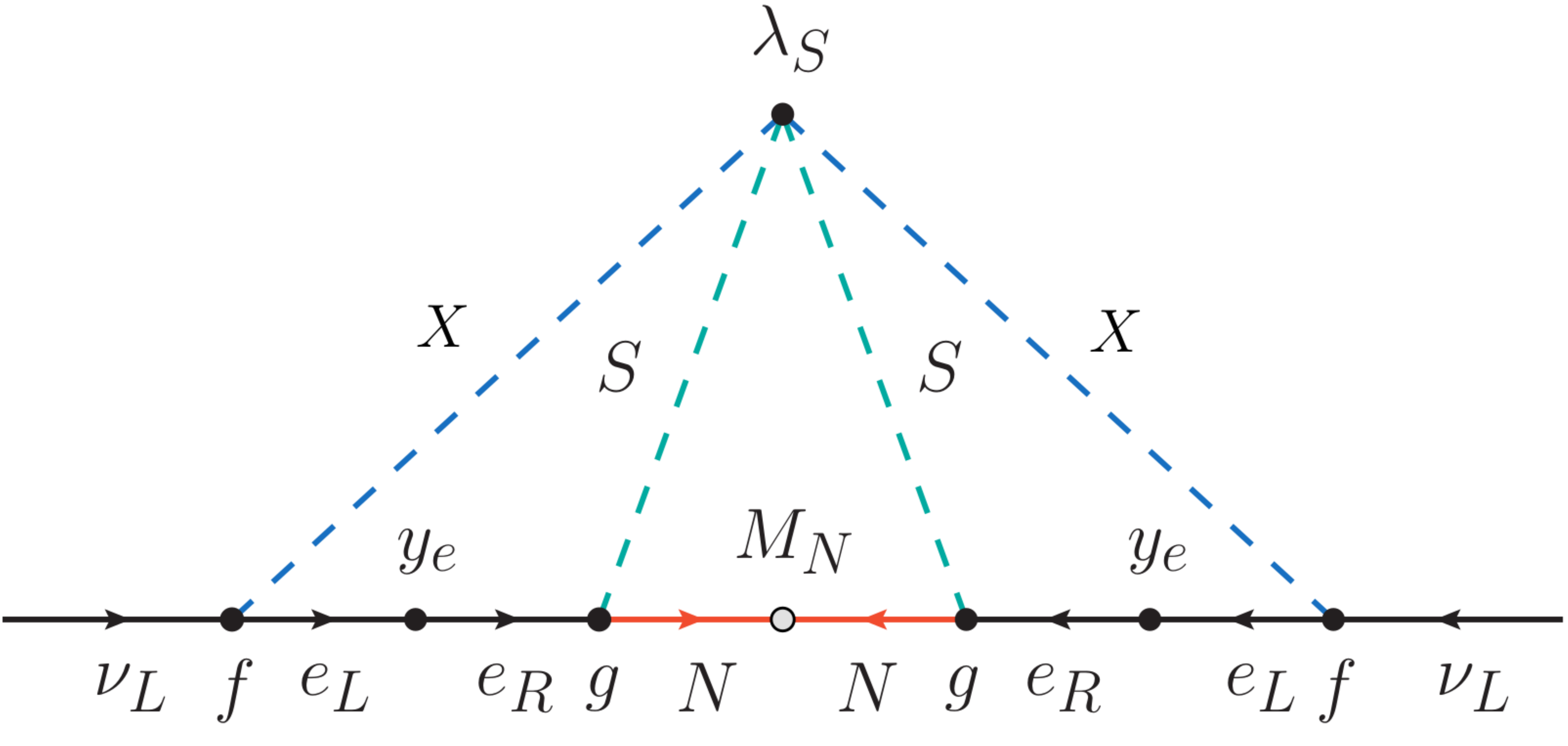}
\caption{3-loop neutrino masses in the KNT model.
  \label{fig:KNT}
  }
\end{figure}

The $\mathbb{Z}_2$ symmetry forbids the standard Higgs Yukawa coupling
with the lepton doublet $L$ and the $N$ singlets. Therefore, the usual
type-I seesaw contribution at tree-level is absent. Instead, neutrino
masses are generated at 3-loop order as shown in
Fig.~\ref{fig:KNT}. The neutrino mass matrix is given by
\begin{align}\label{eq:MnuKNT}
  \left(\mathcal{M}_\nu\right)_{ij} = \frac{2 \lambda_S}{(16 \pi^2)^3} \sum_{\alpha \beta a}  
  \frac{m_\alpha m_\beta}{M_{N_a}} f_{i\alpha} f_{j\beta}
  g_{\alpha a} g_{\beta a} \, F_{\rm KNT}\,,
\end{align}
where $m_\alpha$ is the mass of the $\ell_\alpha$ charged lepton and
$F_{\rm KNT}$ is a loop function that depends on the masses of the
scalars and fermions running in the loops. More information about this
function can be found in Appendix~\ref{app:int-KNT}.

It is important to stress that in the KNT model, each entry in
$(\mathcal{M}_\nu)_{ij}$ contains the sum over the SM charged lepton
masses. Therefore, different from the other models discussed in this
paper, the suppression of the entries in $\mathcal{M}_\nu$ is at most
$m_\mu^2$. The neutrino fit can then reproduce experimental data
with Yukawas which are considerably smaller than in the cocktail or
AKS models.

\subsection{Results}
\label{subsec:results-KNT}

We start this section again with a discussion of the neutrino mass
fit. The coupling $f$ in Eq.~\eqref{eq:YukKNT} is antisymmetric, thus
the determinant of the neutrino mass matrix in Eq.~\eqref{eq:MnuKNT}
is zero, implying that one neutrino is massless. This is reminiscent
of the 2-loop Babu-Zee model of neutrino
mass~\cite{Zee:1985id,Babu:1988ki}, where the same singly charged
scalar is used. In our fitting procedure we use therefore an adapted
version of the solution found
in~\cite{Babu:2002uu,Herrero-Garcia:2014hfa} for the Babu-Zee model.

The procedure consists of two steps. First, because det$(f)= 0$,
the matrix has one eigenvector ${\bf a}= (f_{23},-f_{13},f_{12})$, which
is also an eigenvector of $\mathcal{M}_\nu$:
\begin{equation}\label{eq:defa}
  \widehat{\cal M}_\nu \, U^T{\bf a}=0 \, .
\end{equation}
This implies three equations, one of which is trivial, while the
other two allow to express the ratios $(f_{13}/f_{12},f_{23}/f_{12})$
as functions of the neutrino angles and phases only. These solutions
depend on the neutrino mass hierarchy.

Next, we can write the neutrino mass matrix as
\begin{equation}\label{eq:Maux}
  \mathcal{M}_\nu = -c \, f \, M_{\rm aux} \, f \, .
\end{equation}
$c$ contains all global constants, we have used $f^T=-f$ and $M_{\rm
  aux}$ is an auxiliary matrix, which is complex symmetric.  This
defines a set of 6 complex equations relating the entries in $M_{\rm
  aux}$ to neutrino data. With three independent entries $f_{ij}$, we
can use three of the six equations to express three entries in $M_{\rm
  aux}$ as a function of the remaining ones, neutrino data and
$f_{ij}$. The resulting equations are very lengthy and not at all
illuminating, so we do not present them here.

The definition of $M_{\rm aux}$ in Eq.~\eqref{eq:Maux} shows that
\begin{equation}\label{eq:Maux2}
  M_{\rm aux} = {\widehat{\cal M}_{e}} \, g \left({\widehat M^{\rm eff}}\right)^{-1}
  g^T {\widehat{\cal M}_{e}}
\end{equation}
where
\begin{equation}\label{eq:MNeff}
  \left({\widehat M^{\rm eff}}\right)^{-1} =
   \begin{pmatrix}
 \frac{F_{KNT}(r^X_1,r^S_1)}{M_{N_1}} & 0 & 0 \\
 0 & \frac{F_{KNT}(r^X_2,r^S_2)}{M_{N_2}} & 0  \\
 0 & 0 & \frac{F_{KNT}(r^X_3,r^S_3)}{M_{N_3}} 
 \end{pmatrix} \, ,
\end{equation}
and $r^X_i=(m_{s_1}/M_{N_i})^2$, $r^S_i=(m_{s_2}/M_{N_i})^2$. With
$M_{\rm aux}$ being complex symmetric, we can use a suitably
modified~\cite{Cordero-Carrion:2018xre,Cordero-Carrion:2019qtu}
Casas-Ibarra parametrization~\cite{Casas:2001sr} to express the matrix
$g$ as
\begin{equation}\label{eq:MNeff}
g = \sqrt{{\widehat M^{\rm eff}}}{\cal R} \, \sqrt{\hat M_{\rm aux}} \, U_{\rm aux}^T
    \left({\widehat{\cal M}_{e}}\right)^{-1}.
\end{equation}
${\hat M_{\rm aux}}$ and $U_{\rm aux}$ are the eigenvalues and
eigenvectors of the auxiliary matrix $M_{\rm aux}$. Although in
principle it would be possible to determine ${\hat M_{\rm aux}}$ and
$U_{\rm aux}$ in terms of the input neutrino data analytically, in
practice we find these two matrices numerically for any input point of
experimental data and choice of free parameters. Finally, ${\cal R}$
is a $3 \times 3$ orthogonal matrix.

In summary, neutrino oscillation data provides 6 constraints: $\Delta
m_{\rm Atm}^2$, $\Delta m_{\odot}^2$, three angles and the CP-phase
$\delta$. A number of free parameters can then be scanned over, using
the above procedure. In the neutrino sector we still have
$\alpha_{12}$.\footnote{Since one neutrino is massless, only one of
  the two Majorana phases, i.e. $\alpha_{12}$, is physical.} The
matrix $f$ is fixed from experimental data, up to the overall scale of
the matrix. We choose $f_{12}$ as the free parameter.  The matrix
${\cal R}$, in the most general case, contains 3 complex angles. There
are 3 right-handed neutrino masses, $M_{N_i}$, and 2 scalar masses,
$m_{s_{1,2}}$. And, finally, we can use 3 of the 6 equations for
$M_{\rm aux}$ to eliminate some particularly chosen $(M_{\rm
  aux})_{ij}$.  This leaves as free inputs the remaining 3 entries
in $M_{\rm aux}$.

Up to now, we have been completely general in our discussion.
However, there is still a certain freedom as to which 3 entries in
$(M_{\rm aux})_{ij}$ we fix via 3 of the equations defined by
Eq.~\eqref{eq:Maux}. In practice, we choose to solve for $(M_{\rm
  aux})_{22}$, $(M_{\rm aux})_{23}$ and $(M_{\rm aux})_{33}$ and
assume $(M_{\rm aux})_{1k} = 0$. This particular choice is motivated
by the observation that in this limit all terms in $g \propto 1/m_e$
disappear. In other words, this solution guarantees that the
contribution to $\mu\to e \gamma$ and $\tau\to e \gamma$ from loops
involving $s_2$ and $N_i$ are automatically absent in our scans, due
to $g_{1k}=0$ $\forall k$. Our ansatz is therefore the optimal choice
for minimizing fine-tuning on the other parameters in
$g$.~\footnote{We have explored other ans\"atze but concluded that this
  is indeed the optimal one. For instance, we can generate textures
  with either the 2nd or 3rd column of $g$ vanishing. These will make
  the $\tau \to e \gamma$ or $\tau \to \mu \gamma$ branching ratio
  vanish, but at the cost of a large $\mu\to e \gamma$ branching
  ratio. We have also considered a solution with any or all of
  $(M_{\rm aux})_{1k}\ne 0$. However, in this case we did not find any
  configuration for the remaining free parameters that induces a
  cancellation that suppresses the $\mu\to e \gamma$ branching ratio.}

Before we explore the remaining parameter space of the model, we must
consider lower limits on the masses of the charged scalars from
accelerator searches. LEP provides a lower limit on charged particles
decaying to leptons plus missing momentum, which will essentially rule
out all values of $m_{s_1}$ below 100 GeV~\cite{Tanabashi:2018oca} and
similarly for $m_{s_2}$, unless $M_{N_1}$ is close to
$m_{s_2}$.~\footnote{There are no accelerator limits on $N$, since the
  $\mathbb{Z}_2$ symmetry prohibits their mixing with the active
  neutrinos.} At the LHC there is currently no specific search for
particles with the quantum numbers of $S$ and $X$. However, slepton
pair production with their subsequent decays to a lepton plus a
neutralino provide the same signal and thus, we can make a
reinterpretation of the corresponding searches at
CMS~\cite{Sirunyan:2018nwe} and ATLAS~\cite{Aad:2019vnb}.  The CMS
slepton search~\cite{Sirunyan:2018nwe} is based on $35.9/$fb, while
ATLAS' chargino and slepton search~\cite{Aad:2019vnb} uses
$139/$fb. The ATLAS limits are correspondingly more stringent and we
will therefore discuss these. We have implemented the KNT model in
\texttt{SARAH}~\cite{Staub:2012pb,Staub:2013tta} and generated
\texttt{SPheno} routines~\cite{Porod:2003um,Porod:2011nf} and model
files for
\texttt{MadGraph}~\cite{Alwall:2007st,Alwall:2011uj,Alwall:2014hca}. We
have then calculated cross sections with \texttt{MadGraph} to recast
the results of~\cite{Aad:2019vnb}. For $s_1$ the mass range between
(very roughly) $m_{s_1}=(250-400)$ GeV is excluded by this search. The
range $m_{s_1}=(100-250)$ is currently unconstrained, due to large
backgrounds in~\cite{Aad:2019vnb}. For $s_2$ the limits are even
weaker, unless $m_{s_2}-M_{N_1}$ is larger than ($50-70$) GeV,
depending on $m_{s_2}$.

Let us now turn to the discussion of CLFV. Consider first the
antisymmetric Yukawa coupling $f$. Neutrino data requires all three
elements of $f$ to be non-zero, thus there will always be a non-zero
value for the three possible decays $\mu\to e\gamma$, $\tau\to
e\gamma$ and $\tau\to\mu\gamma$. The constraint from $\mu\to e\gamma$
is the most stringent one. However, since we still have the overall
scale of $f$ as a free parameter, in our choice the value of $f_{12}$,
we can use it to fix Br($\mu\to e\gamma$) to the upper limit (present
or future) for any point in the parameter space.  Since the neutrino
mass matrix is proportional to the square of the matrices $f$ and $g$
however, once this choice is made there is no longer any overall
scaling freedom in the coupling $g$. Putting the calculated Br($\mu\to
e\gamma$) to equal the experimental bound will generate the smallest
values for the entries of $g$ allowed in the model parameter space. A
smaller upper limit on Br($\mu\to e\gamma$) will lead to larger $g$
and thus more stringent constraints from $\tau\to\mu\gamma$.

\begin{figure}[t]
\centering
\includegraphics[width=0.49\textwidth]{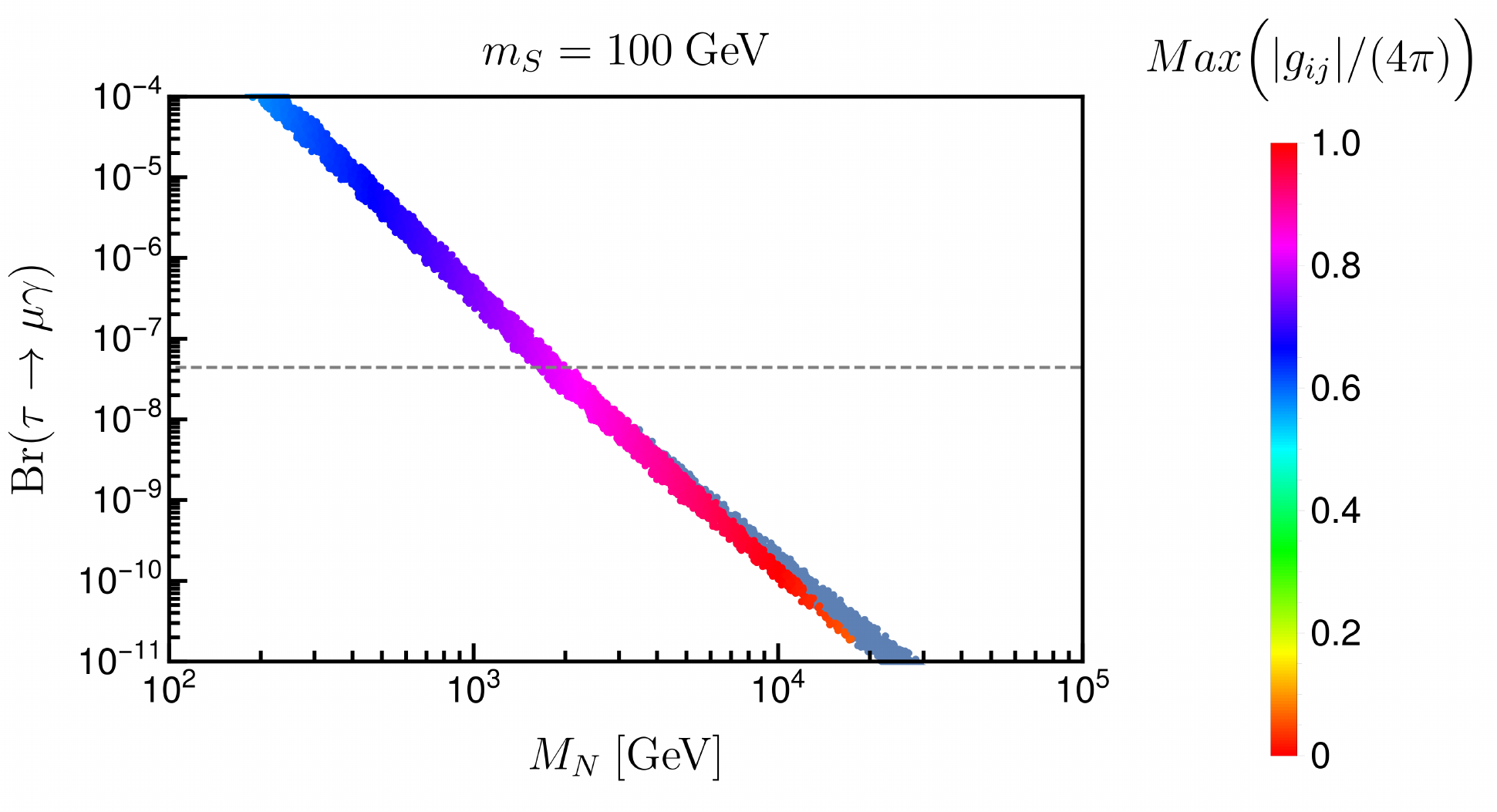}
\includegraphics[width=0.49\textwidth]{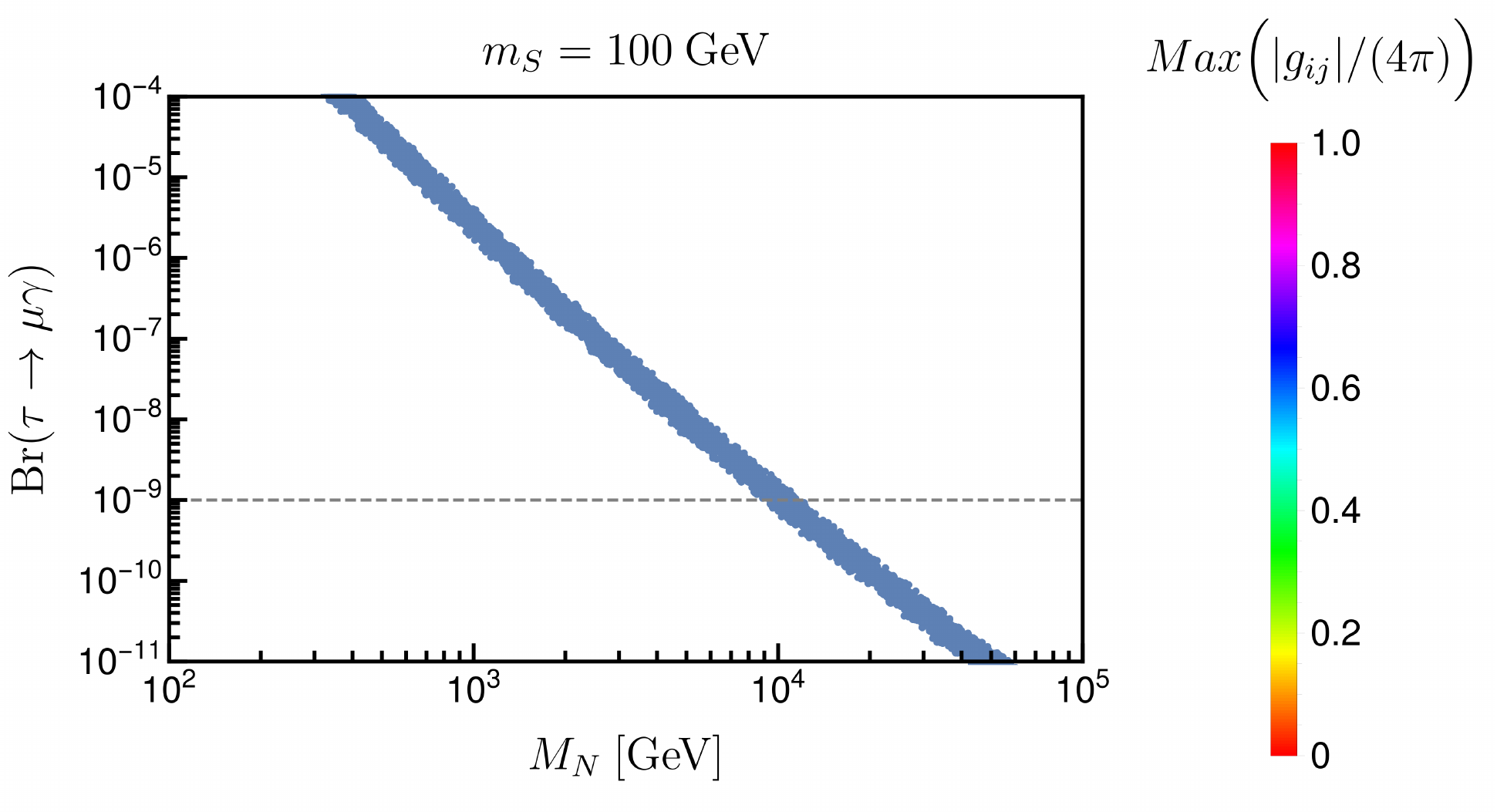}
\caption{Calculated values for Br($\tau\to\mu\gamma$) as function of
  the common fermion mass $M_N$, for the current lower limit on
  $m_{s_1}=m_{s_2}=100$ GeV.  To the left, with current experimental
  limits; to the right for the expected future experimental
  limits. The required size of the Yukawas is color-coded. Bluish-grey
  points mean that at least one entry in $g$ is larger than $4\pi$.
    \label{fig:KNTTauMuG}
  }
\end{figure}

We then scanned over the remaining parameters of the model
numerically. Consider first the case of NH. Some examples for
Br($\tau\to\mu\gamma$) are shown in Fig.~\ref{fig:KNTTauMuG}. In this
plot we have chosen the fixed value $m_{s_{1,2}}=100$ GeV,
corresponding to the experimental lower limit, and the three
right-handed neutrino masses all equal to a common
$M_N$.~\footnote{For degenerate $M_{N_i}$ ${\cal R}$ becomes
  unphysical and drops out of the calculation.} The points are scanned
over the allowed $3 \, \sigma$ ranges for the neutrino data for
NH. The size of the largest entry in $g$ is color-coded in the points.
The plot to the left has been calculated for the current experimental
limit on Br($\mu\to e\gamma$)$< 4.2 \times
10^{-13}$~\cite{TheMEG:2016wtm}, the plot to the right is for the
expected future limit Br($\mu\to e\gamma$)$< 6 \times
10^{-14}$~\cite{Baldini:2013ke}. For the choice of $m_{s_{1,2}}=100$
GeV no valid point with $g_{ij}\le 4\pi$~$\forall ij$ remains in the
parameter space. Constraints are more stringent for IH, and therefore
the same conclusion is reached.

\begin{figure}[t]
\centering
\includegraphics[width=0.49\textwidth]{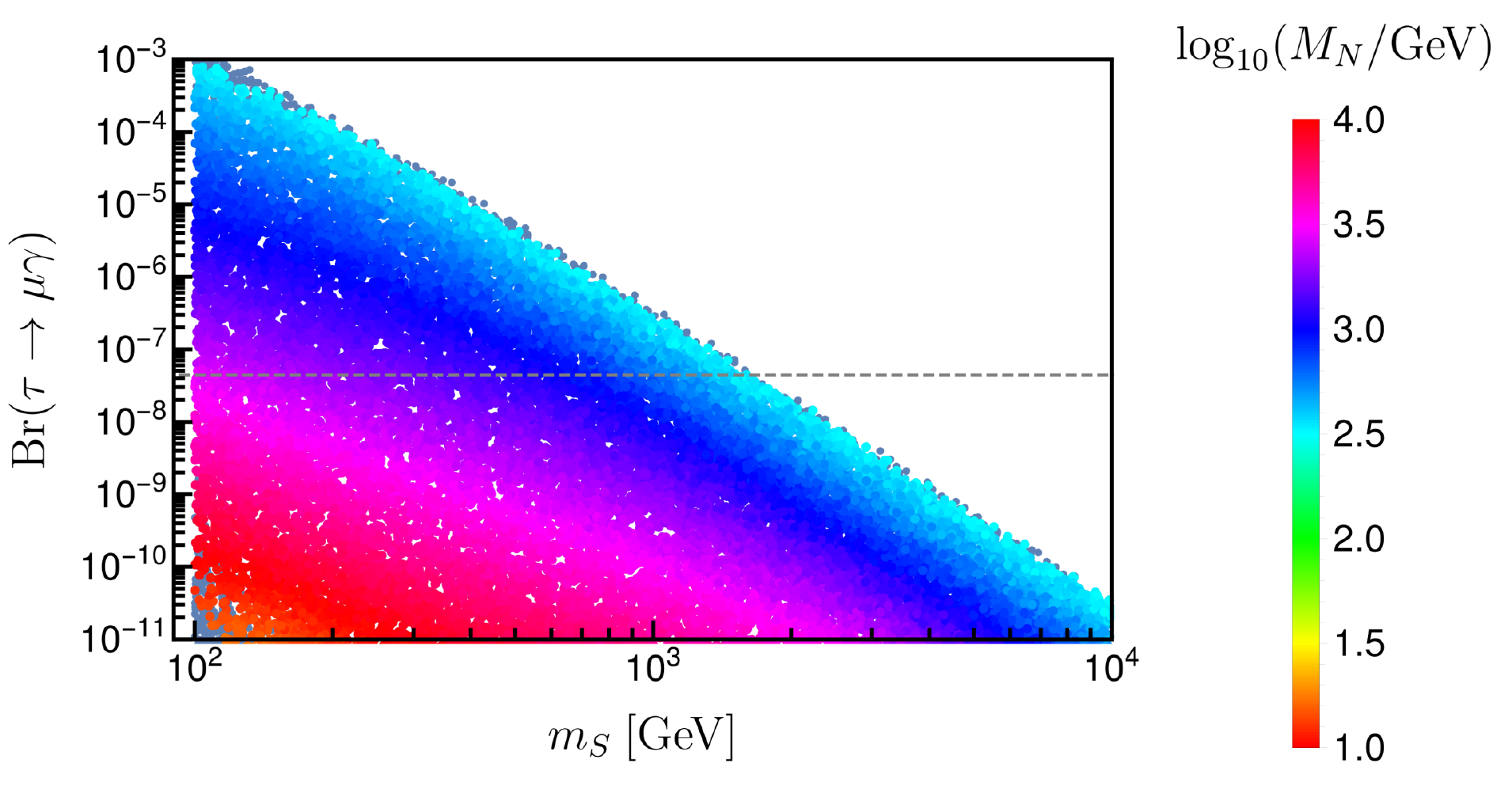}
\includegraphics[width=0.49\textwidth]{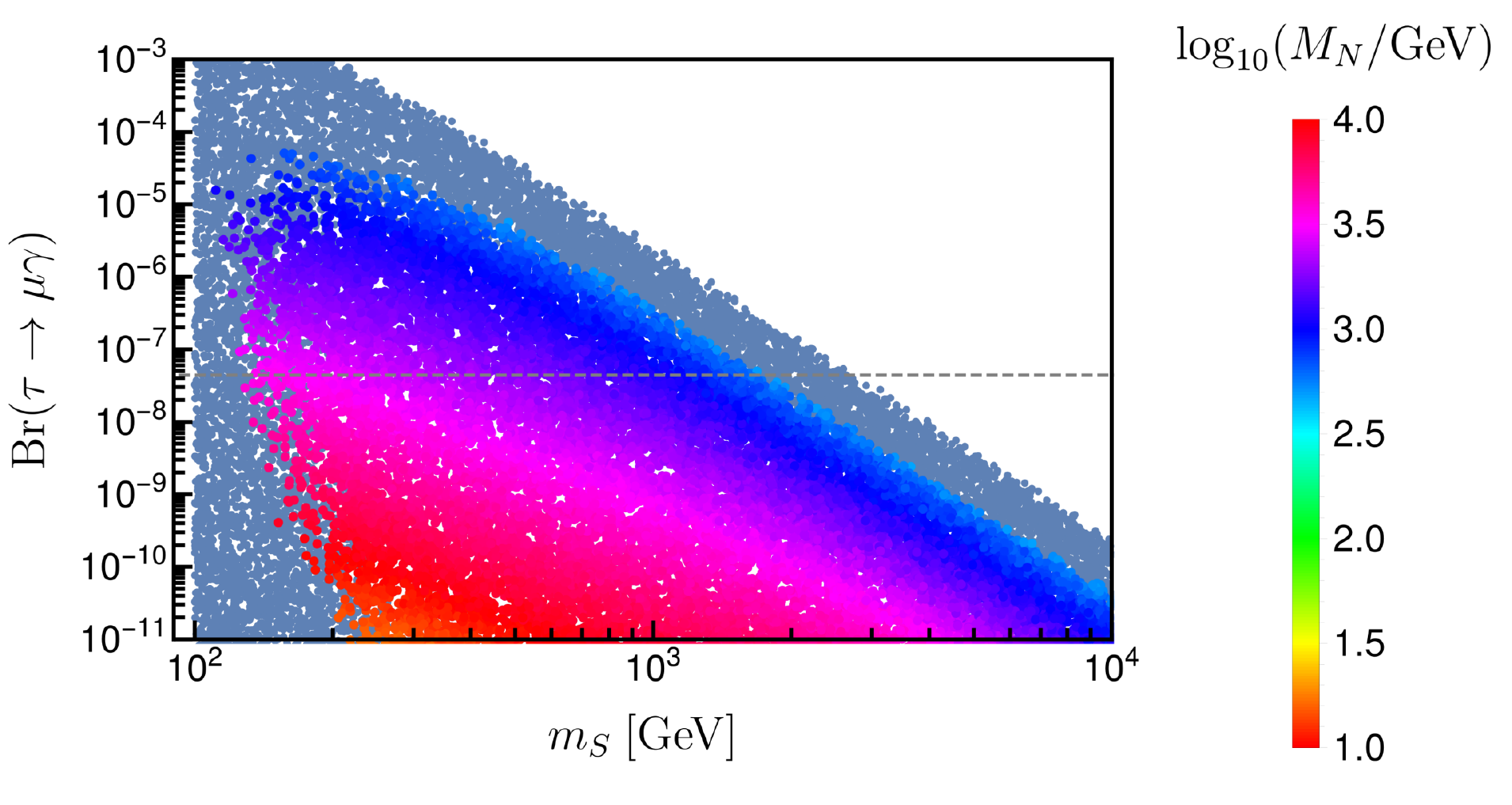}
\caption{Calculated values for Br($\tau\to\mu\gamma$) as function of
  the scalar mass $m_s$ for different values of $M_N$ (color
  coded). Bluish points are ruled out by non-perturbative
  couplings. To the left, with current experimental limits; to the
  right for the expected future experimental limit on Br($\mu\to
  e\gamma$).
    \label{fig:KNTTauMuG2}
  }
\end{figure}

We therefore scanned over $m_{s_{1,2}} \equiv m_S$ and $M_{N_i}$
simultaneously.  The results are shown in
Fig.~\ref{fig:KNTTauMuG2}. Here, $M_{N_i}$ are varied within $20\%$ of
a common $M_N$. The range of $M_N$ is color-coded in the
points. Again, the plot to the left is for the current bound on
Br($\mu\to e\gamma$), while the plot to the right is for the future
bound.  In these plots, points with non-perturbative couplings are
shown in bluish color. This bound eliminates all points below roughly
$M_N = {\cal O}(100)$ GeV already with the current experimental bound
on Br($\mu\to e\gamma$), see however the discussion below.  We show
only the cases with a trivial ${\cal R}$ matrix. For non-zero angles
in ${\cal R}$ the results look similar, although fewer points lie in
the perturbative regime.

The combined constraints of perturbativity and future limits
from CLFV searches would put a lower bound on $m_S$ roughly of
order ($180-200$) GeV. This limit becomes stronger for lower
values of $M_N$, as the plots shows.

\begin{figure}[t]
\centering
\includegraphics[width=0.49\textwidth]{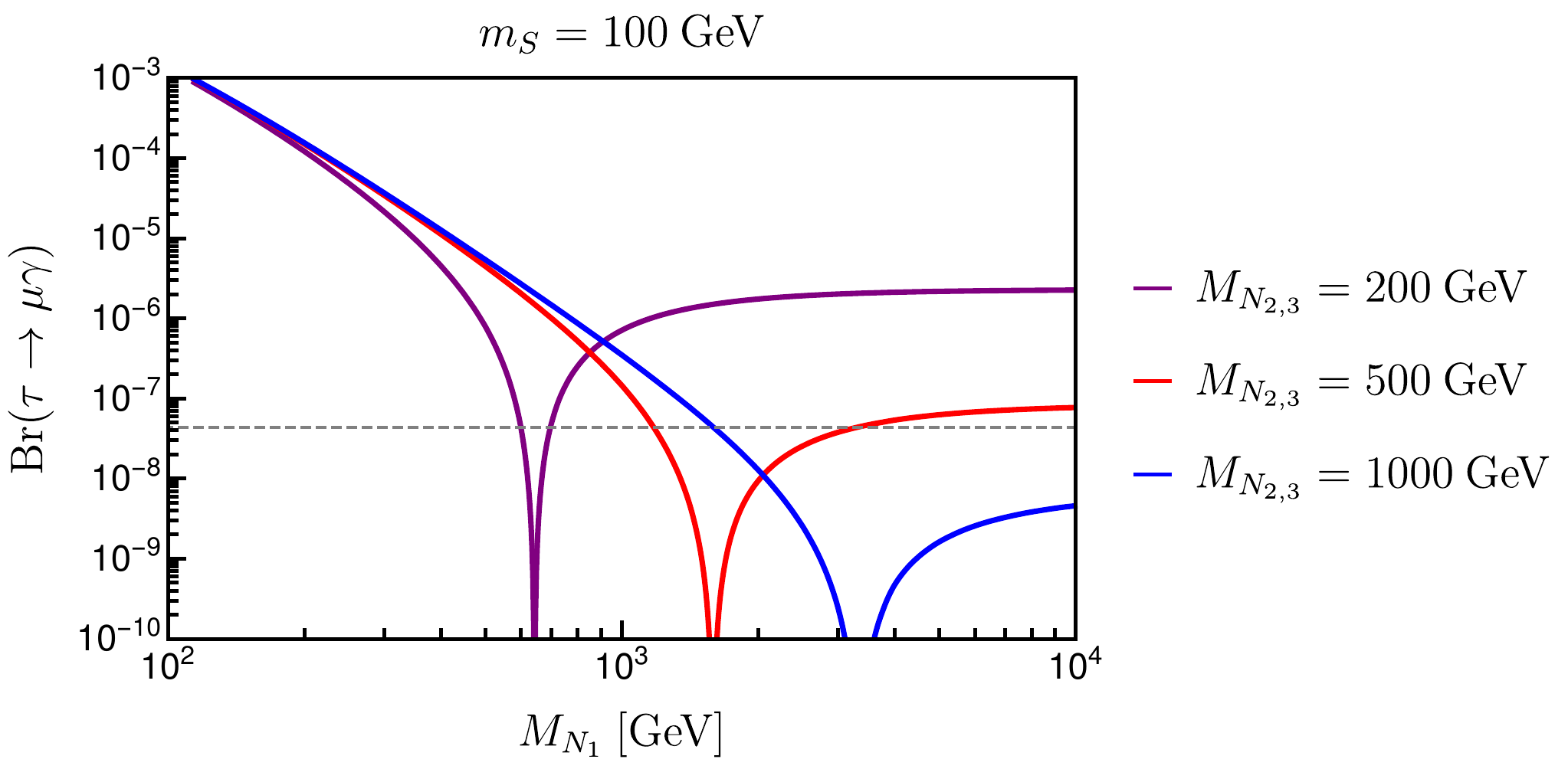}
\includegraphics[width=0.49\textwidth]{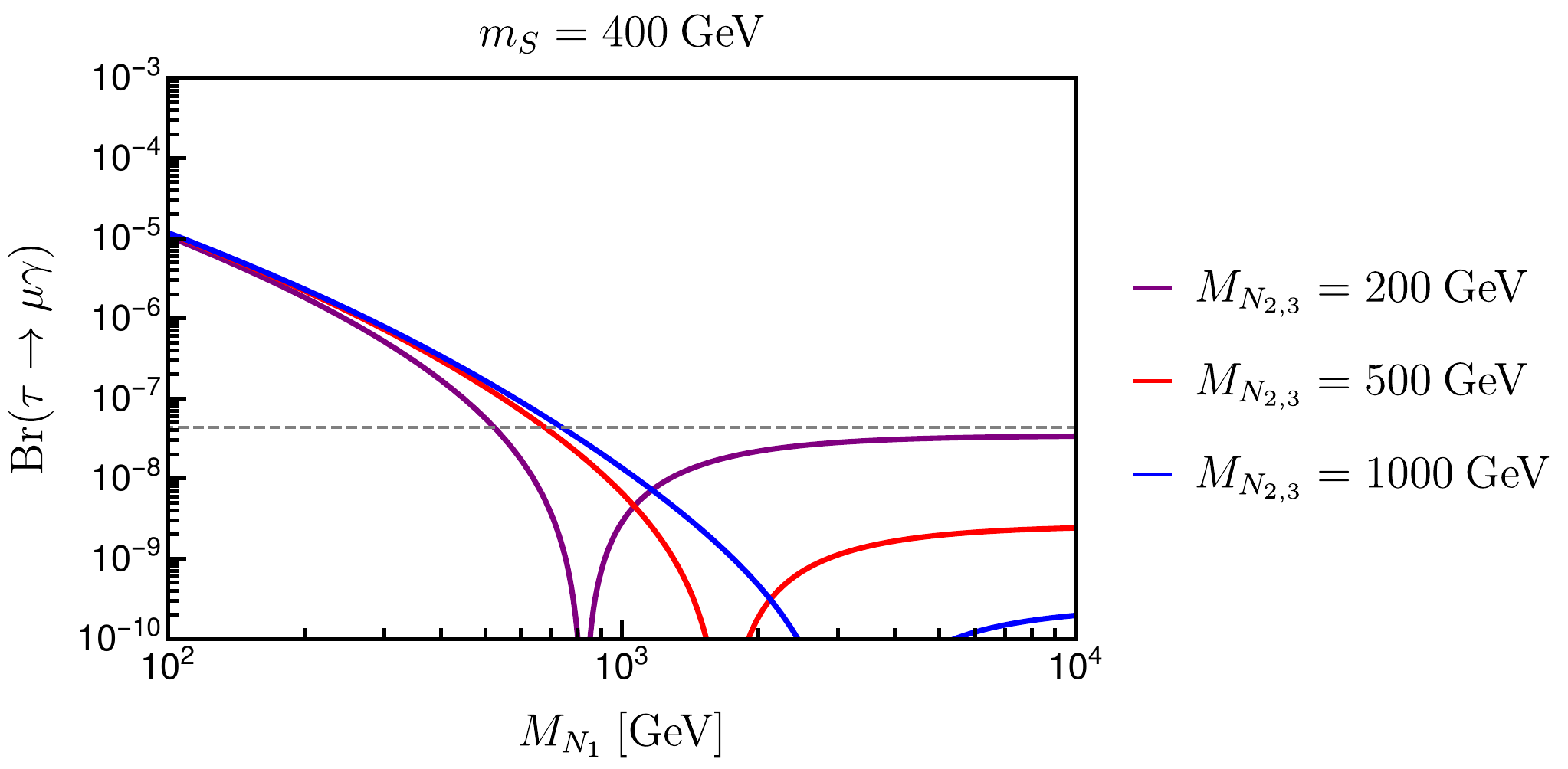}
\caption{Calculated values for Br($\tau\to\mu\gamma$) as function
  of the lightest right-handed neutrino mass $M_{N_1}$ for different
  choices of $M_{N_{2,3}}$ and two different scalar masses $m_S$
  (left and right). The calculation uses for simplicity the b.f.p.
  for neutrino data.
    \label{fig:KNTTauMuG3}
  }
\end{figure}

The above discussion is strictly valid only for the case where the
three right-handed neutrinos have similar masses. For hierarchical
right-handed neutrinos the constraints are usually dominated by the
lightest of these. There exist, however, exceptional points in the
parameter space, where the contributions to Br($\mu\to e\gamma$) from
the three different neutrinos conspire to (nearly) cancel each
other. This is shown in Fig.~\ref{fig:KNTTauMuG3}. The figure shows
Br($\tau\to \mu\gamma$) as a function of the ``lightest'' right-handed
neutrino mass, for different choices of $M_{N_{2,3}}$.  Br($\tau\to
\mu\gamma$) is dominated by the lightest mass eigenstate, except in
some particular points, where cancellations
occur. Figs.~\ref{fig:KNTTauMuG} and \ref{fig:KNTTauMuG2} do not cover
these exceptional combinations of parameters.

\begin{figure}[t]
\centering
\includegraphics[width=0.49\textwidth]{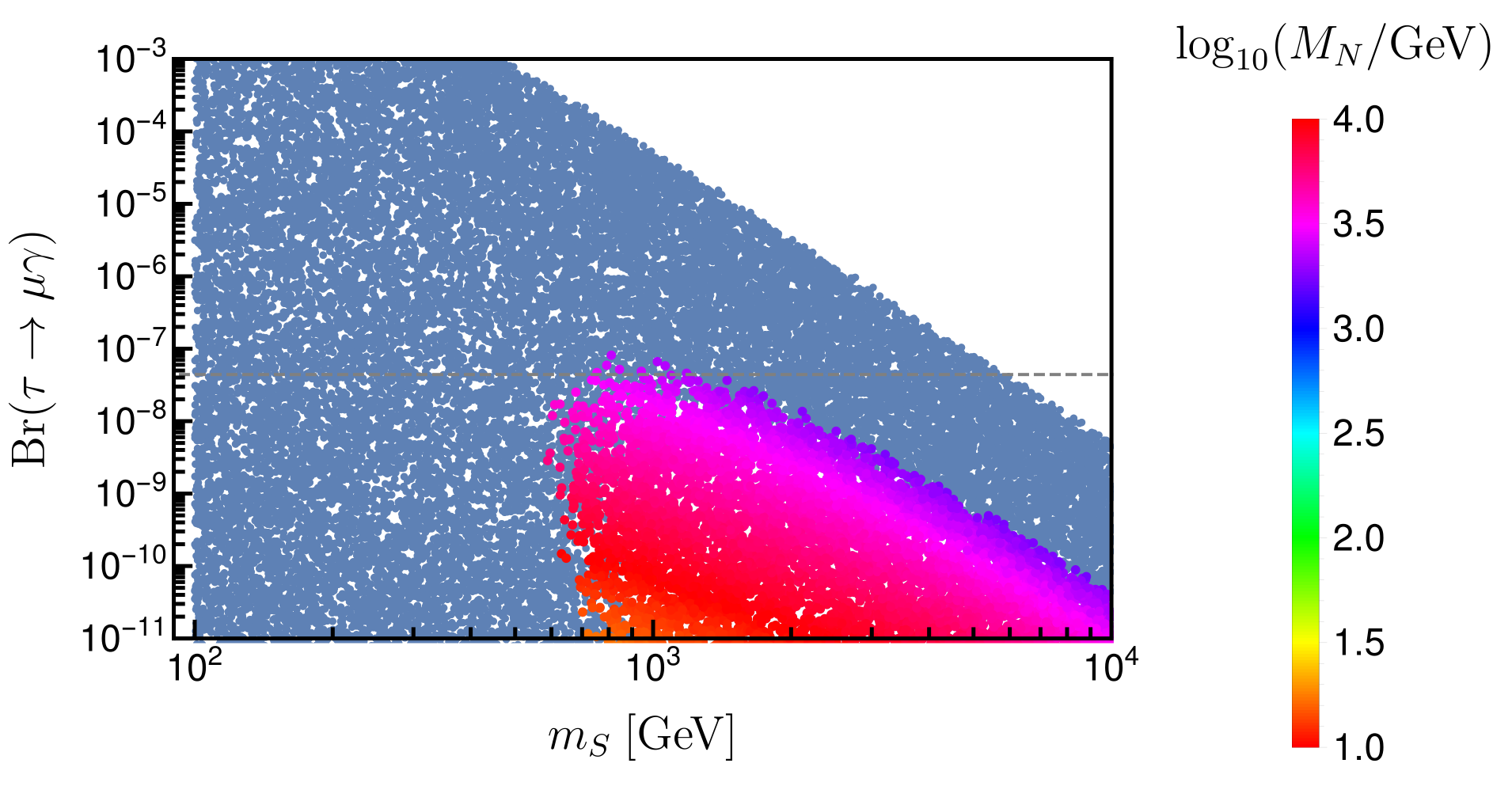}
\includegraphics[width=0.49\textwidth]{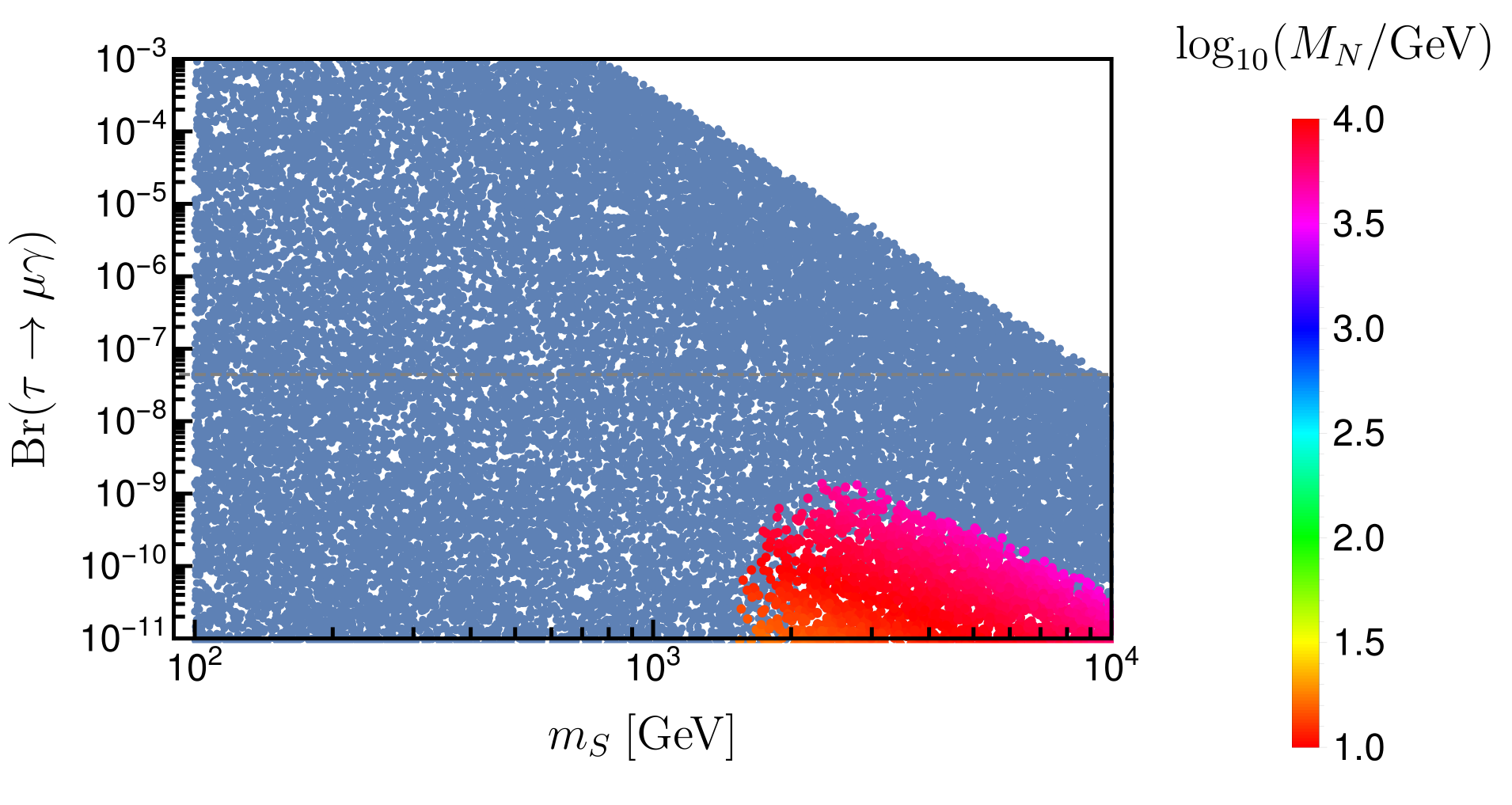}
\caption{Same as Fig.~\ref{fig:KNTTauMuG2} but for IH for the neutrino
  masses. Bluish points are excluded due to perturbativity arguments.
    \label{fig:KNTTauMuG2IH}
  }
\end{figure}

We have repeated the scans discussed above also for the case of IH. An
example is shown in Fig.~\ref{fig:KNTTauMuG2IH}.  IH requires larger
Yukawa couplings, since now two neutrino have masses of order
$\sqrt{\Delta m^2_{\rm Atm}}$. Thus, many more points in the parameter
space are ruled out due to the perturbativity constraint. This pushes
both, fermion as well as the scalar, masses to larger values. Indeed,
already with current constraints there are no points with $m_S$ below
roughly 600 GeV.

Finally, let us mention that in the KNT model there is no short-range
diagram contributing to $0\nu\beta\beta$ decay. Given that the KNT
model predicts one (nearly)\footnote{A tiny lightest neutrino mass
  will be generated at higher loop order.} massless neutrino, it
predicts both, an upper and a lower limit for $0\nu\beta\beta$
decay. For normal [inverted] hierarchy the allowed range is roughly
$m_{ee} \sim (1-5)$ meV [$(20-50)$ meV]. Observing $0\nu\beta\beta$
decay outside this range would rule out the KNT model as an
explanation for the experimental neutrino oscillation data.

\section{AKS model}
\label{sec:AKS}

A general class of models is represented by the AKS
model~\cite{Aoki:2008av}. In this case the particle content is
extended to include new scalars and fermions.

\subsection{The model}
\label{subsec:model-AKS}

The AKS model extends the usual 2HDM with the real scalar singlet
$\varphi$, the singly charged scalar $S$ and three generations of
singlet fermions $N$. Even though a more minimal version with only two
generations of $N$ is possible, we will consider three in the
following. The fields $S$, $\varphi$ and $N$ are assumed to be odd
under the $\mathbb{Z}_2$ parity, while the rest of the particles are
even. The quantum numbers of the new particles in the AKS model are
given in Table~\ref{tab:AKS}.

\begin{table}
\centering
\begin{tabular}{| c c c c c c |}
\hline  
 & generations & $\mathrm{SU(3)}_c$ & $\mathrm{SU(2)}_L$ & $\mathrm{U(1)}_Y$ & $\mathbb{Z}_2$ \\
\hline
\hline 
$\varphi$ & 1 & ${\bf 1}$ & ${\bf 1}$ & $0$ & $-$ \\
$S$ & 1 & ${\bf 1}$ & ${\bf 1}$ & $1$ & $-$ \\
\hline
\hline    
$N$ & 3 & ${\bf 1}$ & ${\bf 1}$ & $0$ & $-$ \\  
\hline
\hline
\end{tabular}
\caption{New particles in the AKS model with respect to the 2HDM.}
\label{tab:AKS}
\end{table}

As explained in Sec.~\ref{sec:notation}, an additional softly-broken
$\mathbb{Z}_2$ symmetry is introduced to avoid dangerous flavor
changing neutral currents. We choose to follow~\cite{Aoki:2008av} and
use this symmetry to couple one of the scalar doublets ($\Phi_1$) only
to leptons, and the other ($\Phi_2$) only to quarks. Due to this
choice, the Yukawa couplings of the model are given in
Eq.~\eqref{eq:2HDMYuk}, along with the Yukawa
\begin{equation} 
    -\mathcal L \supset Y^\ast \, \overline{N^c} \, e_R \, S + \hc \, .
\end{equation}
One can also write Majorana masses for the $N$
singlets,
\begin{equation}
- \mathcal L_N = \frac{1}{2} M_N \overline{N^c} N + \hc \, ,
\end{equation}
with $M_N$ a symmetric matrix. The scalar potential of the model is
given by
\begin{align} \label{eq:PotAKS}
\mathcal V &\supset m_{1}^2 |\Phi_1|^2 + m_{2}^2 |\Phi_2|^2 + \left( \mu_{12}^2 \, \Phi_1^\dagger \Phi_2 + \hc \right) + \frac{1}{2} \lambda_1 \, |\Phi_1|^4 + \frac{1}{2} \lambda_2 \, |\Phi_2|^4 \nn \\
&+ \lambda_{3} \, |\Phi_1|^2 |\Phi_2|^2 + \lambda_4 \, |\Phi_1^\dagger \Phi_2|^2  + \frac{1}{2} \left[ \lambda_5 \, (\Phi_1^\dagger \Phi_2)^2 + \hc \right] \nn \\
&+ \lambda_{\Phi S}^{(1)} \, |\Phi_1|^2 |S|^2 + \lambda_{\Phi S}^{(2)} \, |\Phi_2|^2 |S|^2 + \frac{1}{2} \lambda_{\Phi \varphi}^{(1)} \, |\Phi_1|^2 \varphi^2 + \frac{1}{2} \lambda_{\Phi \varphi}^{(2)} \,
 |\Phi_2|^2 \varphi^2 + \left[ \kappa \, \Phi_1 \, \Phi_2 \, S^\ast \, \varphi + \hc \right] \nn \\
&+ \frac{M_{\varphi}^2}{2} \varphi^2 + M_{S}^2 |S|^2 + \frac{1}{2} \lambda_S \, |S|^4 + \frac{1}{4!} \lambda_{\varphi} \varphi^4 + \frac{1}{2} \xi \, \varphi^2 |S|^2 \, .
\end{align}
As usual, we have omitted $\mathrm{SU(2)}_L$ indices in the previous
expression. We point out that lepton number would be restored in the
limit $\kappa \to 0$. The presence of this coupling breaks lepton
number in one unit.

After electroweak symmetry breaking, the doublet scalars $\Phi_1$ and
$\Phi_2$ get mixed. The mass eigenstates resulting from this mixing
are the SM Higgs, another Higgs, a new charged scalar, and a
pseudoscalar.  The charged and neutral Goldstone bosons are absorbed
by the $Z$ and $W$ gauge bosons. The mass matrix for the CP-even
neutral states in the basis $\mathcal{H}^0 = \text{Re} \,(\Phi_1^0,
\Phi_2^0)^T$ is given by
\begin{equation}
 \mathcal{M}^2_{\mathcal{H}^0} = \left( \begin{array}{cc}
\lambda_1 v_1^2 - \mu_{12}^2 \, \tan \beta  &   v_1 v_2 \left( \lambda_3 + \lambda_4 + \lambda_5\right) + \mu_{12}^2   \\
 v_1 v_2 \left( \lambda_3 + \lambda_4 + \lambda_5\right) + \mu_{12}^2 & \lambda_2 v_2^2 - \mu_{12}^2 \, \cot\beta 
\end{array} \right) \, .
\end{equation}
The CP-odd neutral scalar mass matrix in the basis $\mathcal{A}^0 =
\text{Im} \,(\Phi_1^0, \Phi_2^0)^T$ is
\begin{equation}
 \mathcal{M}^2_{\mathcal{A}^0} = \left( \begin{array}{cc}
- v_2^2 \lambda_5 - \mu_{12}^2 \, \tan \beta  &  v_1 v_2 \lambda_5 + \mu_{12}^2   \\
v_1 v_2 \lambda_5 + \mu_{12}^2 & - v_1^2 \lambda_5 - \mu_{12}^2 \, \cot\beta 
\end{array} \right) \, .
\end{equation}
One finds a massless state, the Goldstone boson that becomes the
longitudinal component of the $Z$ boson. The other state has a mass
\begin{equation}
  m^2_{\mathcal{A}^0}= - \left(v_1 v_2 \lambda_5 + \mu_{12}^2\right) \, \frac{v^2}{v_1 v_2} \, ,
\end{equation}
while the mass of the $Z$ boson is $m_Z^2=\frac{1}{4} v^2 (g_1^2 +
g_2^2)$. The mass matrix for the charged states in the
$\mathcal{H}^\pm = (\Phi_1^\pm, \Phi_2^\pm)^T$ basis is
\begin{equation}
 \mathcal{M}^2_{\mathcal{H}^\pm} = \left( \begin{array}{cc}
-\frac{1}{2} v_2^2 \, \left(\lambda_4 + \lambda_5 \right) - \mu_{12}^2 \, \tan \beta  &   \frac{1}{2} v_1 v_2 \, \left(\lambda_4 + \lambda_5 \right)+ \mu_{12}^2 \\
\frac{1}{2} v_1 v_2 \, \left(\lambda_4 + \lambda_5 \right)+ \mu_{12}^2 & -\frac{1}{2} v_1^2 \, \left(\lambda_4 + \lambda_5 \right) - \mu_{12}^2 \, \cot\beta 
\end{array} \right) \, .
\end{equation}
Again, after diagonalization one obtains a massless state, identified
with the Goldstone boson that becomes the longitudinal part of the $W$
boson, and a massive physical charged scalar with mass
\begin{equation}
m^2_{\mathcal{H}^\pm}=-\left(\frac{\mu_{12}^2}{v_1 v_2}+\frac{\lambda_4+\lambda_5}{2} \right)
v^2 \, .
\end{equation}
The mass of the $W$ boson is given by the standard expression
$m_{W^\pm}^2=\frac{1}{4} g_2^2 v^2$.  Finally, the masses of the
singlet scalars $\varphi$ and $S$ are
\begin{align}
m_{\varphi}^2 &= M_{\varphi}^2 + \frac{1}{2}\left(\lambda_{\Phi \varphi}^{(1)} \, v_1^2+\lambda_{\Phi \varphi}^{(2)} \, v_2^2 \right) \, , \\
m_{S^+}^2 &= M_{S}^2 + \frac{1}{2}\left(\lambda_{\Phi S}^{(1)} \, v_1^2+\lambda_{\Phi S}^{(2)} \, v_2^2\right) \, .
\end{align}

As in the cocktail model, the lightest $\mathbb{Z}_2$-odd state in the
AKS model is stable and can constitute a DM candidate.

\subsubsection*{Neutrino masses}

\begin{figure}[t]
\centering
\includegraphics[width=0.48\textwidth]{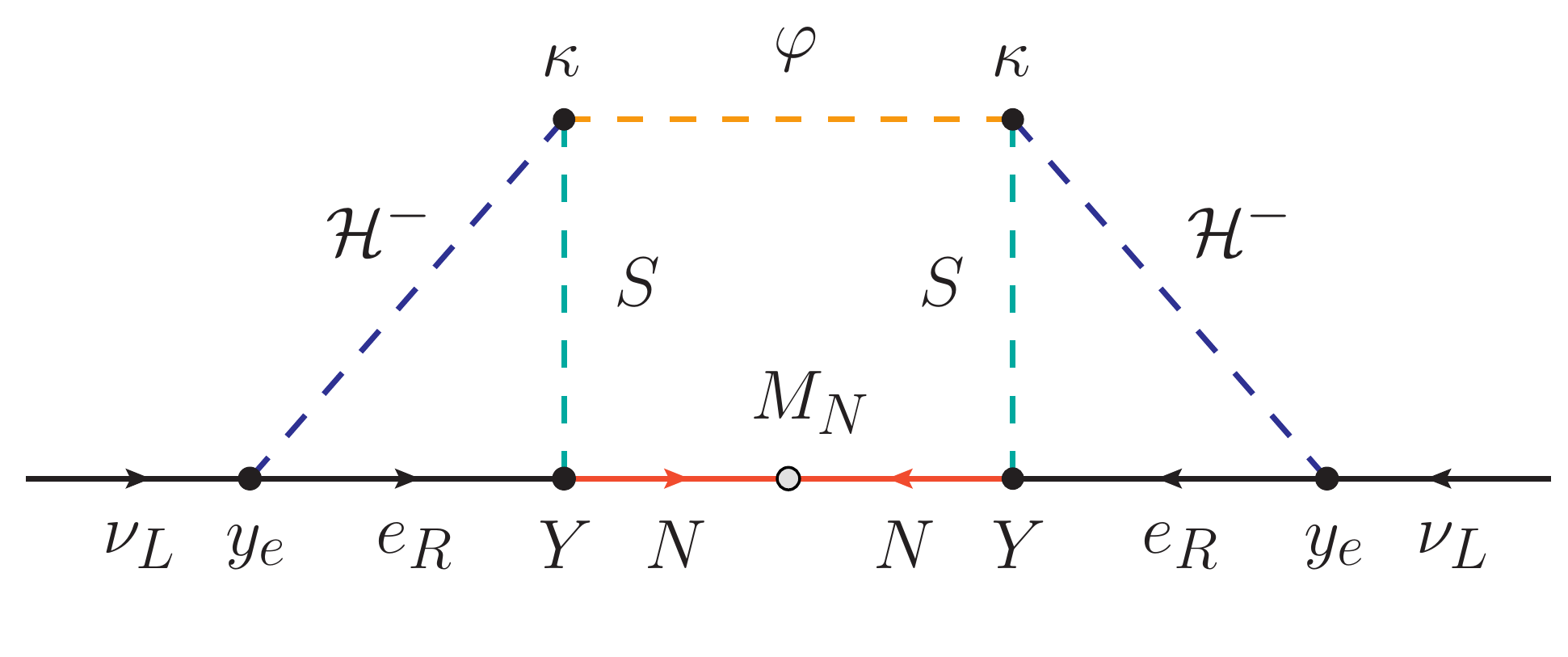}
\includegraphics[width=0.48\textwidth]{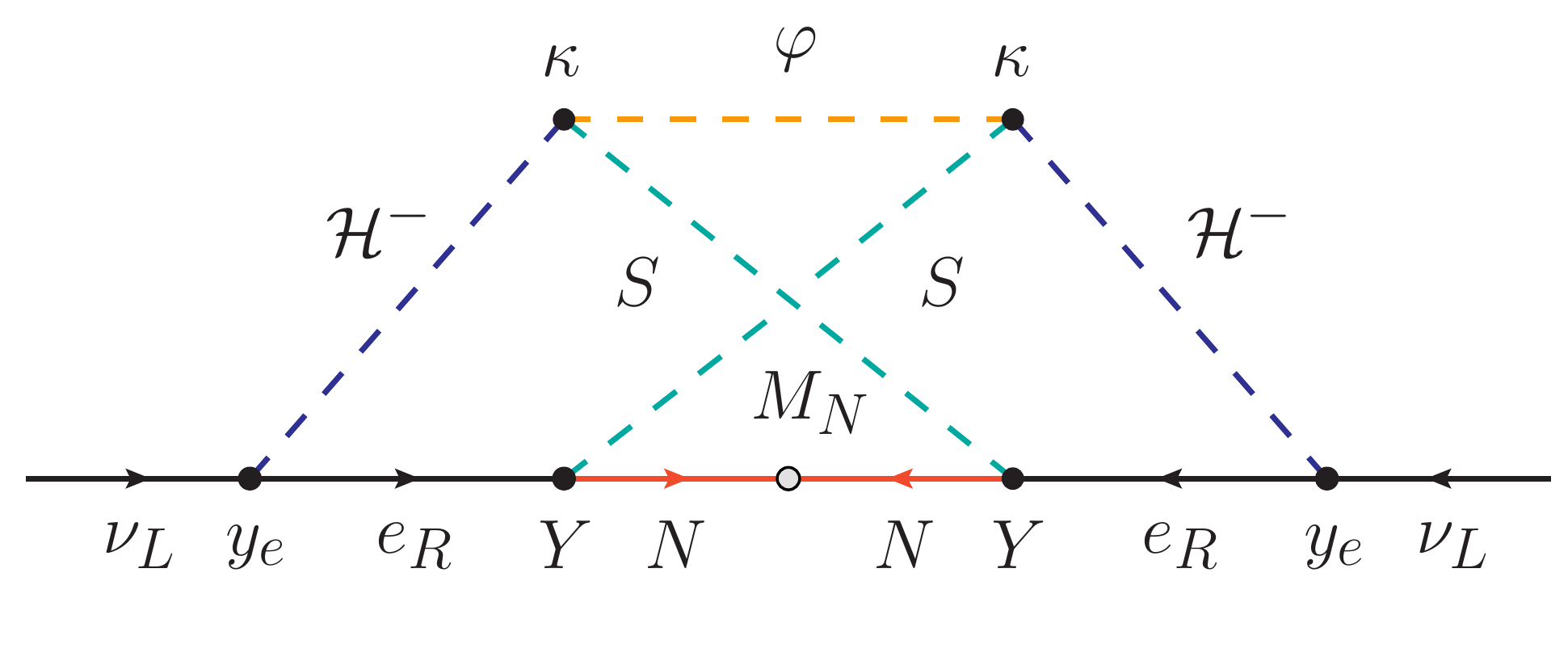}
\caption{3-loop neutrino masses in the AKS model. $\mathcal{H}^- \equiv
  \mathcal{H}^-_{1,2}$ represent the singly charged scalars in the
  model, obtained after diagonalizing the mass matrix of the $\left\{
  \Phi_1^- , \Phi_2^- \right\}$ states.
  \label{fig:AKS}
  }
\end{figure} 

In the AKS model, neutrino masses are induced at 3-loop order, as
shown in the diagrams of Fig.~\ref{fig:AKS}. The resulting neutrino mass
matrix is given by
\begin{align} \label{eq:Mnu AKS}
\left(\mathcal{M}_\nu\right)_{ij} = \frac{\kappa^2 \, \tan^2 \beta}{(16\pi^2)^3} \, \sum_{\alpha \beta} \frac{m_i \, Y_{i \alpha} Y_{j \beta} \, m_j}{(M_N)_{\alpha \beta}} \, F_{\rm AKS} \,,
\end{align}
where $m_i$ is the mass of the $i$-th charged lepton and $F_{\rm AKS}$
is a dimensionless loop function that depends on the masses of the
scalars and fermions in the loop. More details about the calculation
of this loop function can be found in Appendix~\ref{app:int-AKS}.

The Yukawa matrix $Y$ in the AKS model does not have any specific
symmetry. Therefore, this model represents the general class of models
in which the Yukawa matrices can be described by using a
generalization of the Casas-Ibarra parametrization~\cite{Casas:2001sr}
(see also \cite{Cordero-Carrion:2018xre,Cordero-Carrion:2019qtu}).

\subsection{Results}
\label{subsec:results-AKS}

The Yukawa structure of the neutrino mass matrix shown in
Eq.~\eqref{eq:Mnu AKS} resembles that of the type-I seesaw. In order
to fit the experimental oscillation data, we use the Casas-Ibarra
parametrization introducing the neutrino mass matrix in the flavor
basis given in Eq.~\eqref{eq:Mnu}. We find that
\begin{equation} \label{eq:YAKS_fit}
    Y = \frac{i\, (16 \pi^2)^{3/2}}{\kappa \, \tan \beta} \, {\cal R} \, \sqrt{M_N / F_{\rm AKS}} \, \sqrt{\widehat{\cal M}_\nu} \, U^\dagger \, \widehat{\cal M}^{-1}_e \, ,
\end{equation}
where $M_N$ has been taken to be diagonal and ${\cal R}$ is an
arbitrary complex $3\times 3$ orthogonal matrix. We include $F_{\rm
  AKS}$ as it is, in general, a function of the eigenvalues of $M_N$,
see Appendix~\ref{app:int-AKS}. Similar to the KNT model, the presence
of $\widehat{\cal M}^{-1}_e$ in the fit, implies the enhancement of
each column of the Yukawa matrix in terms of the charged lepton
masses, i.e. $Y_{\alpha i} \propto 1/m_i$. This leads to unacceptably
large Yukawa entries in the first column. For instance, choosing NH
with $m_{\nu_1}=0.1$ eV, setting all the phases to $0$ for simplicity,
${\cal R} = \mathbb{I}$ and $(M_N)_{ii} = m_{N}$, we find
\begin{equation} \label{eq:Y_example}
  Y \simeq
 \begin{pmatrix}
 320  &  -0.88  &   0.038 \\
 220  &   0.93  &  -0.074 \\
 160  &   1.45  &   0.079
 \end{pmatrix}
   \Big(\frac{1}{\kappa \, \tan \beta}\Big)
   \Big(\frac{m_{N}}{100\,\text{GeV}}\Big)^{1/2}
   \Big(\frac{1}{F_{\rm AKS}}\Big)^{1/2} \, ,
\end{equation}
clearly in the non-perturbative regime. Insisting on perturbative
Yukawa couplings thus calls for cancellations, especially in the first
column, proportional to $1/m_e$. Moreover, even if for a choice of
parameters, the Yukawa lives at the edge of perturbativity, one should
take care of the constraints coming from CLFV. Especially $\mu \to e
\gamma$, given the hierarchy among the entries of the Yukawa matrix
$Y$.

In order to avoid non-perturbativity and CLFV constraints, first we
exploit the freedom in ${\cal R}$. We fix two of the complex angles to
make two entries of the Yukawa matrix zero or close to zero. We choose
$Y_{21}$ and $Y_{31}$. With this, we find that the third free angle of
${\cal R}$ is not enough to cancel another entry in the Yukawa
matrix. Therefore, we can only fix the values of the phases and
$m_{\nu_1}$ to minimize or cancel $Y_{11}$, similarly to the cocktail
model, or $Y_{12}$, to live below the experimental limit on Br(\mueg),
proportional to $|Y_{k1} Y_{k2}^\ast|^2$. From now on, we also
consider $\kappa = 4\pi$, at the edge of perturbativity, and $\tan
\beta = 1$.

\begin{figure}
\centering
\includegraphics[width=0.47\textwidth]{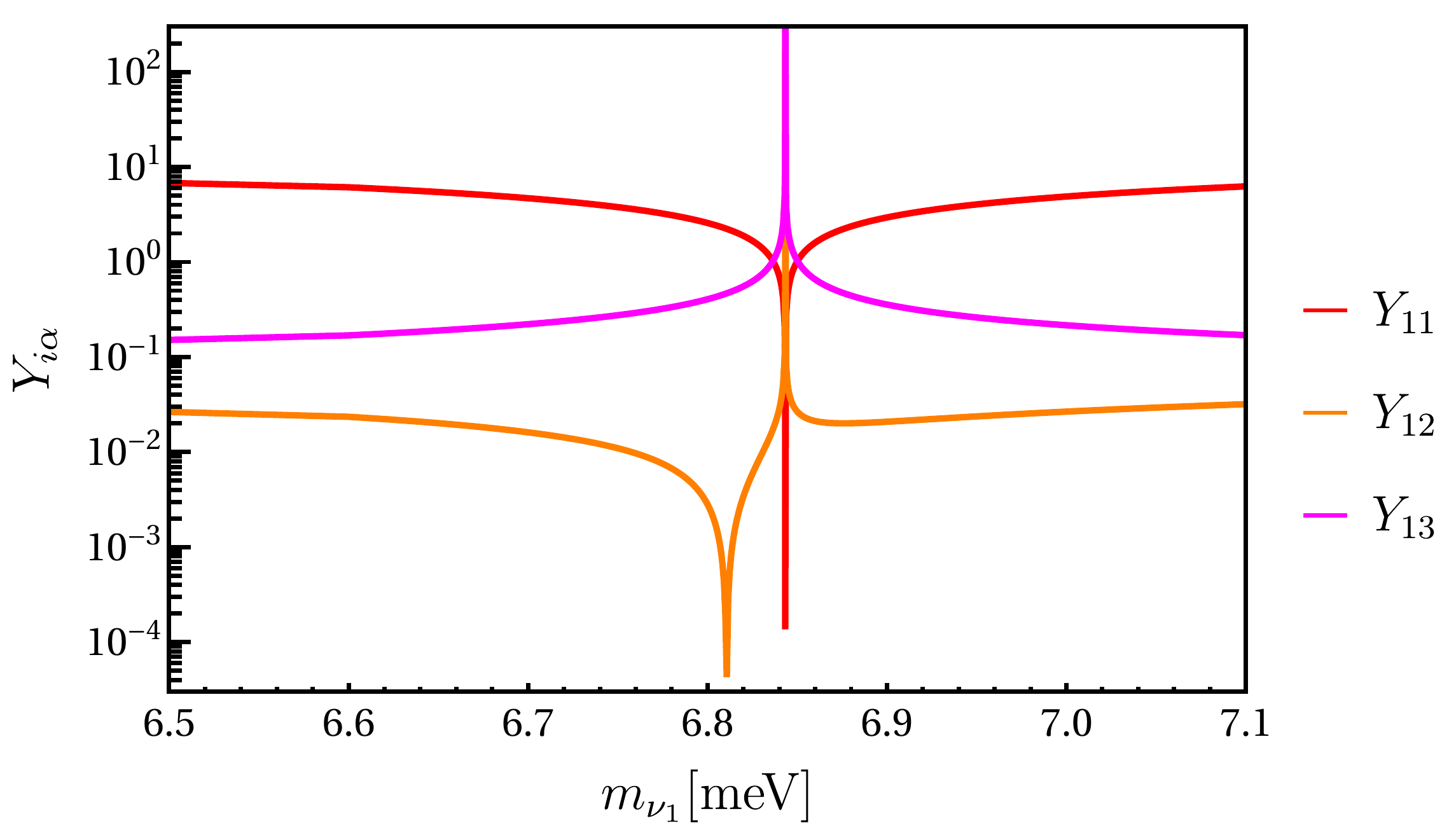}
\includegraphics[width=0.51\textwidth]{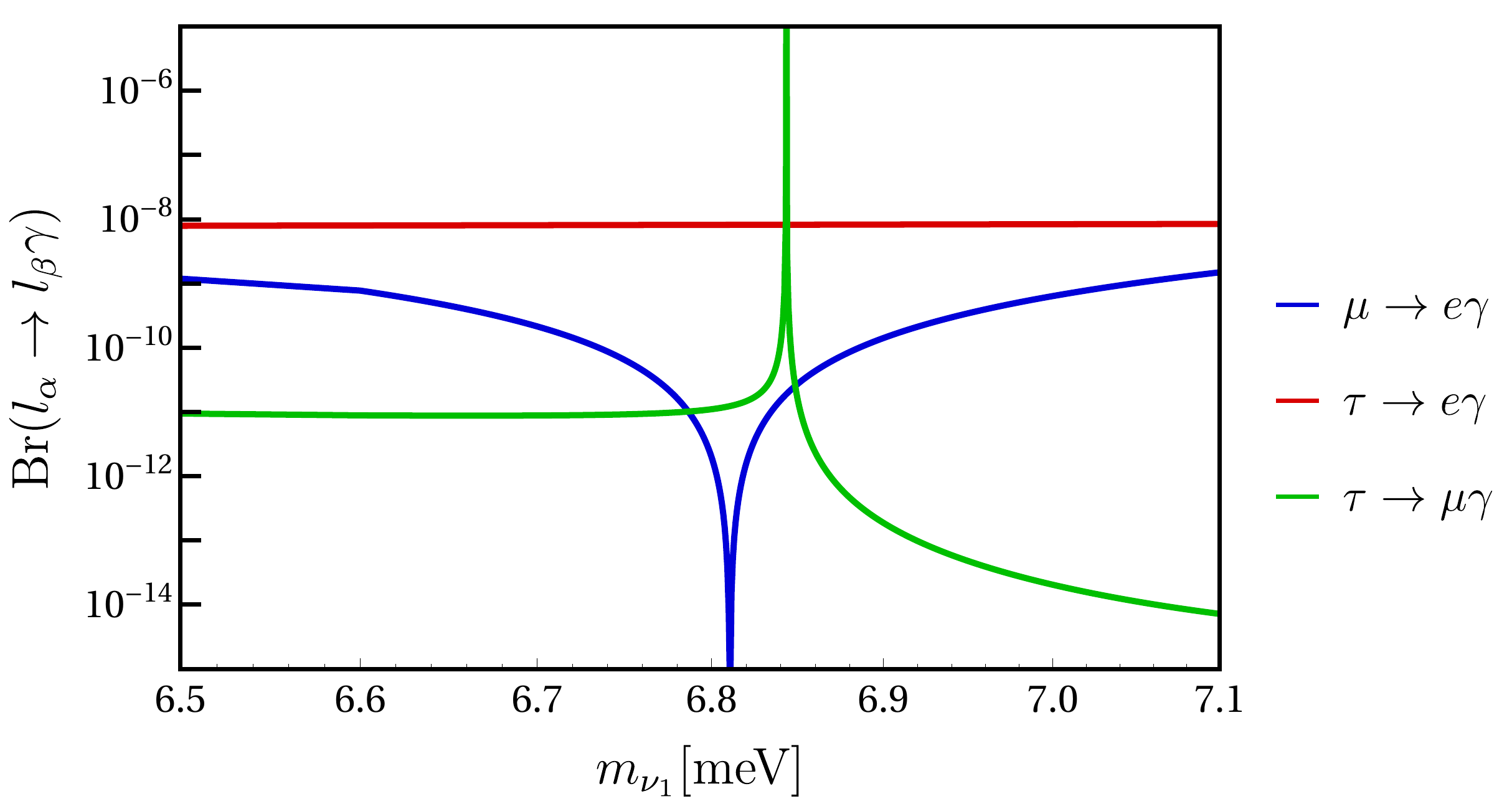}
\caption{To the left, the entries of the first row of the Yukawa
  coupling matrix $Y$ \textit{zoomed} around the poles for $Y_{11}$
  and $Y_{12}$ for $\alpha_{12}=\alpha_{13}=\delta=\pi$. To the right,
  the calculated Br($l_\alpha \to l_\beta \gamma$). Both computed
  fixing ${\cal R}$ for $Y_{21} = Y_{31} = 0$ and at the minimum
  allowed value of $m_N / F_{\rm AKS}$. While a pole for $Y_{11}$
  exists, no pole for Br(\mueg) or Br(\taueg) is associated to it, due
  to the divergence of $Y_{12}$ and $Y_{13}$ on the pole. Note that
  only near the pole for $Y_{12}$ Br(\mueg) is below the experimental
  limit.
  \label{fig:AKS_yuks}
  }
\end{figure}

In Fig.~\ref{fig:AKS_yuks} to the left, we show the behavior of the
first row of $Y$ for $\alpha_{12}=\alpha_{13}=\delta=\pi$. We
considered for simplicity that all the scalar masses are equal to
$m_{S^+} = m_\varphi = m_{\mathcal{H}^\pm} \equiv m_S$ and all the $N$
singlet fermion masses to be degenerate, $m_N$, and minimize $m_N /
F_{\rm AKS}$ to find the lowest value of $Y$, see
Eq.~\eqref{eq:YAKS_fit}. We found this minimum for $m_N = 272$ GeV and
$m_{\mathcal{S}} = 100$ GeV, where $F_{\rm AKS} \approx 0.44$,
compatible with the limit on scalar masses from
LEP~\cite{Tanabashi:2018oca}. Here we do not show the other four
non-zero Yukawas for simplicity. They are nearly constant and of order
$0.1$. Similar to the cocktail model (Fig.~\ref{fig:Yuks}), poles
exist in the different Yukawa entries for particular values of the
phases and $m_{\nu_1}$. The main difference lies in the divergence
that appears when $Y_{11}=0$. This is caused by our choice of
${\cal R}$ matrix, such that $Y_{21} = Y_{31} = 0$. In this case, the
pole in $Y_{11}$ does not imply a pole in Br(\mueg) or Br(\taueg), as
it can be seen in Fig.~\ref{fig:AKS_yuks} to the right. In fact, the
product of $|Y_{11} Y_{13}^*|$ remains constant over the pole and very
close to the current experimental limit of $3.3 \times
10^{-8}$~\cite{Aubert:2009ag}. Only the region around the pole in
$Y_{12}$ is allowed by the experimental limit Br(\mueg)$< 4.2 \times
10^{-13}$~\cite{TheMEG:2016wtm}.

To sum up, the parameter space of the AKS model is constrained mainly
by perturbativity and Br(\mueg). The former can be addressed with the
freedom in ${\cal R}$ to set $Y_{21}$ and $Y_{31}$ to zero. As well as by
fixing the Majorana and Dirac phases, and the lightest neutrino mass,
to be near the pole of $Y_{11}$, where its value is lower than
$4\pi$. On the other hand, to be below the experimental limit on
Br(\mueg), a similar fine-tuning of the phases and $m_{\nu_1}$ should
be done to be around the \textit{narrow} pole of $Y_{12}$. The
parameter space is then restricted to those values of the phases and
$m_{\nu_1}$ where the poles of $Y_{11}$ and $Y_{12}$ exist, and they
are close enough to each other to avoid the limit on Br(\mueg) while
$Y_{11}$ is still perturbative.

\begin{figure}
\centering
\includegraphics[width=0.7\textwidth]{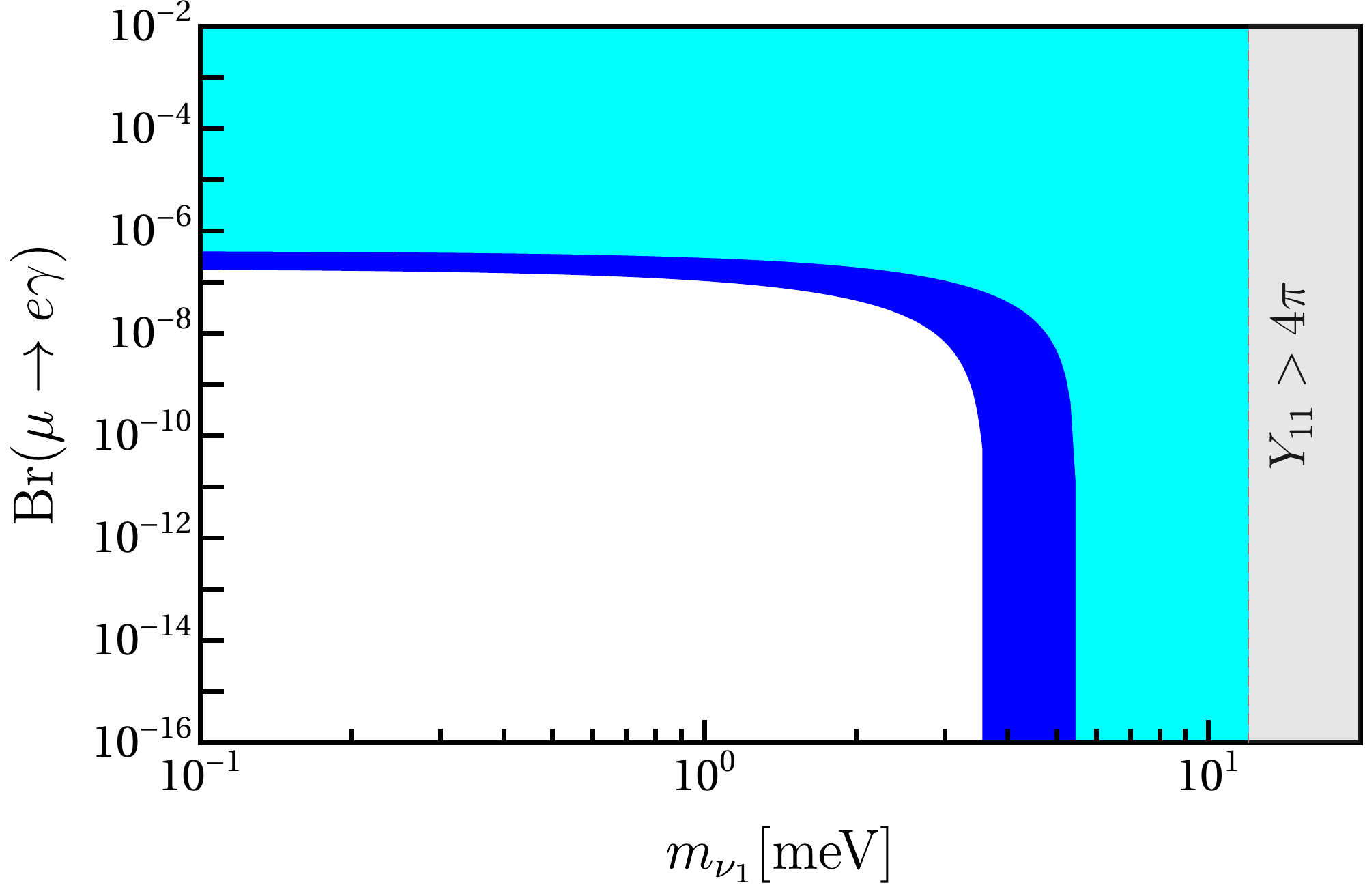}
\caption{Br(\mueg) scanned over neutrino oscillation data in $1\sigma$
  (light blue) and $3\sigma$ (dark blue) ranges. This plot scans over
  the Majorana phases. The shaded gray area corresponds to the most
  conservative limit to non-perturbative Yukawas.
  \label{fig:AKS_meg}
  }
\end{figure}

In Fig.~\ref{fig:AKS_meg} we show the value of Br(\mueg) scanning over
the complete range of oscillation parameters (NH) and phases. On the
right, we give the limit due to perturbativity of $Y_{11}$, reducing
the parameter space to a small window of $m_{\nu_1} = (4.5 - 20)$
meV. Note that like in the cocktail model, $Y_{11}$ behaves as
$m_{ee}$, and for $m_{\nu_1} \gsim 10$ meV, $m_{ee}$ has no pole, so
$Y_{11}$ is in the non-perturbative region. Moreover, the cancellation
of $Y_{11}$ and $Y_{12}$ only occurs for NH, so the model can only
explain neutrino data with this neutrino mass ordering. In the
following, we shall consider only NH.

\begin{figure}
\centering
\includegraphics[width=0.7\textwidth]{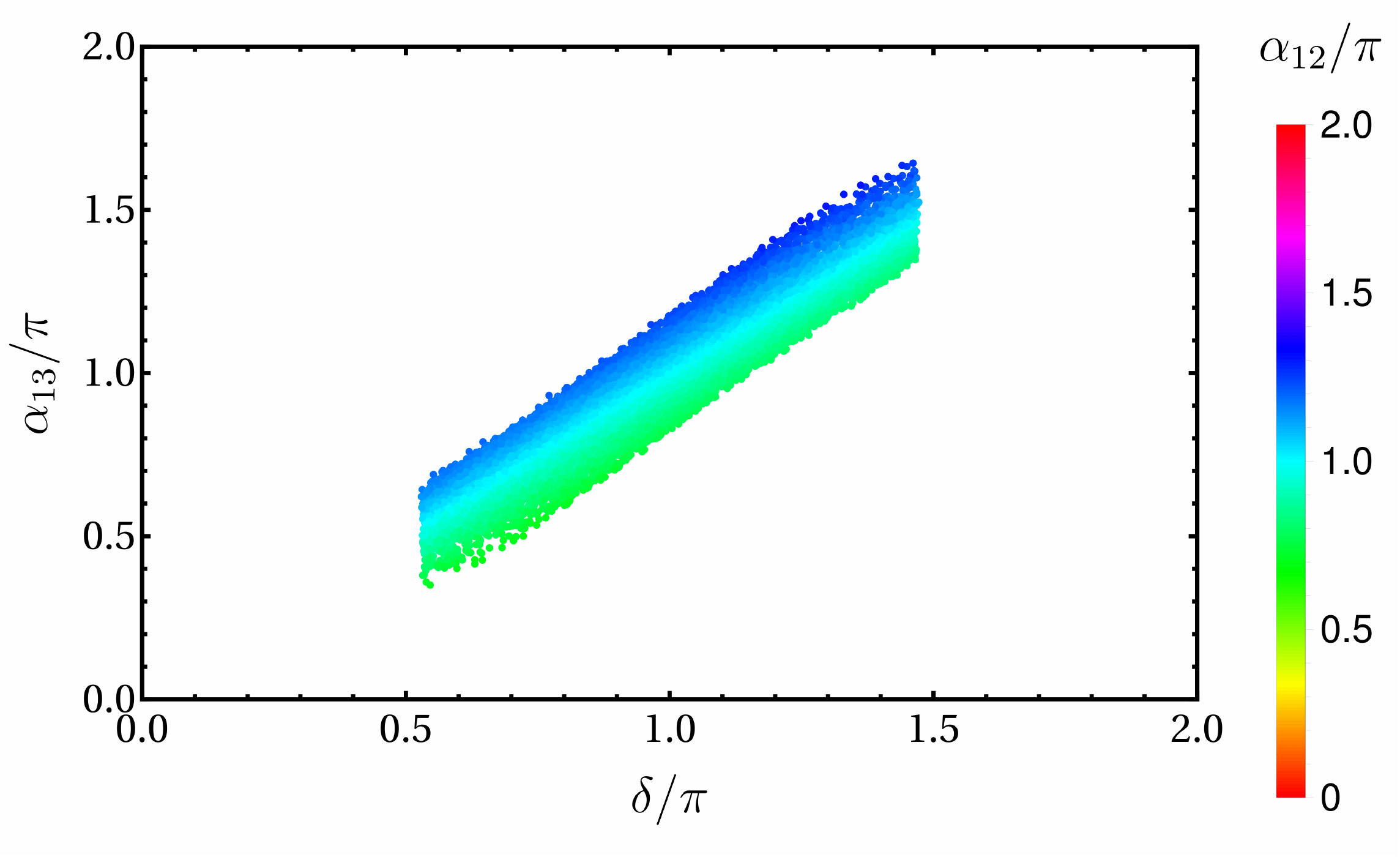}
\caption{Allowed parameter space for $\alpha_{12}$, $\alpha_{13}$ and
  $\delta$. Neutrino oscillation data was scanned over the $3\sigma$
  uncertainties, except $\delta$ which was left free. For points
  outside this region, either $Y_{11}$ is non-perturbative or
  Br(\mueg) is above the experimental limit.
  \label{fig:AKS_phases}
  }
\end{figure}

Fig.~\ref{fig:AKS_meg} not only implies a constraint on $m_{\nu_1}$,
but also on the phases. In Fig.~\ref{fig:AKS_phases} we show the
points allowed by perturbativity and the experimental limits on
Br(\mueg), Br(\taueg) and Br(\taumug), for the values of the three
phases. We scanned over the phases and masses, with $m_S > 100$ GeV,
allowing oscillation data to vary in $3\sigma$. As it can be seen,
$\alpha_{12}$ should be closely around $\pi$, while $\delta$ is
constrained to values between roughly $\pi/2$ and $3\pi/2$. For
$\delta$ outside this window, there is no cancellation of $Y_{12}$. In
the following, we restrict the results shown to the region where
Br(\mueg)$< 4.2 \times 10^{-13}$.

\begin{figure}
\centering
\includegraphics[width=0.48\textwidth]{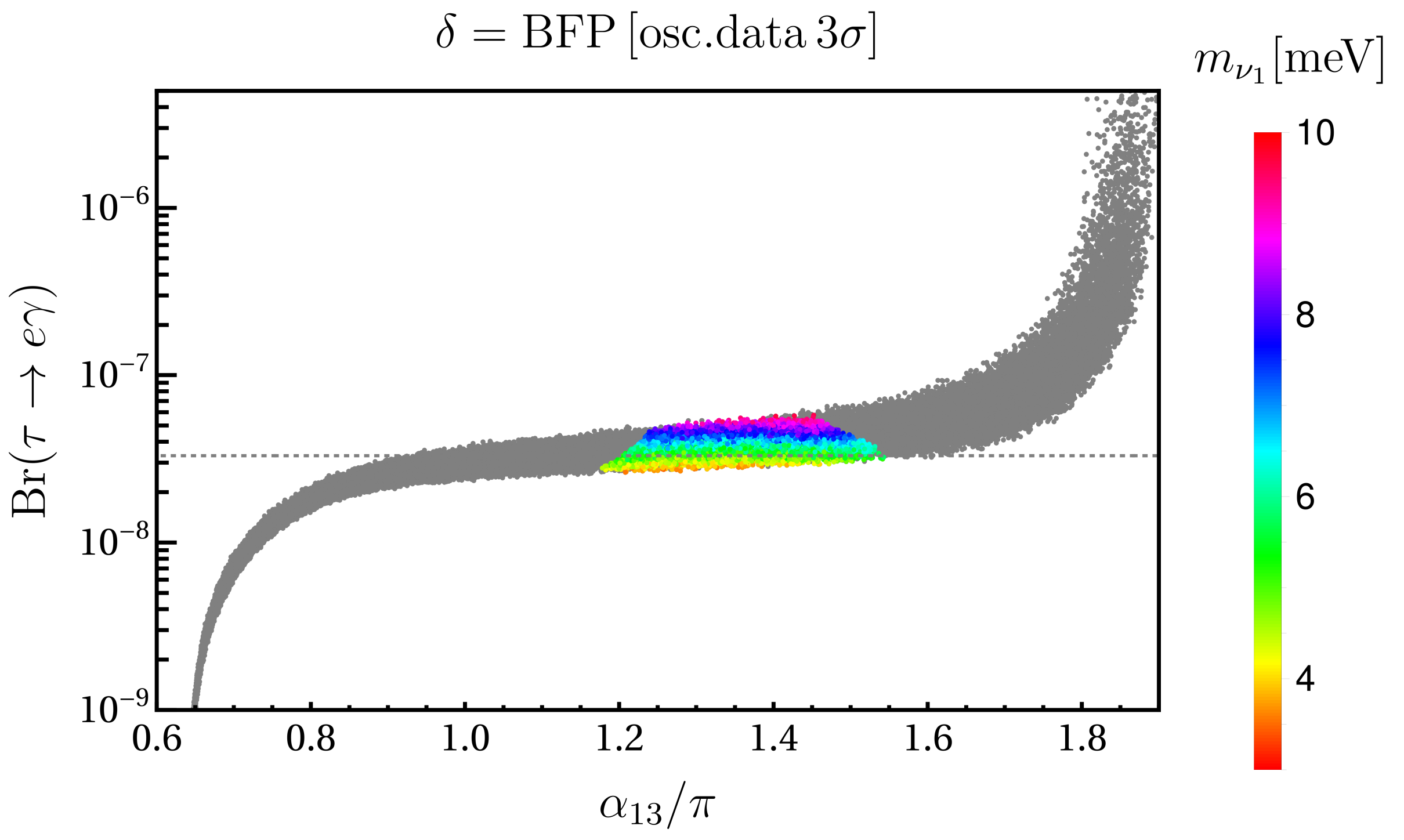}
\includegraphics[width=0.48\textwidth]{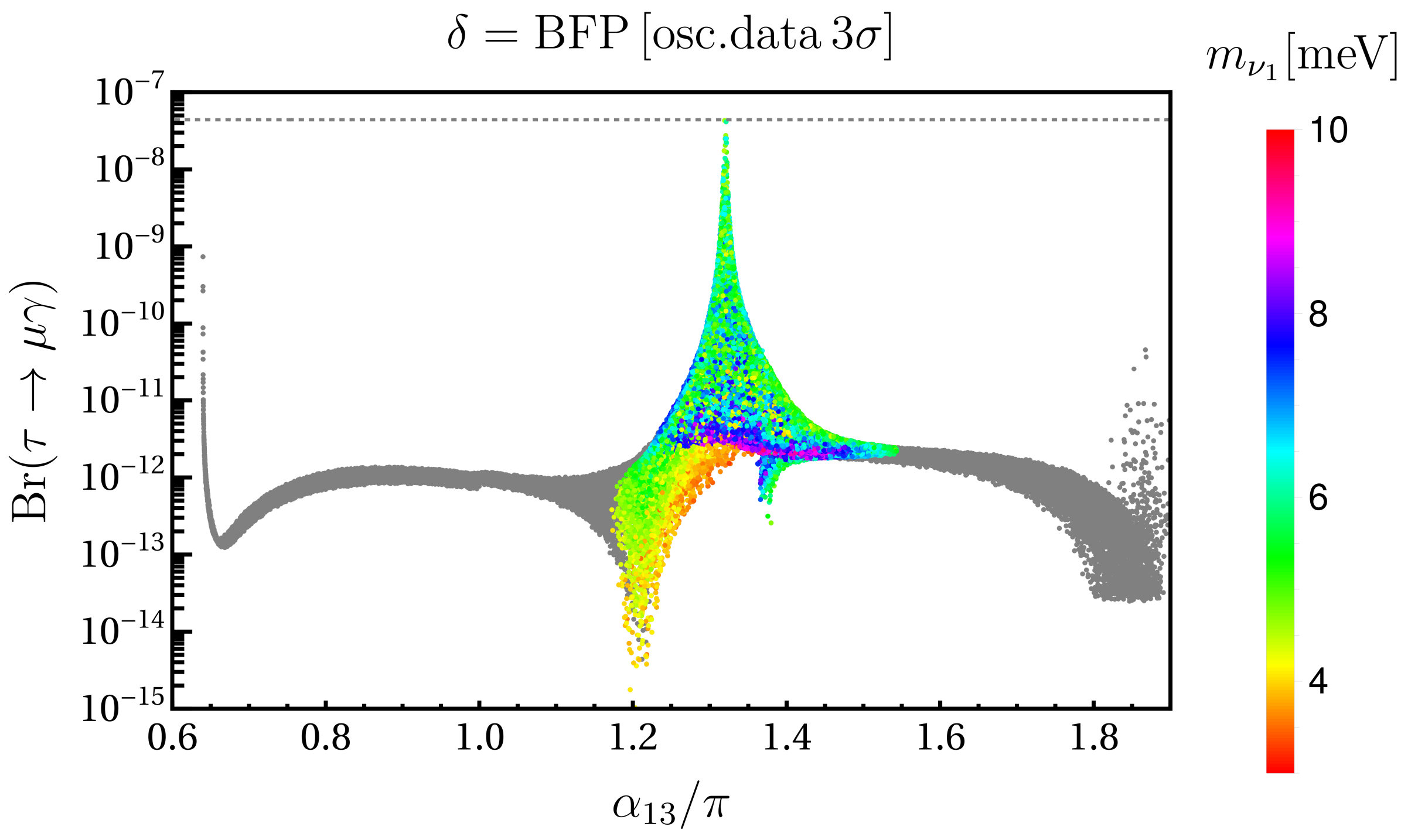}
\caption{Br(\taueg) and Br(\taumug) as functions of $\alpha_{13}$ for
  different values of the lightest neutrino mass (color-coded) along
  with the current experimental limits (dotted line). We scanned over
  $3\sigma$ uncertainties of the oscillation data, except for $\delta$
  which was fixed to the b.f.p. Gray points are excluded due to
  perturbativity arguments.
  \label{fig:AKS_tau}
  }
\end{figure}

\begin{figure}
\centering
\includegraphics[width=0.48\textwidth]{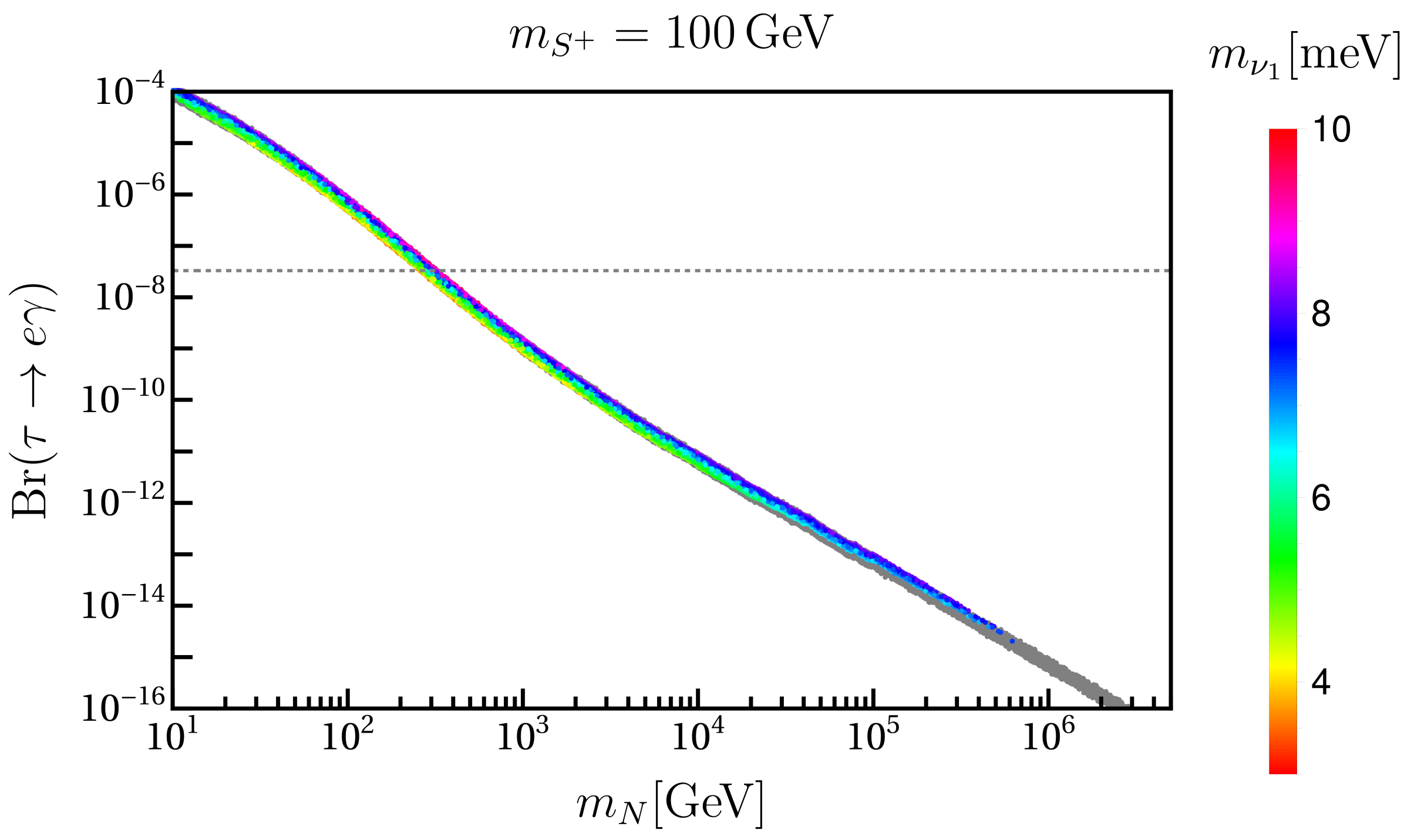}
\includegraphics[width=0.48\textwidth]{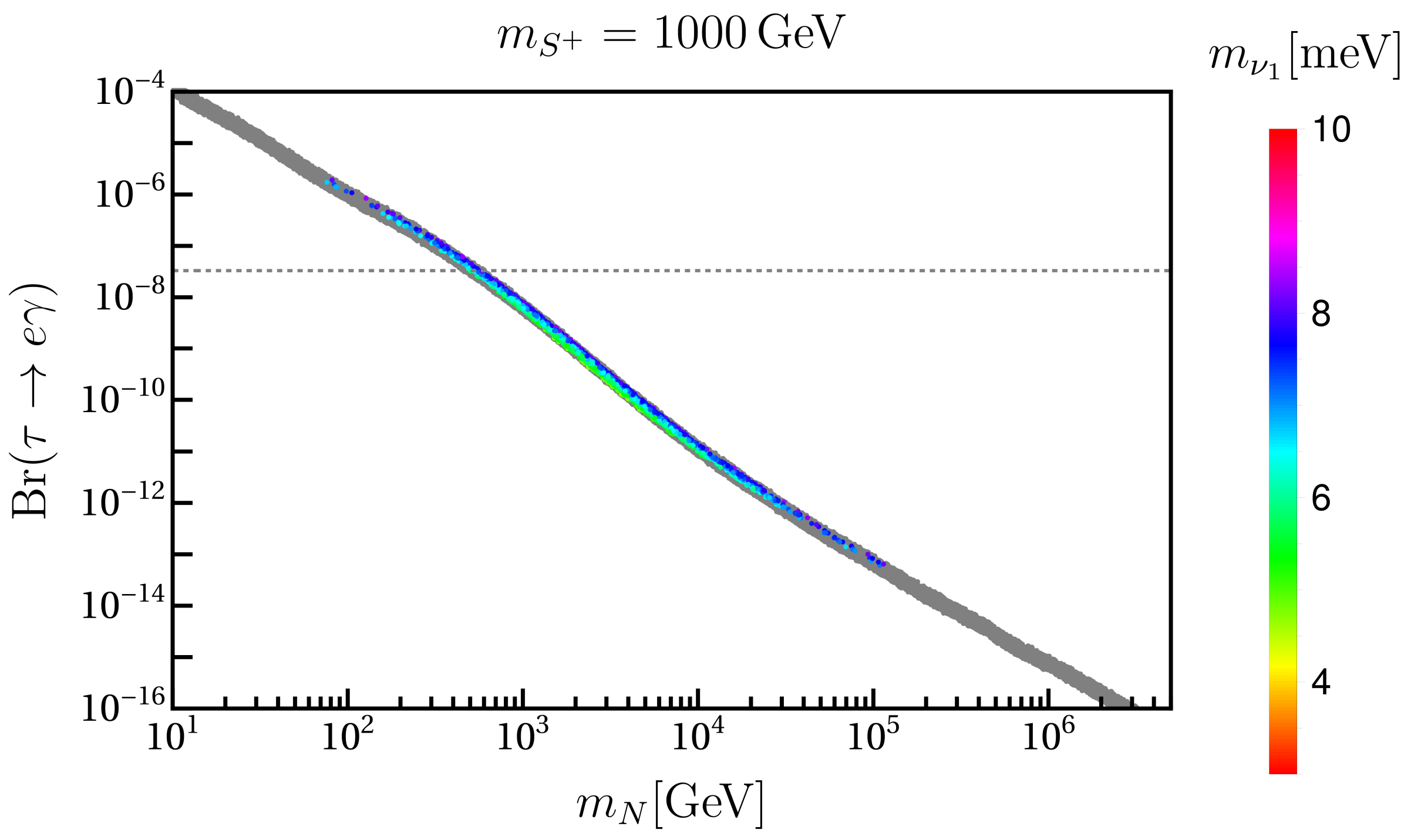}
\caption{Calculated values for Br(\taueg) as function of $m_N$ for
  different values of $m_{S^+}$. Here, we are maximizing the allowed
  parameter space in terms of $\alpha_{13}$. For the gray points, at
  least one entry in $Y$ is larger than $4\pi$.
  \label{fig:AKS_mnms}
}
\end{figure}

Now we move to analyse $\tau \to e \gamma$ and \taumug. As shown in
Fig.~\ref{fig:AKS_yuks} (right), while Br(\taumug) is below the
experimental limit, except on the pole of $Y_{11}$, Br(\taueg) is
mainly constant and close to the experimental limit. In
Fig.~\ref{fig:AKS_tau} we give both branching ratios fixing $\delta$
to the b.f.p. and scanning over the uncertainties in the rest of the
oscillation parameters. We consider $(m_N / F_{\rm AKS})_{min}$ with
$m_S = 100$ GeV and $m_N = 272$ GeV. Points colored in gray correspond
to non-perturbative Yukawas. We see that while Br(\taumug) is
\textit{safe}, the allowed region on the left plot is severely
constrained by the experimental limit Br(\taueg)$<3.3 \times
10^{-8}$. This tension can be mitigated by raising the masses, see
Fig.~\ref{fig:AKS_mnms}. For the AKS model, the dominant contribution
to Br($l_\alpha \to l_\beta \gamma$) is approximately proportional to
$1/M^4$, with $M$ the dominant scale~\cite{Lavoura:2003xp}. On the
other hand, $m_N / F_{\rm AKS}$ is minimal for masses around
$m_S = 100$ GeV and $m_N = 272$ GeV. So for masses away from these
values, $m_N / F_{\rm AKS}$ increases and, consequently, the absolute
scale of the Yukawas increases as well (see Eq.~\eqref{eq:YAKS_fit}),
hence narrowing the region where the Yukawas are perturbative. For
$m_N (m_{S^+}) \sim 10^6$ GeV, we found no points allowed by
perturbativity and the experimental limit on Br(\mueg). In
Fig.~\ref{fig:AKS_mnms}, in order to minimize the Yukawas, we fixed
$m_\varphi = m_{\mathcal{H}^{\pm}} = 100$ GeV and change $m_{S^+}$ and
$m_N$, which enter in the calculation of Br($l_\alpha \to l_\beta
\gamma$).

\begin{figure}
\centering
\includegraphics[width=0.565\textwidth]{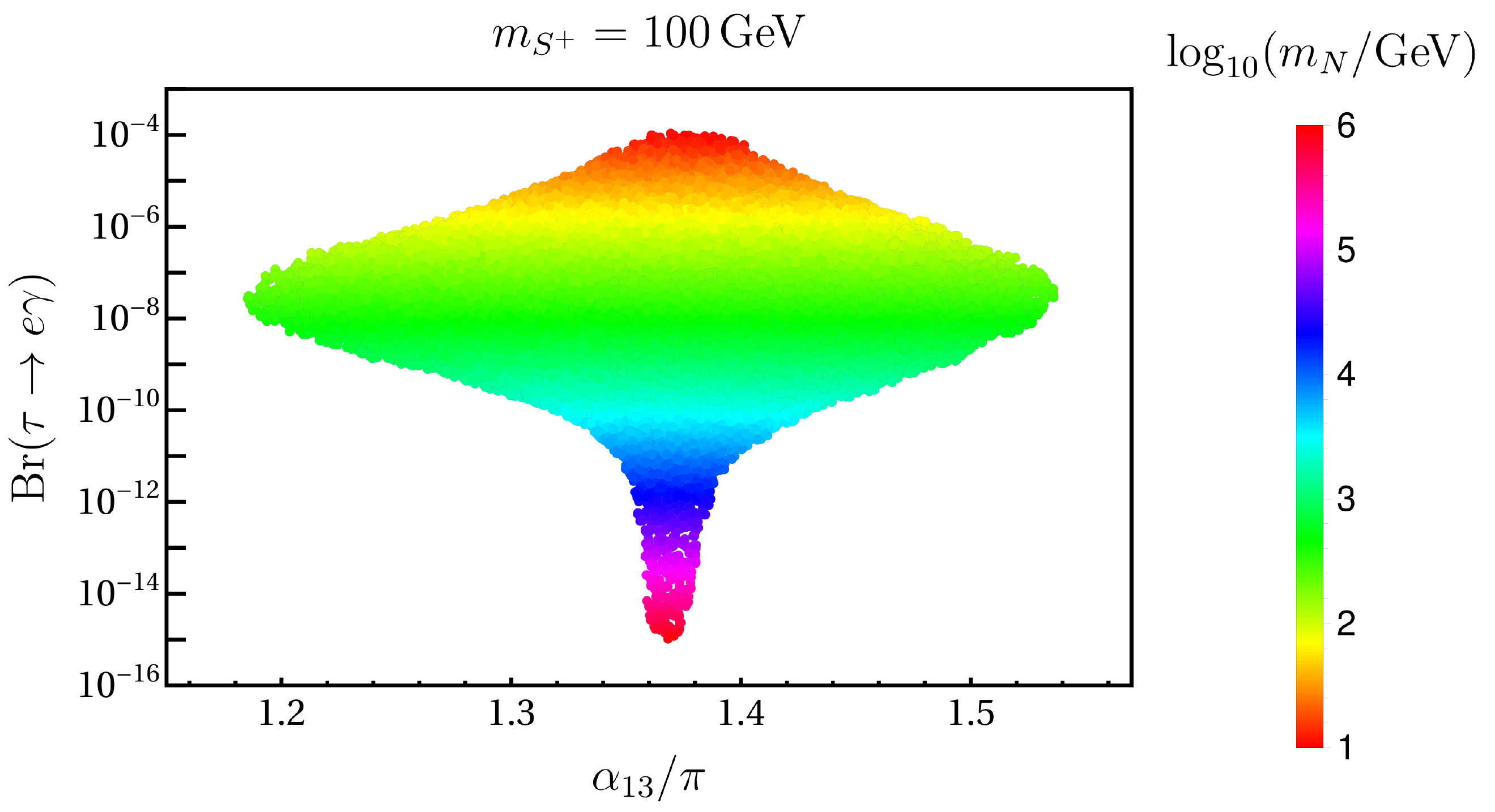}
\includegraphics[width=0.427\textwidth]{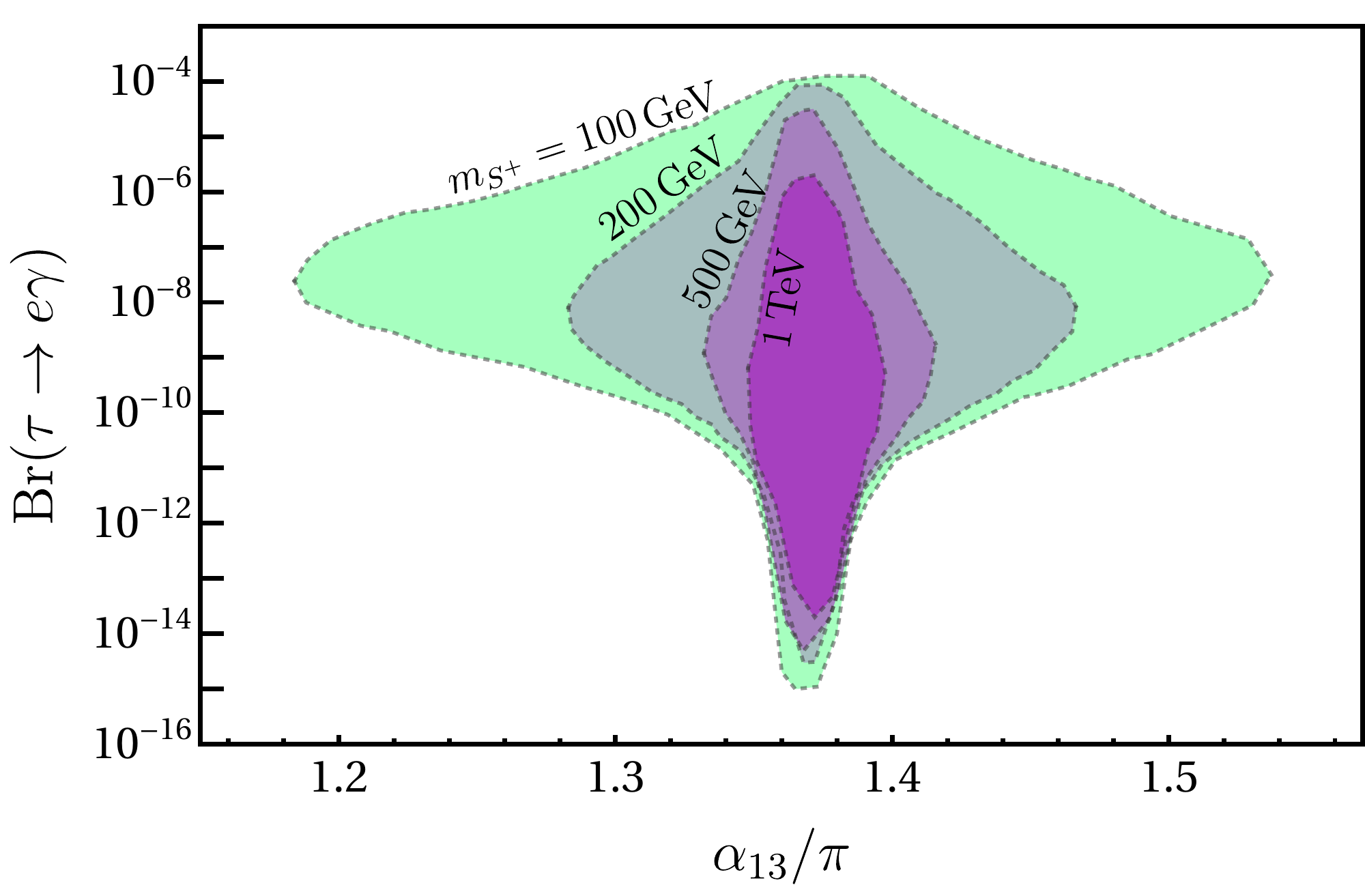}
\caption{Br(\taueg) for different values of $m_N$ and $m_{S^+}$. To
  the left, we fixed $m_{S^+}$ and see how by modifying $m_N$ the
  parameter space widens or narrows due to perturbativity arguments. A
  similar behavior can be observed for the plot on the right, where we
  show contour lines for different values of $m_{S^+}$ scanning over
  $m_N$.
  \label{fig:AKS_alp13_m}
  }
\end{figure}

A similar analysis can be done scanning over the Majorana phases
too. Fig.~\ref{fig:AKS_alp13_m} shows Br(\taueg) as a function of
$\alpha_{13}$ for different fermion and scalar masses. The allowed
parameter space is \textit{bigger} for $m_N$ around $272$ GeV, where
$m_N / F_{\rm AKS}$ is minimal. For different masses the parameter
space \textit{narrows}, because $m_N / F_{\rm AKS}$ increases, as
explained before. The upper limit is due to the phenomenological limit
$m_{S^+} > 100$ GeV, as for $m_N \ll m_{S^+}$, Br(\taueg) is dominated
by $m_{S^+}$. On the other side, while going to larger $m_N$ reduces
considerably Br(\taueg), a lower limit always exists due to
perturbativity.\\

We close this discussion with a short comment on $0\nu\beta\beta$
decay.  There is no short-range diagram for $0\nu\beta\beta$ decay in
the AKS model. Since, as discussed above, the AKS model survives only
for normal hierarchy and in the part of parameter space where
$m_{ee}$ is largely cancelled, observation of  $0\nu\beta\beta$
decay in the next round of experiments would definitely rule out
AKS as an explanation of neutrino masses.

\section{Discussion}
\label{sec:discussion}

In this paper we have considered the cocktail, KNT and AKS models and
studied their CLFV phenomenology. In these models, Majorana neutrino
masses are generated at the 3-loop order, which naturally implies that
large Yukawa couplings are required in order to reproduce the mass
scales observed in neutrino oscillation experiments. As a result of
this, perturbativity is typically lost. We have shown that one can
decrease the Yukawa couplings by tuning some of the free parameters of
these scenarios, such as the lightest neutrino mass $m_{\nu_1}$ or the
Dirac and Majorana phases contained in the leptonic mixing matrix
$U$. However, even after these parameters are tuned to recover
perturbativity, the resulting CLFV branching ratios tend to largely
exceed the existing bounds. In order to reduce the CLFV rates further
tuning is needed. Our main conclusion is that the three models survive
only in tiny correlated regions of their parameter spaces.

One should note that CLFV alone cannot exclude any of these
models. The reason is that one can always reduce the CLFV rates as
much as necessary by tuning the parameters of the model more
finely. However, additional experimental handles exist. First,
perturbativity imposes upper limits on the masses of some of the
particles running in the loops. The reason is simple: larger mediator
masses would imply a stronger suppression of the loop functions and
then require larger Yukawa couplings.  Thus, also future searches at
the LHC in the high-luminosity phase would further restrict the
available parameter space. An important experimental handle on the
models is $0\nu\beta\beta$ decay.  Since $m_{\nu_1}$ and the Majorana
phases must be tuned for the models to survive, the effective
$0\nu\beta\beta$ neutrino mass $m_{ee}$ becomes strongly constrained
and definite predictions for the $0\nu\beta\beta$ rates are obtained
for the AKS and KNT models.  Any observation of $0\nu\beta\beta$ decay
with the next generation of experiments would definitely rule out the
AKS model. For the KNT model, because $m_{\nu_1}\simeq 0$, $m_{ee}$
has to be either in the range $m_{ee} \simeq (2-6)$ meV or $(15-50)$
meV for normal hierarchy or inverted hierarchy. Only the cocktail
model is more flexible in its predictions for $0\nu\beta\beta$ decay,
due to additional contributions from a sizable short-range diagram.

We mention also that only the KNT model can explain neutrino data for
both hierarchies. Neither the cocktail nor the AKS model has any
acceptable point in all of their parameter space in the case of
inverse hierarchy.

A crucial ingredient in our analysis is the allowed size for the
quartic scalar potential couplings that play a role in the neutrino
mass generation mechanism, for example $\lambda_5$ in the cocktail model,
$\lambda_S$ in the KNT model and $\kappa$ in the AKS model. Since
neutrino masses are proportional to (some power of) these couplings,
the larger they are, the smaller the Yukawa couplings can be. In our
analysis, scalar couplings as large as $4 \pi$ have been allowed. A
more restrictive choice, with couplings at most of $\mathcal{O}(1)$,
would alter the conclusions dramatically. In fact, all three models
would already be ruled out, if all their couplings are restricted to
be not larger than $\mathcal{O}(1)$.

Finally, we emphasize again that our strong claims only apply to the
three minimal models considered here. There are several ways to modify
these models so that they can evade the perturbativity and flavor
constraints. For instance, one can introduce new exotic states in
order to get rid of the proportionality to the charged lepton masses,
at the origin of the problems discussed in our paper. Also, one may
enhance the contributions to the neutrino mass matrix by using colored
states.  Nevertheless, we also note that there may be many other
3-loop (or 4-loop) neutrino mass models with the same issues.

\appendix

\section{Loop integrals}
\label{app:integrals}

In this Appendix we discuss the calculation of the loop integrals in the cocktail, KNT and AKS models. Here, we derive the loop functions used in the previous sections. For their computation, we did not rely on approximations, but implemented the full integral numerically using {\tt pySecDec}~\cite{Borowka:2017idc}.

Moreover, all the integrals shown here, can be factorized in terms of five master integrals (see for example~\cite{Martin:2016bgz}), as normally done. Nevertheless, we decided not to do it, because there is still no analytical general solution to all the 3-loop master integrals and their factorization could lead to numerical precision issues. Note that these five master integrals have divergent parts, while the full integral is finite.

\subsection{Cocktail model}
\label{app:int-cocktail}

To compute the dimensionless integral $F_{\rm Cocktail}$ in Eq.~\eqref{eq:mnu cocktail}, we choose the Feynman-'t Hooft gauge $\xi=1$. In this gauge, the propagator of the $W_\mu$ boson has no momenta structure in the numerator, while the standard Goldstone $H^+$ contribution with a mass $m_W^2$ should be included. We decided to show the diagrams in the gauge basis to be able to identify the different contributions that enter in $F_{\rm Cocktail}$.

\begin{figure}
    \centering
    \includegraphics[width=0.4\textwidth]{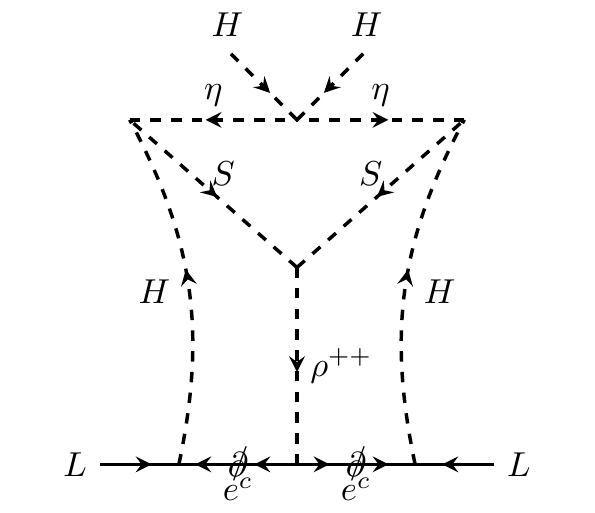}
    \includegraphics[width=0.4\textwidth]{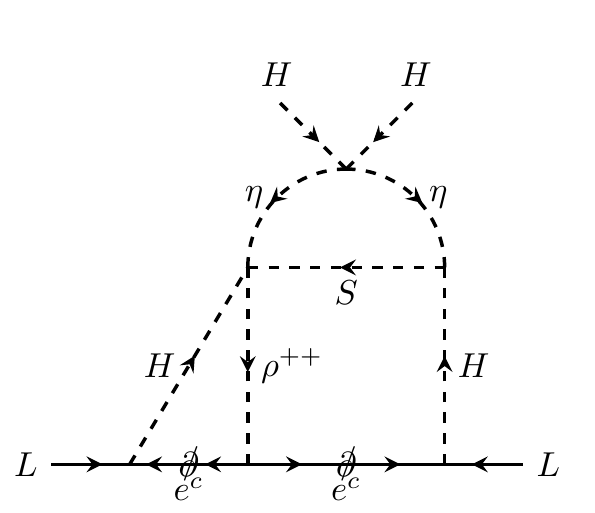}
    \caption{Dimension $5$ mass diagrams in the gauge basis. Note that
      given the chirality, the corresponding integral has two momenta
      in the numerator. This is denoted with
      $\slashed{\partial}$. When referring to these diagrams we will
      use the notation $\mathfrak{I}^{(5)}_i$ with $i=1,2$ following
      the order of the figures.
      \label{fig:CT_diag_d5}
    }
\end{figure}

\begin{figure}[t]
    \centering
    \includegraphics[width=0.4\textwidth]{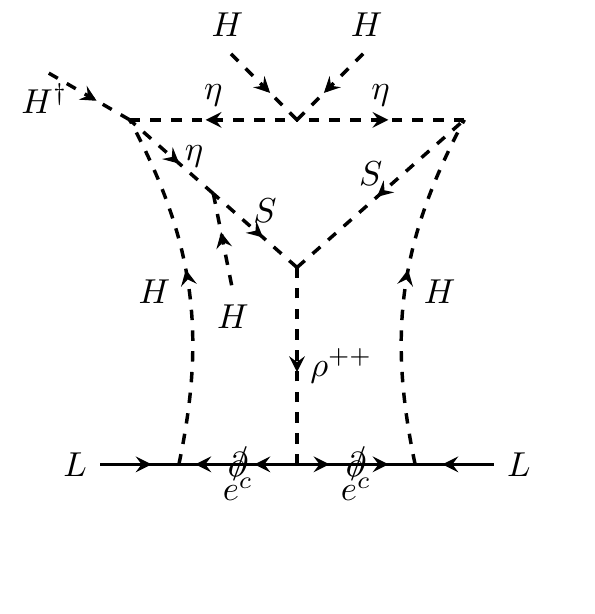}
    \includegraphics[width=0.4\textwidth]{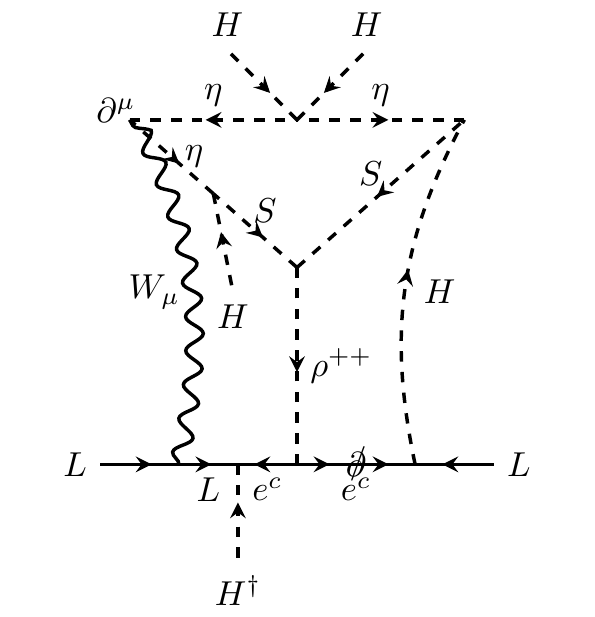}
    \\
    \includegraphics[width=0.4\textwidth]{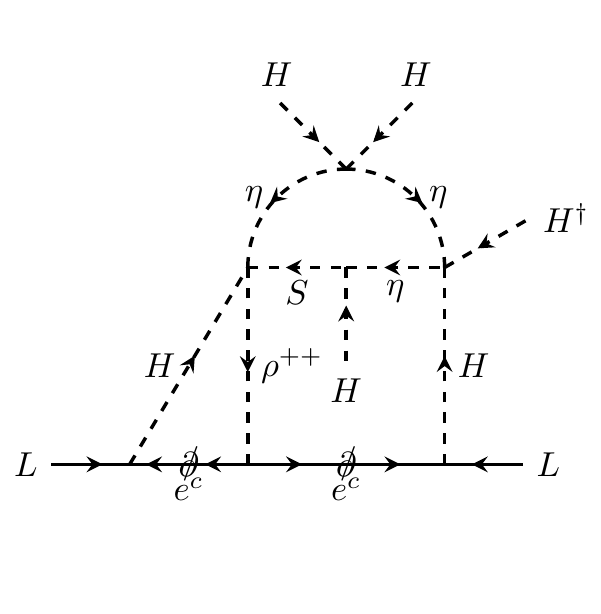}
    \includegraphics[width=0.4\textwidth]{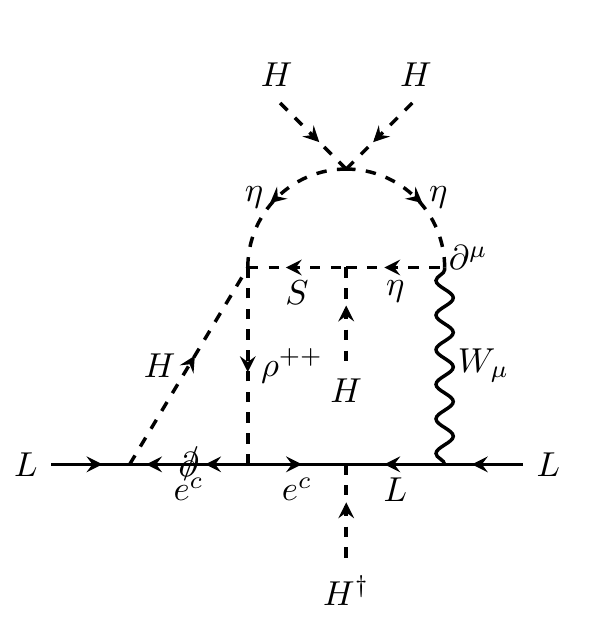}
    \caption{Dimension $7$ mass diagrams in the gauge basis. $W_\mu$
      couples with a derivative to the scalars, so every diagram has
      the same number of derivatives. We will use the notation
      $\mathfrak{I}^{(7)}_i$ with $i=1,4$ following the usual order
      from left to right and top to bottom.
      \label{fig:CT_diag_d7}
      }
\end{figure}

\begin{figure}[t]
    \centering
    \includegraphics[width=0.32\textwidth]{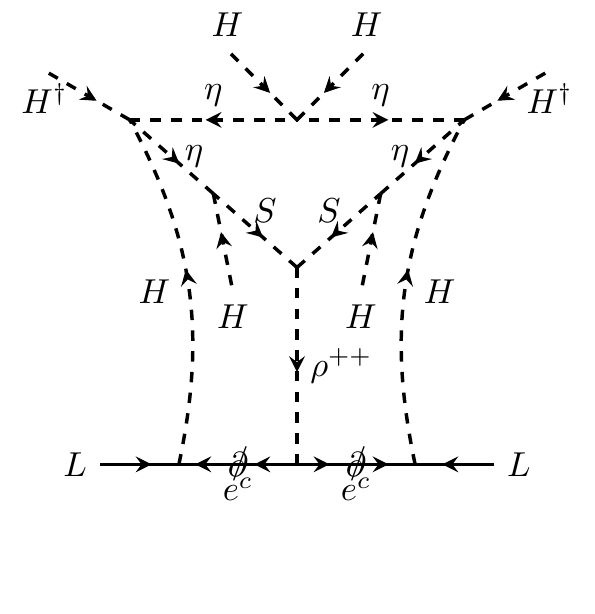}
    \includegraphics[width=0.32\textwidth]{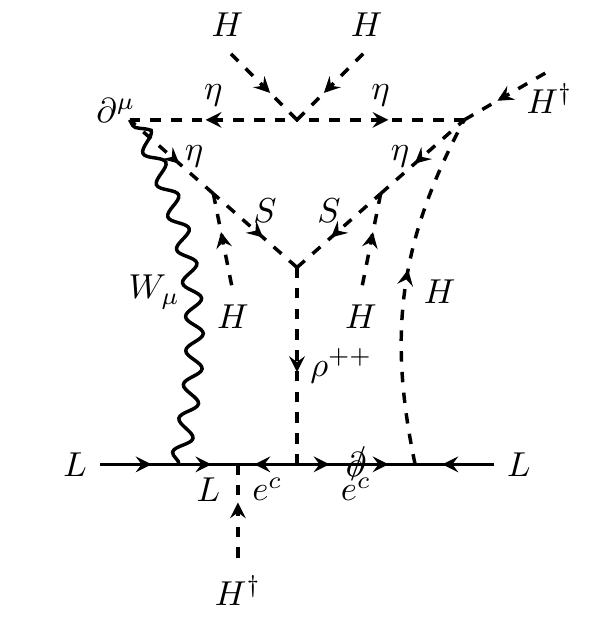}
    \includegraphics[width=0.32\textwidth]{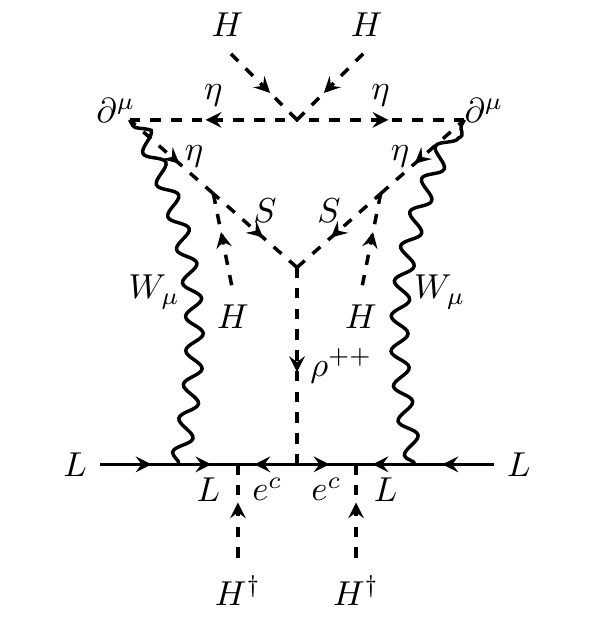}
    \\
    \includegraphics[width=0.32\textwidth]{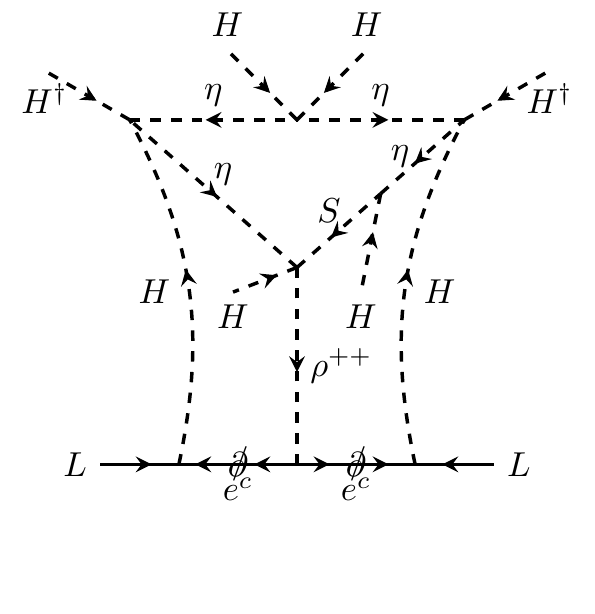}
    \includegraphics[width=0.32\textwidth]{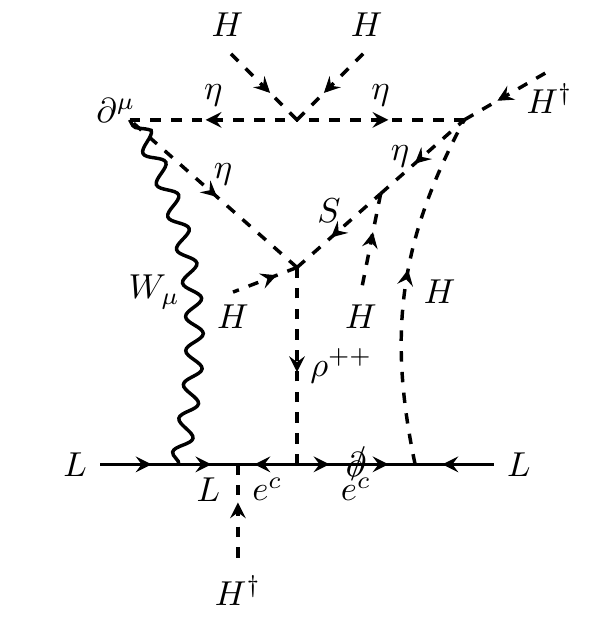}
    \includegraphics[width=0.32\textwidth]{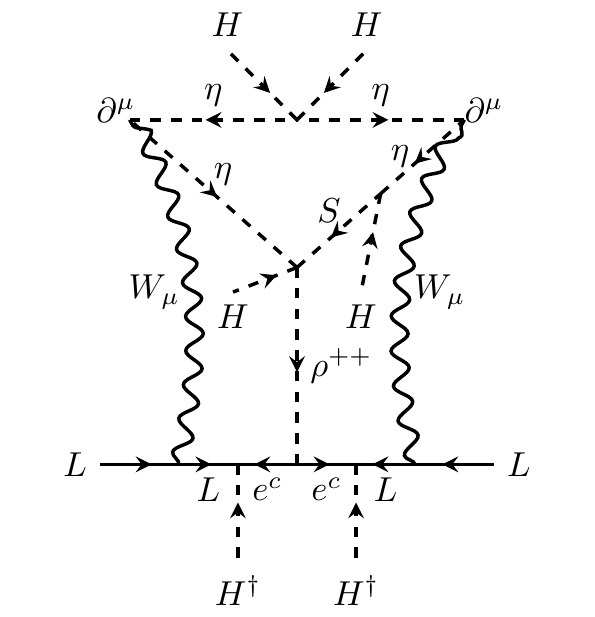}
    \caption{Dimension $9$ mass diagrams in the Feynman-'t Hooft
      gauge. Denoted as $\mathfrak{I}^{(9)}_i$ with $i=1,6$ when
      required, following the standard ordering (left to right and top
      to bottom).
      \label{fig:CT_diag_d9}
      }
\end{figure}

We identified $12$ different diagrams in the gauge basis with dimensions $5$, $7$ and $9$, see Figs.~\ref{fig:CT_diag_d5}-\ref{fig:CT_diag_d9}. All of them are proportional to the mass of the charged leptons squared and with two derivatives. Naively, one could expect that the dominant contribution comes from the dimension $5$ diagrams. However, as we are considering $4\pi$ couplings and lowering the new physics scale as much as possible, all $12$ diagrams could be in principle relevant.

\begin{figure}[t]
    \centering
    \includegraphics[width=0.4\textwidth]{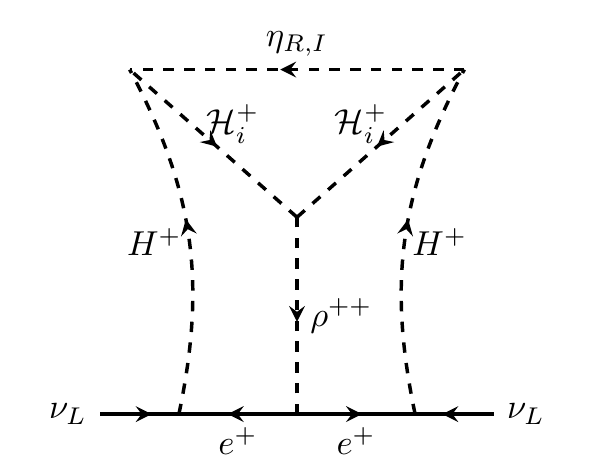}
    \caption{Mass diagram after EWSB in the Feynman-'t Hooft
      gauge. $H^+$ is the Goldstone boson associated to $W_\mu$ with
      mass $m_W$.
      \label{fig:CT_diag_mass}
      }
\end{figure}

We shall show in detail how we derived the integral of the first diagram in Fig.~\ref{fig:CT_diag_d9} as an example, denoted as $\mathfrak{I}^{(9)}_1$, and give just the results for the rest. We chose this diagram as it gives a similar prefactor as in the original work~\cite{Gustafsson:2012vj}. After electroweak symmetry breaking (EWSB), we rotate the diagram to the mass basis, see Fig.~\ref{fig:CT_diag_mass}. $H^+$ is the Goldstone boson associated to $W_\mu$, which appears explicitly with mass $m_W$ in the Feynman-'t Hooft gauge. $\mathcal{H}^+$ are the two mass eigenstates with eigenvalues $m_+^2$ coming form the mixing of $S^+$ and $\eta^+$,
\begin{equation} 
    \mathcal{M}^2_{\mathcal{H}^+} = \left( \begin{array}{cc}
        M_S^2 + \frac 12 \lambda_{SH} v^2  &  \frac{1}{\sqrt{2}} \mu_1 v 
        \\
        \frac{1}{\sqrt{2}} \mu_1 v  &  M_\eta^2 + \frac 12 \left(\lambda_{\eta H}^{(1)} + \lambda_{\eta H}^{(3)} \right) v^2
\end{array} \right) \,,
\end{equation}
which can be trivially diagonalized by a $2 \times 2$ rotation matrix $R_{\mathcal{H}^+}$ with angle $\beta$. $\eta_{R,I}$ are the CP-even and CP-odd components of $\eta^0$, with masses $m_{R,I}^2 = M_\eta^2 \mp \frac 12 \lambda_5 v^2$.

Defining $\int_k \equiv (16\pi^2) \int d^4 k /(2\pi)^3$ and assigning momenta in the loop, the integral of the diagram in Fig.~\ref{fig:CT_diag_mass} in the mass insertion approximation is given by
\begin{multline} \label{eq:I9}
    \mathcal{I}^{(9)}_1 =  (R_{\mathcal{H}^+})_{1i} (R_{\mathcal{H}^+})_{2i} (R_{\mathcal{H}^+})_{1j} (R_{\mathcal{H}^+})_{2j} \,\,\, \times
    \\
   \medmath{
    \iiint\limits_{k_1\, k_2 \, k_3} \, \frac{k_1 \cdot k_2}{ (k_1^2) (k_2^2) (k_1^2+m_W^2) (k_2^2+m_W^2) ((k_1+k_2)^2+m_{\rho^{++}}^2) (k_3^2 + m_{a}^2) ((k_1+k_3)^2 + m_{+i}^2) ((k_2-k_3)^2 + m_{+j}^2) } } \, ,
\end{multline}
where $a=R,I$ and we have neglected the masses of the charged SM fermions. The sum over free indices can be explicitly done enlarging the denominator of the integral. For example,
\begin{equation} 
    \sum_{a=1}^2 \frac{1}{k^2-m_a^2} = (m_1^2 - m_2^2) \frac{1}{(k^2-m_1^2)(k^2-m_2^2)} \, .
\end{equation}
Defining $\Delta m_0^2 = m_R^2- m_I^2$ and $\Delta m_+^2 = m_{+1}^2- m_{+2}^2$, Eq.~\eqref{eq:I9} can be written as,
\begin{equation} 
    \mathcal{I}^{(9)}_1 = \frac 14 \sin^2\! 2\beta \, \Delta m_0^2 \, (\Delta m_+^2)^2 \frac{1}{m_{\rho^{++}}^8} \, \widehat{\mathcal{I}}^{(1)}_1 \, ,
\end{equation}
with $\widehat{\mathcal{I}}^{(1)}_1$ a dimensionless integral defined in Eq.~\eqref{eq:CT_integrals}, which depends only on mass ratios with $m_{\rho^{++}}$,
\begin{equation} \label{eq:CT_massratios}
    x_W = \frac{m_W^2}{m_{\rho^{++}}^2}, \quad x_R = \frac{m_R^2}{m_{\rho^{++}}^2}, \quad x_I = \frac{m_I^2}{m_{\rho^{++}}^2}, \quad x_1 = \frac{m_{+1}^2}{m_{\rho^{++}}^2}, \quad x_2 = \frac{m_{+2}^2}{m_{\rho^{++}}^2} \, .
\end{equation}

Finally, including the corresponding couplings from the potential Eq.~\eqref{eq:PotCocktail}, the expression for the diagram in Fig.~\ref{fig:CT_diag_mass} reads
\begin{equation} 
    \mathfrak{I}^{(9)}_1 = \frac{1}{4} {\lambda_{\eta H}^{(3)}}^2 \, \sin^2\! 2\beta \, \frac{ \mu_2 \,  \Delta m_0^2 \, (\Delta m_+^2)^2}{m_{\rho^{++}}^8} \, \widehat{\mathcal{I}}^{(1)}_1 \, ,
\end{equation}
where the Yukawa $h$ and the SM charged fermion masses have been omitted.

The computation of the rest of the diagrams in Figs.~\ref{fig:CT_diag_d5}-\ref{fig:CT_diag_d9} is very similar to the example shown. We only give the results here and omit their calculation. The function $F_{\rm Cocktail}$ in Eq.~\eqref{eq:mnu cocktail} is given by the sum of the different contributions from the diagrams, i.e.
\begin{equation} 
    F_{\rm Cocktail} = \frac{m_{\rho^{++}}}{\lambda_5} \sum_{d,i} \, \mathfrak{I}^{(d)}_i \, ,
\end{equation}
with $d=5,7,9$ using the notation in Figs.~\ref{fig:CT_diag_d5}-\ref{fig:CT_diag_d9}. The prefactor originates from the normalization of Eq.~\eqref{eq:mnu cocktail}. The corresponding $12$ contributions from each diagram are
\begingroup
\allowdisplaybreaks
\begin{eqnarray} 
    \mathfrak{I}^{(5)}_1 &=& 2 \, \frac{\mu_1^2 \, \mu_2 \, \Delta m_0^2}{v^2 m_{\rho^{++}}^4 } \left[ \cos^4\!\beta \, \frac{(\Delta m_+^2)^2}{m_{\rho^{++}}^4} \, \widehat{\mathcal{I}}^{(1)}_1 \, + 2 \cos 2\beta \, \frac{\Delta m_+^2}{m_{\rho^{++}}^2} \, \widehat{\mathcal{I}}^{(1)}_2 \, + \, \widehat{\mathcal{I}}^{(1)}_3 \right] ,
    \\
    \mathfrak{I}^{(5)}_2 &=& 4 \, \kappa \, \frac{\mu_1 \, \Delta m_0^2}{v^2 m_{\rho^{++}}^2 } \left[ \cos^2\!\beta \, \frac{\Delta m_+^2}{m_{\rho^{++}}^2} \, \widehat{\mathcal{I}}^{(1)}_4 \, + \, \widehat{\mathcal{I}}^{(1)}_5 \right] ,
    \\
    \mathfrak{I}^{(7)}_1 &=& \sqrt{2} \, \lambda_{\eta H}^{(3)} \, \frac{\mu_1 \, \mu_2 \, \Delta m_0^2 \, \Delta m_+^2}{v \, m_{\rho^{++}}^6 } \, \sin\!2\beta \, \left[ \cos\!\beta \, \sin\!2\beta \, \frac{\Delta m_+^2}{m_{\rho^{++}}^2} \, \widehat{\mathcal{I}}^{(1)}_1 \, + \, \widehat{\mathcal{I}}^{(1)}_2 \right] ,
    \\
    \mathfrak{I}^{(7)}_2 &=& - \frac 12 \, g_2^2 \, \frac{\mu_1 \, \mu_2 \, \Delta m_0^2 \, \Delta m_+^2}{v \, m_{\rho^{++}}^6 } \, \sin\!2\beta \, \left[ \cos\!\beta \, \sin\!2\beta \, \frac{\Delta m_+^2}{m_{\rho^{++}}^2} \, \widehat{\mathcal{I}}^{(1)}_1 \, + \, \widehat{\mathcal{I}}^{(1)}_2 \right] ,
    \\
    \mathfrak{I}^{(7)}_3 &=& \sqrt{2} \, \lambda_{\eta H}^{(3)} \, \kappa \, \frac{\Delta m_0^2 \, \Delta m_+^2}{v\, m_{\rho^{++}}^4} \, \sin 2\beta \, \widehat{\mathcal{I}}^{(1)}_4 \, ,
    \\
    \mathfrak{I}^{(7)}_4 &=& - \frac 12 \, g_2^2 \, \kappa \, \frac{\Delta m_0^2 \, \Delta m_+^2}{v\, m_{\rho^{++}}^4} \, \sin 2\beta \, \widehat{\mathcal{I}}^{(2)}_4 \, ,
    \\
    \mathfrak{I}^{(9)}_1 &=& \frac 14 {\lambda_{\eta H}^{(3)}}^2 \, \frac{ \mu_2 \,  \Delta m_0^2 \, (\Delta m_+^2)^2}{m_{\rho^{++}}^8} \, \sin^2\! 2\beta \, \widehat{\mathcal{I}}^{(1)}_1 \, ,
    \\
    \mathfrak{I}^{(9)}_2 &=& - \frac{1}{4\sqrt{2}} \, g_2^2 \, \lambda_{\eta H}^{(3)} \, \frac{ \mu_2 \, \Delta m_0^2 \, (\Delta m_+^2)^2}{m_{\rho^{++}}^8} \, \sin^2\! 2\beta \, \widehat{\mathcal{I}}^{(2)}_1 \, ,
    \\
    \mathfrak{I}^{(9)}_3 &=& \frac{1}{32} \, g_2^4 \, \frac{ \mu_2 \,  \Delta m_0^2 \, (\Delta m_+^2)^2}{m_{\rho^{++}}^8} \, \sin^2\! 2\beta \, \widehat{\mathcal{I}}^{(3)}_1 \, ,
    \\
    \mathfrak{I}^{(9)}_4 &=& -\frac{1}{\sqrt{2}} \, {\lambda_{\eta H}^{(3)}}^2 \, \kappa \, \frac{v \, \Delta m_0^2 \, \Delta m_+^2}{m_{\rho^{++}}^6 } \, \sin\!2\beta \, \left[ \cos\!\beta \, \sin\!2\beta \, \frac{\Delta m_+^2}{m_{\rho^{++}}^2} \, \widehat{\mathcal{I}}^{(1)}_1 \, + \, \widehat{\mathcal{I}}^{(1)}_2 \right] ,
    \\
    \mathfrak{I}^{(9)}_5 &=& \frac 14 \, g_2^2 \, \lambda_{\eta H}^{(3)} \, \kappa \, \frac{v \, \Delta m_0^2 \, \Delta m_+^2}{m_{\rho^{++}}^6 } \, \sin\!2\beta \, \left[ \cos\!\beta \, \sin\!2\beta \, \frac{\Delta m_+^2}{m_{\rho^{++}}^2} \, \widehat{\mathcal{I}}^{(1)}_1 \, + \, \widehat{\mathcal{I}}^{(1)}_2 \right] ,
    \\
    \mathfrak{I}^{(9)}_6 &=& -\frac{1}{4\sqrt{2}} \, g_2^4 \, \kappa \, \frac{v \, \Delta m_0^2 \, \Delta m_+^2}{m_{\rho^{++}}^6 } \, \sin\!2\beta \, \left[ \cos\!\beta \, \sin\!2\beta \, \frac{\Delta m_+^2}{m_{\rho^{++}}^2} \, \widehat{\mathcal{I}}^{(1)}_1 \, + \, \widehat{\mathcal{I}}^{(1)}_2 \right] ,
\end{eqnarray}
\endgroup
For simplicity we introduced the notation
\begin{equation} \label{eq:CT_integrals}
    \widehat{\mathcal{I}}^{(a)}_i = \iiint\limits_{k_1\, k_2 \, k_3} \, \frac{ \mathcal{N}_a }{\mathcal{D}_i} \, ,
\end{equation}
where each numerator, associated to the derivatives depicted in the diagrams, is defined as
\begin{eqnarray} 
    \mathcal{N}_1 &=& k_1 \cdot k_2 \, ,
    \nonumber
    \\
    \mathcal{N}_2 &=& k_2 \cdot (2 k_3 + k_1 ) \, ,
    \\ \nonumber
    \mathcal{N}_3 &=& (2 k_3 + k_1) \cdot (2 k_3 + k_2 ) \, ,
\end{eqnarray}
while for the denominators,
\begin{eqnarray} 
    \mathcal{D}_0 &=& (k_1^2) (k_2^2) (k_1^2+x_W) (k_2^2+x_W) ((k_1+k_2)^2+1) (k_3^2 + x_R) (k_3^2 + x_I) \, ,
    \nonumber
    \\ \nonumber
    \mathcal{D}_1 &=& \mathcal{D}_0 \times ((k_1+k_3)^2 + x_1) ((k_2-k_3)^2 + x_1) ((k_1+k_3)^2 + x_2) ((k_2-k_3)^2 + x_2) \, ,
    \\ \nonumber
    \mathcal{D}_2 &=& \mathcal{D}_0 \times ((k_1+k_3)^2 + x_1) ((k_1+k_3)^2 + x_2) ((k_2-k_3)^2 + x_2) \, ,
    \\ 
    \mathcal{D}_3 &=& \mathcal{D}_0 \times ((k_1+k_3)^2 + x_1) ((k_2-k_3)^2 + x_2) \, ,
    \\ \nonumber
    \mathcal{D}_4 &=& \mathcal{D}_0 \times ((k_2-k_3)^2 + x_1) ((k_2-k_3)^2 + x_2) \, ,
    \\ \nonumber
    \mathcal{D}_5 &=& \mathcal{D}_0 \times ((k_2-k_3)^2 + x_2) \, ,
\end{eqnarray}
with the mass ratios $x$ defined in Eq.~\eqref{eq:CT_massratios}. The
integrals in Eq.~\eqref{eq:CT_integrals} are evaluated numerically
using {\tt pySecDec}.  \\

We now discuss the maximization of $F_{\rm Cocktail}$. We are
interested in this case, because the Yukawa $h$ in Eq.~\eqref{eq:hfit}
is inversely proportional to $F_{\rm Cocktail}$, and we want to
explore the parameter space where $h$ is small enough to be
perturbative and avoid CLFV constraints.  In general, every integral
$\widehat{\mathcal{I}}^{(a)}_i$ gets larger for smaller masses or,
equivalently, smaller ratios. We shall fix a limit of $100$ GeV on the
scalar masses of $\eta_{R,I}$ and $\mathcal{H}^+$, and $800$ GeV for
the doubly charged singlet $\rho^{++}$, see Sec.~\ref{sec:cocktail}
for details. We set the dimensionless couplings $\lambda_5$,
$\lambda_{\eta H}^{(3)}$, and $\kappa$ to $4\pi$. For the dimensionful
$\mu$ couplings we impose the limits $\mu_1 < 4 \, \text{max}[m_{+1},
  m_{+2}]$ and $\mu_2 < 4 \, \text{max}[m_{+1}, m_{+2},
  m_{\rho^{++}}]$, required to avoid the radiative generation of
negative quartic scalar couplings~\cite{Babu:2002uu}.

We found the maximum value of $F_{\rm Cocktail}^{\rm max} \simeq 192$ for $m_{R} = m_{+1} = 100$ GeV, $m_{\rho^{++}} = 800$ GeV, $m_{I} = 878$ GeV, and $m_{+2} = 1237$ GeV, with maximal mixing angle $\beta=\pi/4$. $\mu_1 = (\Delta m_+^2)/\sqrt{2} v = 4372$ GeV, while $\mu_2$ is simply $4\,m_{+2}$.

\subsection{KNT model}
\label{app:int-KNT}

\begin{figure}
    \centering
    \includegraphics[width=0.5\textwidth]{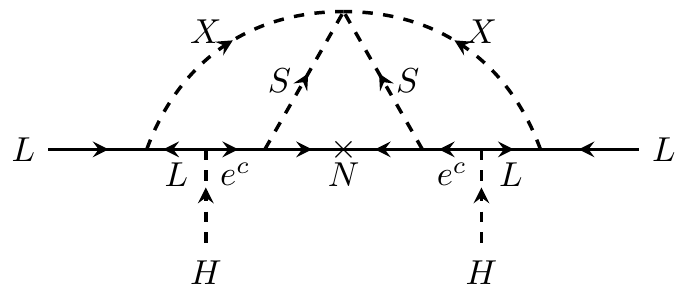}
    \caption{Neutrino mass diagram for the KNT model in the gauge
      basis.
      \label{fig:knt_diag}
      }
\end{figure}

The computation of the mass diagram for the KNT model is much simpler than the one in the cocktail model. The main contribution to the neutrino mass comes only from the diagram in Fig.~\ref{fig:knt_diag} in the electroweak symmetric basis. Moreover, there is no mixing between the scalars participating in the loop, so $F_{\rm KNT}$ in Eq.~\eqref{eq:MnuKNT} is just the 3-loop integral of the diagram in the mass basis shown in Fig.~\ref{fig:KNT}. Neglecting the SM charged fermion masses, one finds 
\begin{equation} \label{eq:FKNT}
    F_{\rm KNT} = \iiint\limits_{k_1\, k_2 \, k_3} \frac{1}{ (k_1^2) (k_2^2) (k_1^2+x_1) (k_2^2+x_1) (k_3^2 + 1) ((k_1-k_3)^2 + x_2) ((k_2-k_3)^2 + x_2) } \, , 
\end{equation}
which is a dimensionless function of the ratios,
\begin{equation} 
    x_1 = \frac{m_{s_1}^2}{M_{N_i}^2} \, , \quad x_2 = \frac{m_{s_2}^2}{M_{N_i}^2} \, .
\end{equation}
Note that Eq.~\eqref{eq:FKNT} is simple enough to be easily decomposed in terms of 3-loop master integrals~\cite{Martin:2016bgz}. Due to the repetitions of the momenta in the denominator, using relation (22) from~\cite{Cepedello:2018rfh}, one has
\begin{multline}
    F_{\rm KNT} = \iiint\limits_{k_1\, k_2\, k_3} \frac{1}{ (k_3^2-1) ((k_1-k_3)^2-x_2) ((k_2-k_3)^2-x_2) } \,\, \times
    \\
    \frac{1}{x_1} \left[ \frac{1}{ (k_1^2-x_1) (k_2^2-x_1) } - \frac{1}{ (k_1^2) (k_2^2-x_1) } - \frac{1}{ (k_1^2-x_1) (k_2^2) } + \frac{1}{ (k_1^2) (k_2^2) } \right] .
\end{multline}
As the second and third terms are identical under the exchange of $k_1$ and $k_2$, one can finally write $F_{\rm KNT}$ in terms of a combination of the master integral $\mathbf{G}$ integral given in~\cite{Martin:2016bgz}. The resulting expression is
\begin{equation}
    F_{\rm KNT} = \frac{1}{x_1} \left[ \mathbf{G}(1,x_1,x_2,x_1,x_2) - 2 \, \mathbf{G}(1,x_1,x_2,0,x_2) + \mathbf{G}(1,0,x_2,0,x_2) \right].
\end{equation}
The integral has an analytical expression for $x_{1i}=x_{2i}=1$.
\\

About the maximum value of $F_{\rm KNT}$, we proceeded analogously to the cocktail model. In this case, we maximized $F_{\rm KNT} / M_{N_i}$, since the neutrino mass matrix in Eq.~\eqref{eq:MnuKNT} is proportional to this ratio. We set a lower limit on the mass of the singly charged scalars of $100$ GeV and let $M_{N_i} = M_N$ free. We found that the maximum is around $F_{\rm KNT} \simeq 60$ with $m_{S_1} = m_{S_2} = 100$ GeV and $M_{N_i} = 840$ GeV.

\subsection{AKS model}
\label{app:int-AKS}

\begin{figure}
    \centering
    \includegraphics[width=0.45\textwidth]{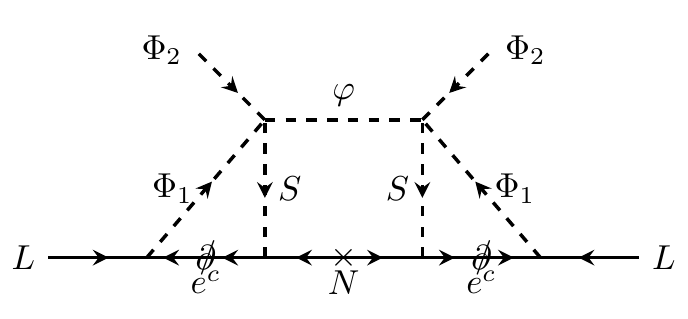}
    \quad
    \includegraphics[width=0.45\textwidth]{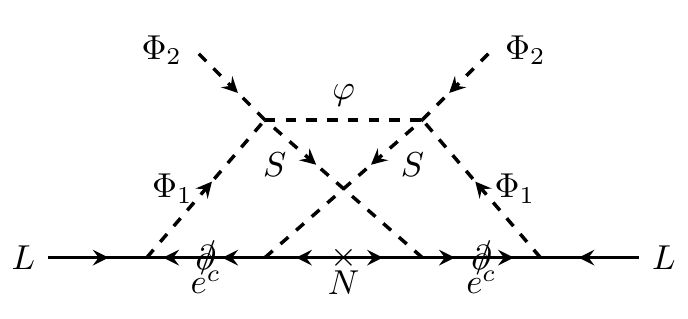}
    \caption{Neutrino mass diagrams for the AKS model in the gauge
      basis.
      \label{fig:aks_diag}
      }
\end{figure}

In this case, there exists two non-equivalent diagrams shown in Fig.~\ref{fig:aks_diag}, which differ by the crossing of the internal $S$-lines. $F_{\rm AKS}$ is then the sum of the integrals from both diagrams with the correct normalization, given in Eq.~\eqref{eq:Mnu AKS},
\begin{equation} 
    F_{\rm AKS} = \mathfrak{I}_1 \, + \, \mathfrak{I}_2 \, .
\end{equation}
By assigning momenta to the internal fields, the two dimensionless integrals can be compactly expressed as,
\begin{equation} 
    \mathfrak{I}_i = \iiint\limits_{k_1\, k_2\, k_3} \frac{k_1 \cdot k_2}{\mathcal{D}_i} \, ,
\end{equation}
with the denominators,
\begin{eqnarray} 
    \mathcal{D}_0 &=& (k_1^2) (k_1^2-x_1) (k_2^2) (k_2^2-x_1) (k_3^2-1) ((k_1-k_3)^2-x_S) ((k_2+k_3)^2-x_S) \, ,
    \nonumber\\
    \mathcal{D}_1 &=& \mathcal{D}_0 \, \times \, (k_3^2-x_\varphi) \, ,
    \\ \nonumber
    \mathcal{D}_2 &=& \mathcal{D}_0 \, \times \, ( (k1+k_2+k_3)^2-x_\varphi) \, ,
\end{eqnarray}
where we have neglected the SM charged fermion masses. The ratios of masses are then defined as
\begin{equation} 
    x_1 = \frac{m_\mathcal{H}^2}{M_{N_i}^2} \, , \quad x_\varphi = \frac{m_\varphi^2}{M_{N_i}^2} \, , \quad x_S = \frac{m_{S^+}^2}{M_{N_i}^2} \,.
\end{equation}
\\

Similar to the previous models, we computed the maximum of the function $F_{\rm AKS}/M_{N_i}$ to minimize the absolute scale of the Yukawa $Y$. We considered $M_{N_i} = m_N$ and set a lower limit of $100$ GeV to the scalar masses. We found the maximum for $m_N = 272$ GeV and $m_\mathcal{H} = m_\varphi = m_S = 100$ GeV where $F_{\rm AKS} \simeq 0.45$.

\section*{Acknowledgements}

Work supported by the Spanish grants FPA2017-85216-P
(MINECO/AEI/FEDER, UE), SEJI/2018/033 and PROMETEO/2018/165 grants
(Generalitat Valenciana), and FPA2017-90566-REDC (Red Consolider
MultiDark).
RC is supported by the FPU15/03158 fellowship.
AV acknowledges financial support from MINECO through the Ramón y Cajal
contract RYC2018-025795-I.

\bibliographystyle{utphys}
\bibliography{refs}

\end{document}